\documentclass[12pt]{article}
\usepackage{amsmath}
\usepackage{amsthm}
\usepackage{amsfonts}
\usepackage{graphicx}
\usepackage{xcolor}
\usepackage{mathtools}
\usepackage{float}
\usepackage[ruled,vlined]{algorithm2e}
\usepackage{enumerate}
\usepackage{subfig}
\usepackage{caption}
\usepackage[hidelinks,colorlinks=true,linkcolor=blue,citecolor=blue, citebordercolor = blue]{hyperref}
\usepackage{colortbl }
\usepackage{pdflscape}
\usepackage{verbatim}
\usepackage{natbib}
\usepackage{footnote}
\usepackage{authblk}

\usepackage{enumitem}
\def\thick#1{\hbox{\rlap{$#1$}\kern0.25pt\rlap{$#1$}\kern0.25pt$#1$}}

\SetKwInput{kwInit}{Init}

\usepackage{graphicx}
\usepackage{float} 
\usepackage{tikz}
\usetikzlibrary{shapes,arrows,decorations.pathmorphing,graphs,positioning,backgrounds}

\restylefloat{table}

\begin{document}
	\input amssym.def
	\input amssym.tex
	
	\title{\textbf{\sffamily Assessing univariate and bivariate risks of late-frost and drought using vine copulas: A historical study for Bavaria}}
	\author{Marija Tepegjozova}
	\author[2]{Benjamin F. Meyer}
	\author[2]{Anja Rammig}
	\author[3]{Christian S. Zang}
	\author[1,4]{Claudia Czado}

	\affil[1]{ Technical University of Munich,  TUM School of Computation, Information, and Technology, Munich, Germany} 
	\affil[2]{Technical University of Munich,  TUM School of Life Sciences, Freising, Germany }
	\affil[3]{Weihenstephan-Triesdorf Universitry of Applied Sciences, Professorship of Forests Climate Change, Freising, Germany}
	\affil[4]{Munich Data Science Institute,  Munich, Germany}
	
	\maketitle
	
	In light of climate change's impacts on forests, including extreme drought and late-frost, leading to vitality decline and regional forest die-back, we assess univariate drought and late-frost risks and perform a joint risk analysis in Bavaria, Germany, from 1952 to 2020. Utilizing a vast dataset with 26 bioclimatic and topographic variables, we employ vine copula models due to the data's non-Gaussian and asymmetric dependencies. We use D-vine regression for univariate and Y-vine regression for bivariate analysis, and propose corresponding univariate and  bivariate conditional probability risk measures. We identify "at-risk" regions, emphasizing the need for forest adaptation due to climate change.


\section{Introduction}\label{climvine:intro}
Since the end of the Industrial Revolution, carbon emissions caused by human activity have increased the concentration of carbon dioxide in the atmosphere by nearly 150\% \citep{friedlingsteinGlobalCarbonBudget2022}. The direct result of this are the shifts in long-term weather patterns more commonly referred to as anthropogenic climate change. Climate change is often associated with a change in the average conditions of temperature and precipitation. However, an additional consequence of climate change is the increase in frequency and intensity of extreme climate \citep{fieldManagingRisksExtreme2012, ipccClimateChange2023}. This poses a threat to forest ecosystems when the processes of physiological acclimation can no longer keep up with the novel climate conditions \citep{grossmanPhenologicalPhysiologySeasonal2023, ordonezMappingClimaticMechanisms2016}. Forests play an important role in the global carbon cycle by storing about 45\% of terrestrial carbon and providing an annual net carbon sink of around 1.1 Pg C \citep{bonanForestsClimateChange2008, panLargePersistentCarbon2011}. In Central Europe, natural forests are dominated by European beech (\emph{Fagus sylvatica} L.) \citep{leuschner2017ecology}. Naturally, this tree species would cover more than 60$\%$ of the land surface area of Germany \citep{bohn2003potenzielle}, and it is also widespread across Europe with its distribution ranging from Sicily in the South up to Bergen in Southern Norway, covering approximately $140,000 km^2$ of forested area in total \citep{durrant2016fagus}. European beech has been promoted as a tree species well adapted to the future climate and as the most efficient broad-leaved tree species for climate change mitigation \citep{yousefpourRealizingMitigationEfficiency2018}.

However, recent evidence points to increased susceptibility of beech forests to increasingly dry and hot summers which have been the main effect of climate change in Central Europe in the past 20 years  \citep{spinoniPaneuropeanSeasonalTrends2017}. In the absence of ample water supply, beech forests are susceptible to growth declines, large-scale damage and mortality \citep{scharnweberDroughtMattersDeclining2011, meyerHigherSusceptibilityBeech2020}. Most recently, this has been observed in the wake of two successive drought events in 2018 and 2019 \citep{burasQuantifyingImpactsDrought2019, schuldtFirstAssessmentImpact2020}. Although these conditions are extreme outliers in the current climate, as climate change progresses they will likely occur more regularly \citep{christidis2015dramatically, ipccClimateChange2023, fieldManagingRisksExtreme2012}. In addition to the adverse effects of increasing frequency and intensity of drought, beech ecosystems are also affected by another climate extreme: late-spring frost. Below freezing temperatures in spring, after trees have begun unfurling their leaves, can result in late-frost damage, defoliating large parts of the canopy \citep{dittmarImpactLateFrost2006, menzelPatternsLateSpring2015}. Consequently, affected trees must expend carbohydrate reserves to grow a second canopy before the physiological processes necessary for photosynthesis can resume \citep{dandreaWinterBiteBeech2019}. Somewhat counter intuitively, increasing temperatures may exacerbate spring late-frost risk:  as (mean) temperature rises, the timing of leaf-out shifts – instead of leaves unfurling near the beginning of May, they can develop as early as the beginning of April when the probability of sub-zero minimum temperatures is higher \citep{zohnerLateSpringFrostRisk2020}.  Both types of disturbance through climate extremes inhibit the regular functioning of beech and force the trees to expend stored resources to recuperate at the cost of forest vitality and productivity.
Consequently, the joint occurrence of spring late-frost and drought poses
a significant threat to forest health, multiplying the detrimental effects in comparison to the isolated effect
of one of these climate extremes alone. However, we currently lack basic understanding of the statistical
coupling between drought and spring late-frost as the necessary underpinning for risk assessment and associated forest management recommendations. Thus, our main objective is to quantify the joint  probability of drought and spring late-frost in the historic domain and identify regions that exhibit the highest joint risk of extreme late-frost and drought conditions.

We approach this objective using dependence modelling with  copulas, which have become more
popular in ecological analysis in recent years due to their ability to deal with non-Gaussian data. Climate data and indices derived from climate data often fall into this category, as they frequently belong to bounded or skewed distributions \citep{scholzelMultivariateNonnormallyDistributed2008}. The copula approach is a multivariate modeling approach that can
handle complex dependence structures. It allows for separate modeling of the copula function and its arguments, the univariate marginal distributions functions. As a result, a wide range of dependence structures can be modeled by utilizing different functional forms for both, the copula function and the marginal distribution functions. 
Our proposed work is a  change from previous applications of copulas in ecology, which so
far have focused only on jointly modelling multiple components of the same climate extreme, for example drought severity and drought duration \citep{sarhadi2016time, kwon2016copula} or frost severity and duration \citep{chatrabgoun2020copula}. 
In our case however, we are interested in joint modeling of two extremes given a set of possible predictors. However, standard multivariate copulas are not able to handle the high-dimensionality of this problem, thus we propose a vine copula based approach. Vine copulas provide a flexible way of high-dimensional copula construction that uses a set of  (conditional) bivariate (pair) copulas as building blocks.  These pair copulas can be chosen independent of each other, based on the (conditional) dependence characteristics of each pair of variables. This flexible model can
handle high-dimensional data with asymmetric dependencies and tail-dependencies,
which is the main benefit of its usage.

When quantifying the joint probability of drought and late spring-frost occurrences, and especially when relying on predictions from this quantification over longer periods, one has to account for extreme  weather events, observed in the tail of a distribution.
To properly quantify these tail events we propose to use joint regression modeling of drought and late spring-frost based on a specific R-vine copula, able to jointly model two responses with a symmetric treatment, the Y-vine copula based regression \citep{tepegjozova2022bivariate}. In addition, we model drought and late spring-frost  separately with a D-vine regression model \citep{kraus2017d}, which is appropriate for univariate responses. By comparing the marginal and joint effects of changes in these variables, we can better understand their individual and combined impacts.
In addition, we propose novel conditional risk measures derived from the vine copula based regression models. These measures help identify spatial and temporal "at-risk" regions for the forest ecosystems. We are also able to estimate survival probabilities to identify spatial regions "at-risk" over longer periods and return periods of extreme events to identify temporal "at-risk" regions. To the best of our knowledge, we are the first to explore the model capabilities of vine copulas in such a climatological context.

This manuscript is based on Chapter 6 of the first author's PhD thesis \cite{dissertationMarija}. The manuscript is organised as follows: Section  \ref{climvine:datasesc} describes the data set utilized.
Section \ref{climvine:modeling} introduces the data modeling approaches we use, the D-vine  and Y-vine copula regression.  It also contains an exploratory dependence analysis of the data at hand and an exploration of the  fitted models, by studying the pair copula families that are fitted and the orders of the predictors the models choose.   
Section  \ref{climvine:riskmeasures} propose novel conditional risk probability measures for the D-vine and Y-vine regression models. These vine copula based risk measures are used to identify high risk years and regions, for both univariate and bivariate responses. We also estimate associated survival probabilities and analyse how these conditional probabilities vary over all locations in Section  \ref{climvine:returntimes}. Based on the survival probabilities, we are also able to estimate return periods for each extreme. All these measures are used for finding temporal and spatial "at-risk" regions. Finally, we conclude and propose possible areas of future research in Section \ref{climvine:conclusionoutlook}.

\section{Data description}\label{climvine:datasesc}
To quantify changing drought and frost risk we use a late-frost index and a drought index rather than raw climate variables. We calculate these indices using the BayObs product, a multivariate, gridded climate data set covering Bavaria at a spatial resolution of 5km by 5km, provided by the Bavarian Environment Agency (LfU). The dataset contains daily minimum air temperature, daily maximum air temperature, daily mean air temperature, and daily precipitation sum from 1952 until 2020 \citep{lfu2020} for all grid cells. \\

\subsection{Late-frost and drought index}

\noindent To quantify frost risk we use a modified version of the Frost Index in April (FI4) proposed by \citet{sanguesa-barredaWarmerSpringsHave2021}. The original FI4 takes into account mean and minimum temperatures between mid-April and mid-May, a time period which generally marks the beginning of leaf-unfolding in European beech. In contrast, our modified index, Frost Index at Leaf-Out (FILO) uses a phenological model to more accurately pinpoint the begin of leaf-unfolding. We use the phenological model outlined in \citet{kramerChillingForcingRequirements2017}. A frost index having a value of 0 indicates average conditions (i.e. average frost risk), positive values indicate a lower frost risk, and negative values indicate a higher frost risk. 

\noindent To determine drought risk we use the Standardized Precipitation Evapotranspiration Index (SPEI). This index describes the relative water availability at a given site and time as a function of precipitation and potential evapotranspiration, i.e. the difference between water supply and water demand \citep{vicente-serranoMultiscalarDroughtIndex2010, begueriaStandardizedPrecipitationEvapotranspiration2014}. Negative SPEI values indicate drier-than-average conditions while positive values indicate wetter-than-average conditions. The SPEI is standardized across the entire period of historical climate data available in the BayObs data set (1952-2020). Here, we focus on the SPEI-6 in August, that is, the SPEI integrated over August and the preceding 5 months. This allows us to take into account medium-term droughts spanning from early spring to the height of summer which have been associated to drought related growth responses in European beech \citep{bhuyanDifferentResponsesMultispecies2017}. 

\subsection{Climatic and topographic predictors}

\noindent Previous studies have identified the possible effect of factors such as elevation, aspect, annual precipitation, and mean annual temperature on the spatial incidence of late-frost events \citep{olanoSatelliteDataMachine2021}. The effect of a lack of precipitation for causing drought is locally modified by temperature conditions and the topographic situation  \citep{bhuyanDifferentResponsesMultispecies2017, vanloonHydrologicalDroughtExplained2015}. To identify factors which influence both late-frost and drought we utilized a set of bioclimatic indices as well as a set of topographic indices. The bioclimatic indices are based on the bioclimatic variables derived from the WordClim database \citep{fickWorldClimNewKm2017, hijmansVeryHighResolution2005}. Since we are interested in intra-annual fluctuations of precipitation and temperature patterns, we derive these indices on a yearly basis. We first aggregated our daily climate data (precipitation, min. temperature, max. temperature, mean temperature) to monthly values. Subsequently, we calculated the annual bioclimatic variables using the R package \texttt{dismo}  \citep{hijmansDismoSpeciesDistribution2021}.

For the topographic predictors (elevation, slope, aspect), we extracted relevant terrain information from the digital surface model (DSM) EU-DEM v1.0 provided by the European Environment Agency (EEA) under the Copernicus program (publicly available at \hyperref[link]{http://land.copernicus.eu/pan-european/satellite-derived-products/eu-dem/eu-dem-v1-0-and-derived-products/eu-dem-v1.0/view}). 
We reprojected the EU-DEM from its native resolution of 25 m and ETRS89 reference system to a resolution of 5 km and a WGS84 reference system to match our climate data. We then extracted slope and aspect information from the DSM using the R package  \texttt{terra} \citep{hijmansTerraSpatialData2022}. Also, we include the gridcell specific location by including latitude and longitude in the model. 
\begin{table}
	\centering
	\begin{tabular}{ r r   } 
		Variable name & Description\\
		\hline
		\hline
		\multicolumn{2}{c}{\textit{Responses}}  \\
		\hline
		\hline
		frost & late-frost index at leaf out\\
		drought & drought index\\
		\hline
		\hline
		\multicolumn{2}{c}{\textit{Bioclimatic variables}}  \\
		\hline
		\hline
		\multicolumn{2}{c}{\textit{Temperature related  variables}}  \\
		\hline
		temp\_mean & Annual mean temperature (°C)\\
		temp\_diu &  Mean diurnal range (mean of monthly max
		temp - min temp) (°C)\\
		isotherm &  Isothermality (°C)\\
		temp\_season &  Temperature seasonality (°C)\\
		temp\_max  &  Temperature of warmest month (°C)\\
		temp\_min  &  Temperature of coldest month (°C)\\
		temp\_season & Temperature annual range (°C)\\
		temp\_wet &  Mean temperature of wettest quarter (°C)\\
		temp\_dry & Mean temperature of driest quarter (°C)\\
		temp\_warm &  Mean temperature of warmest quarter (°C)\\
		temp\_cold & Mean temperature of coldest quarter (°C)\\
		\hline
		\multicolumn{2}{c}{\textit{Precipitation related  variables}}  \\
		\hline
		preci &  Total (annual) precipitation (mm)\\
		preci\_wet\_m &  Precipitation of wettest month (mm)\\
		preci\_dry\_m &  Precipitation of driest month (mm)\\
		preci\_season &  Precipitation seasonality (coefficient of variation)\\
		preci\_wet\_q &  Precipitation of wettest quarter (mm)\\
		preci\_dry\_q &  Precipitation of driest quarter (mm)\\
		preci\_warm &  Precipitation of warmest quarter (mm)\\
		preci\_cold &  Precipitation of coldest quarter (mm)\\
		\hline
		\hline
		\multicolumn{2}{c}{\textit{Topographic variables}} \\
		\hline
		\hline
		elevation& Average elevation above sea level (in meters) \\
		aspect& Aspect of each gridcell in degrees (0° = north) \\
		slope & Average slope of each gridcell in degrees\\
		\hline 
		latitude & latitude of gridcell (WGS 84)\\
		longitude & longitude of gridcell (WGS 84)\\
		\hline
	\end{tabular}
	\caption{\label{tab:widgets}Variable description.}
	
\end{table}

\subsection{Data summary and exploration}

\noindent Overall, we have produced a data set containing annual data for 69 years (1952-2020) for each of the 2867 gridcells considered in the region of Bavaria, Germany. For each year and location (or gridcell) there are 26 available  variables in total (2 responses, 19 bioclimatic predictors and 5 topographic predictors). Thus, in total the data set has a size of $197 823 (=2867 \cdot 69) $ data points. In Table \ref{tab:widgets}  we give a short description of the variables used in our data analysis. Further in Figure S1, given in the supplementary, are exploratory data analysis plots for each variable (apart from the topographic variables). In the left panels, we show the mean of the observations per year over all gridcells (1952-2020), a fitted
moving average model with the associated 95\% confidence interval (CI) for each variable. In the right panels, we give  the empirical 50 \% and  95\% CI per year, together with the annual mean value over all gridcells.

\begin{figure}
	\includegraphics[width = \textwidth]{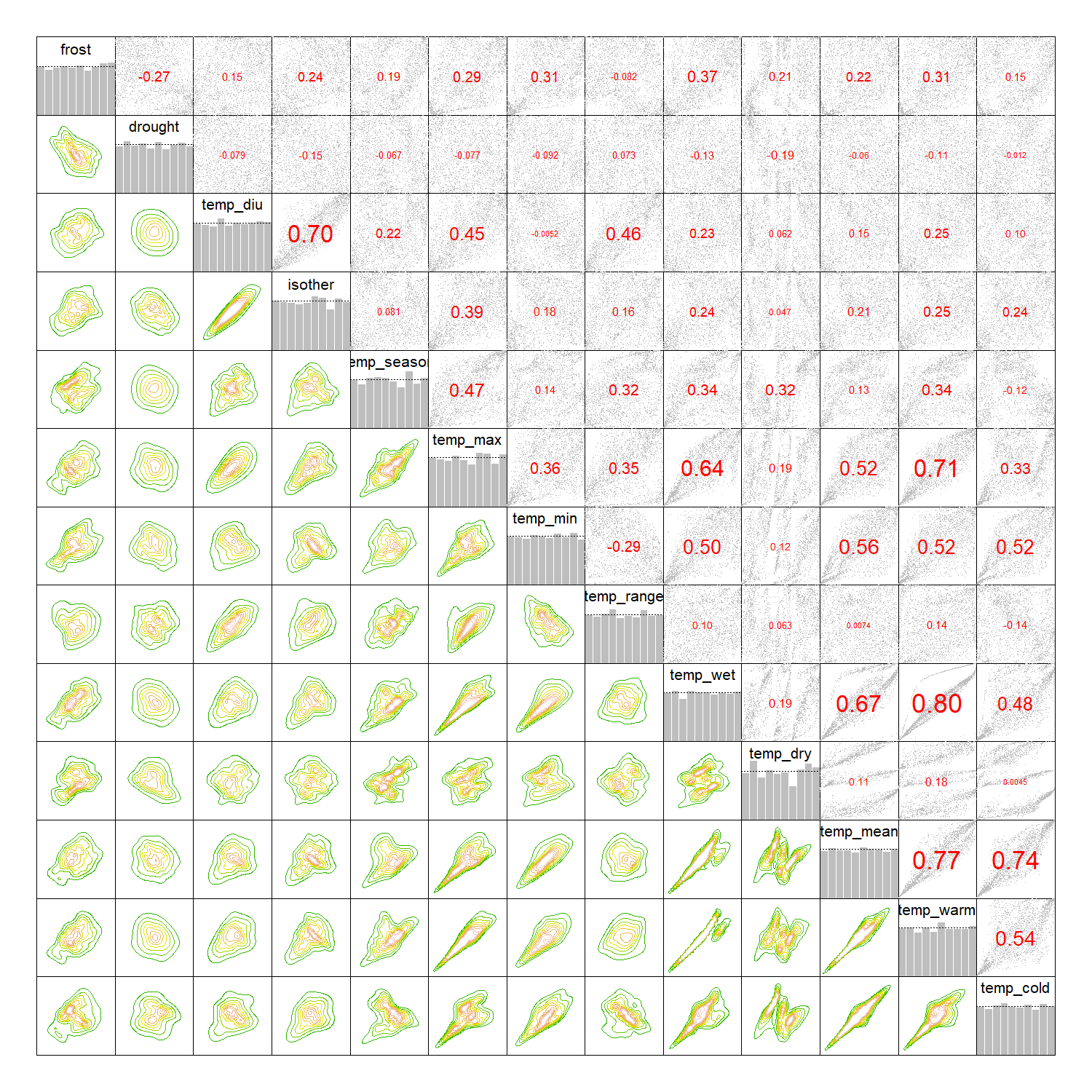}
	\centering
	\caption{Lower diagonal: marginally normalized contour plots, upper diagonal: pairwise scatter plots with the associated empirical Kendall's $\hat{\tau}$ values and on the diagonal: histograms of the u-data, for the two responses (frost, drought) and the \textbf{temperature related} predictor variables  for all 2867 locations in year \textbf{1953}.}
	\label{fig:pairsplot1953_1}
\end{figure}

\begin{figure}
	\includegraphics[width = \textwidth]{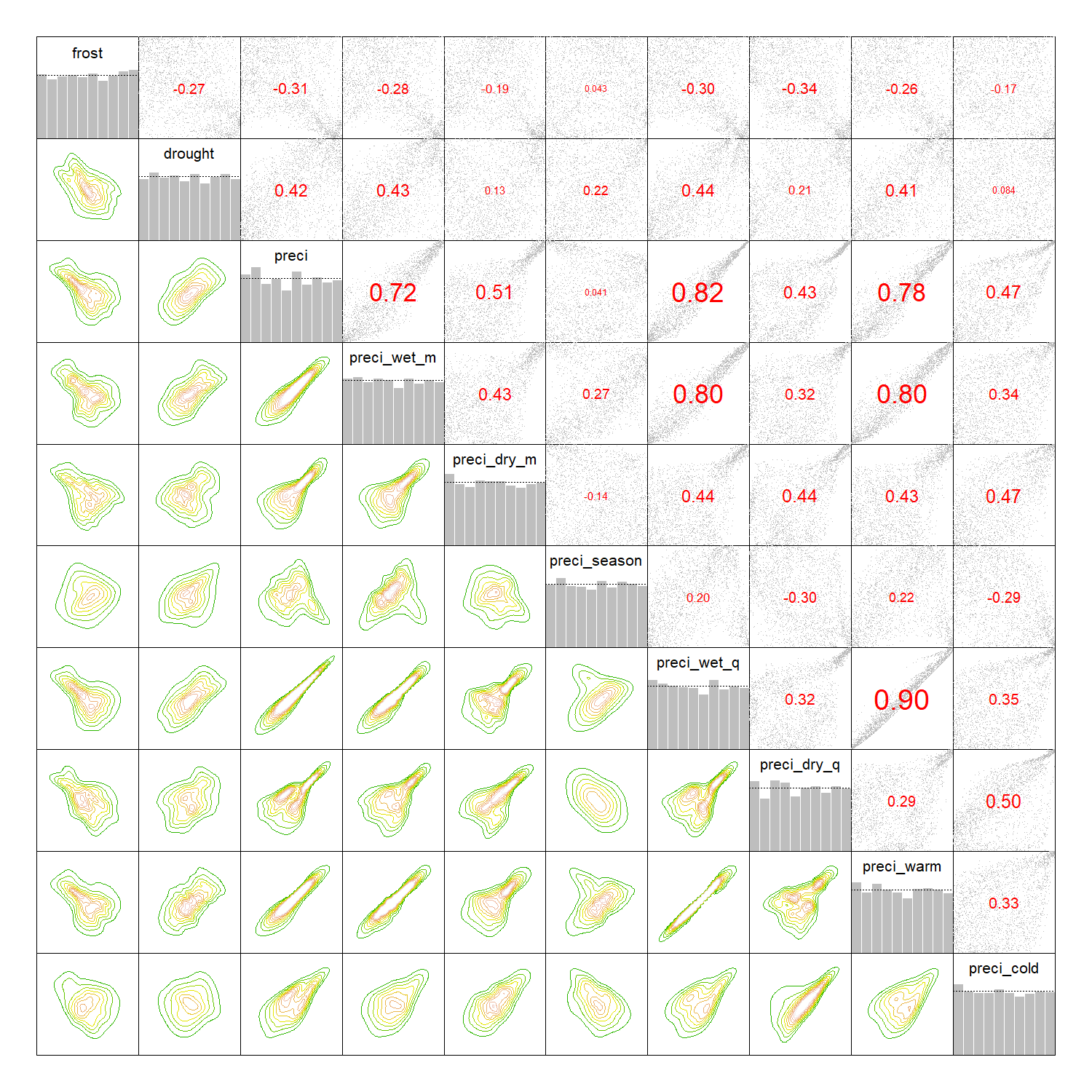}
	\centering
	\caption{Lower diagonal: marginally normalized contour plots, upper diagonal: pairwise scatter plots with the associated empirical Kendall's $\hat{\tau}$ values and on the diagonal: histograms of the u-data,  for the two responses (frost, drought) and the \textbf{precipitation related} predictor variables  for all 2867 locations in year \textbf{1953}.}
	\label{fig:pairsplot1953_2}
\end{figure}

\begin{figure}
	\includegraphics[width = \textwidth]{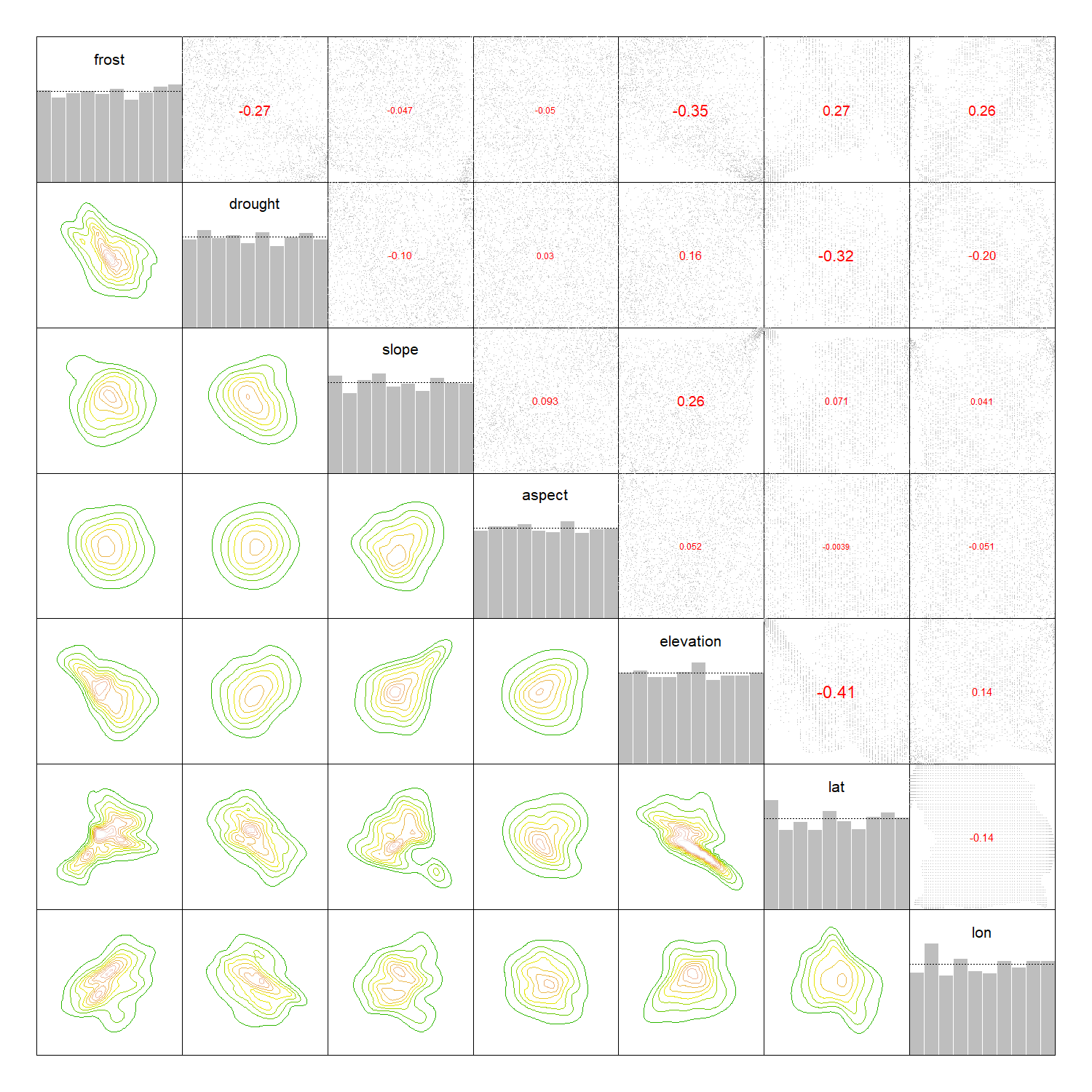}
	\centering
	\caption{Lower diagonal: marginally normalized contour plots, upper diagonal: pairwise scatter plots with the associated empirical Kendall's $\hat{\tau}$ values and on the diagonal: histograms of the u-data,  for the two responses (frost, drought) and the \textbf{topographic} predictor variables  for all 2867 locations in year \textbf{1953}.}
	\label{fig:pairsplot1953_3}
\end{figure}

In addition, Figures \ref{fig:pairsplot1953_1}, \ref{fig:pairsplot1953_2} and \ref{fig:pairsplot1953_3} show the marginally normalized contour plots, where the marginal distributions are fitted nonparametrically, using kernel density smoothing, for a randomly chosen year, at the beginning of our analysis, year 1953.  Each plot is based on all 2867 locations for the two responses and a subset of the predictors. On the lower diagonal, any deviance from elliptical shapes indicates a non-Gaussian dependence structure in the data (see Section 3.8 of \citet{czado2019analyzing} for a precise definition) and we see that almost all panels imply non-Gaussian dependence structures. In addition, on the upper diagonal, we see a scatter plot of the estimated probability integral transformed data (approximately uniformly distributed and denoted as u-data) together with the corresponding estimated
empirical pairwise Kendall’s $\hat{\tau}$. Kendall’s $\tau$ is a rank based dependence measure with range of values in the interval
$[-1, 1]$. Closer values to the boundaries of the interval $[-1, 1]$ mean greater dependence.
Positive values indicate positive dependence, while negative values of the Kendall’s $\tau$ indicate negative dependence between two random variables. In Figures 1-3, we can see many high values of the pairwise Kendall’s $\hat{\tau}$.  Similar results follow for  other years (the associated plots for 2011 are given in the supplement as another example), i.e. majority of non-Gaussian dependence between pairs of variables and high estimated empirical pairwise Kendall’s $\hat{\tau}$ are detectable for all years considered. However, using vine copulas we can efficiently model and capture these high non-Gaussian dependencies between  pairs of variables.

\section{Data modeling}\label{climvine:modeling}

\subsection{Vine copulas}
Let $\mathbf{X}$ be a continuous d-dimensional random vector $\mathbf{X}=\left(X_1,\ldots,X_d\right)^T$ with observed values  $\mathbf{x}=\left(x_1,\ldots,x_d\right)^T$. Assume  that $\mathbf{X}$ has joint distribution function  $F$, joint density $f$ and marginal distributions $F_{X_i},\;i=1,\ldots, d.$ 
The fundamental representation theorem of \citet{sklar1959fonctions} for multivariate distributions in terms of their marginal distributions and a corresponding $d$-dimensional copula $C$, states that
$	F\left(x_1,\ldots,x_d\right) = C\left(   F_{X_1}(x_1),\ldots, F_{X_d}(x_d)\right).
$
The copula $C:\left[0,1\right]^d\mapsto\left[0,1\right]$  corresponds to the distribution function of the random vector $\mathbf{U} = \left(U_1,\ldots ,U_d\right)^T,$ where the components of $\mathbf{U}$  (u-scale) are the probability integral transforms (PITs)  of the components of $\mathbf{X}$  (x-scale),  $U_i = F_{X_i}\left(X_i\right)$ for $i=1,\ldots ,d$. Every $U_i$  is  uniformly distributed and their joint distribution function $C$   is  the  copula  associated  with $\mathbf{X}$.

\citet{joe1996families} has shown that  a $d$-dimensional copula density can be decomposed into $d\left(d-1\right)/2$ bivariate copula densities. However, the decomposition is not unique. 
A graphical model  introduced by \citet{bedford2002vines} called regular vine copulas (R-vines),  organizes all such decompositions that lead to a valid joint density.
Thus, the estimation of any $d$-dimensional copula density can be divided into the estimation of $d (d-1)/2$  two-dimensional pair copula  densities, which  can be chosen completely independent of each other.
A regular vine copula  consists of a regular vine tree sequence (or tree structure), denoted by $\mathcal{V}$, a set of bivariate copula families (also known as pair copulas) $\mathcal{B}\left(\mathcal{V}\right)$, and a set of parameters corresponding to the bivariate copula families $\Theta\left(\mathcal{B}\left(\mathcal{V}\right)\right)$.  Given  $d$ uniformly distributed random variables $U_1,\ldots ,U_d,$ the vine tree sequence  $\mathcal{V}$ consists of a sequence of $d-1$ linked trees, $T_k=\left(N_k,E_k\right),\; k=1,\ldots ,d-1$, satisfying the following conditions:
1. $T_1$ is a tree with node set $N_1=\left\{U_1,\ldots ,U_d\right\}$ and edge set $E_1$;
2. For $k\geq 2$, $T_k$ is a tree with node set $N_k=E_{k-1}$ and edge set $E_k$;
3. (Proximity condition) For $k\geq 2$, two nodes of the tree $T_k$ can  be connected by an edge if the corresponding edges of $T_{k-1}$ have a common node.
The tree sequence uniquely specifies which bivariate (conditional) copula densities occur in the decomposition. To facilitate tractable estimation a simplifying assumption has to be made (see more in \citet{haff2010simplified} and \citet{stoeber2013simplified}). Under this valid joint copulas can be constructed and this assumption we assume throughtout the paper. Foe more details on vine copulas, we refer to \cite{czado2019analyzing} and \cite{joe2014dependence}.

\subsection{Vine copula based regression methods}
Based on regular vine copulas there are associated regression models proposed, for example see \cite{kraus2017d, tepegjozova2021nonparametric, chang2019prediction, zhu2021simplified}. Their main advantages are: no need for transformations or interactions of variables, by construction quantile crossings are avoided and possibility to model complex non-symmetric dependencies. Also, these models do not assume homoscedasticity, or a linear relationship between the response and the predictors.   
For our data, we utilize the D-vine based regression by \cite{kraus2017d} 
for a single response regression and the Y-vine regression developed for a bivariate response  by \cite{tepegjozova2022bivariate}.

\begin{figure}[h!]
	\centering
	\includegraphics[width=0.8\textwidth]{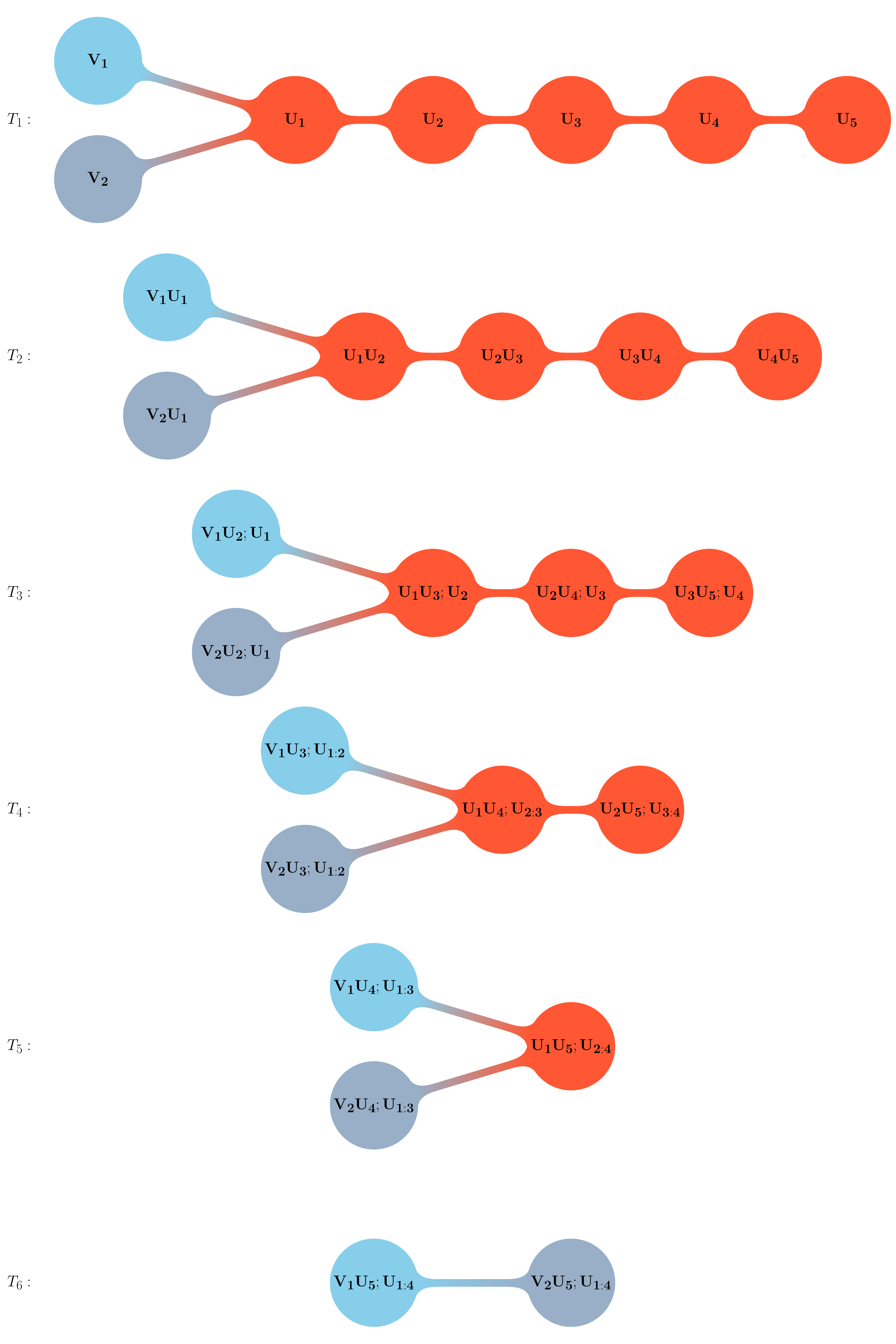} 
	\caption{Y-vine tree sequence with 2 response and 5 predictor variables. }
	\label{fig:yvine}
\end{figure}

\noindent Let $\mathbf{Y} = \left( Y_1,Y_2\right)^T$ and $\mathbf{X} = \left(X_1,\ldots ,X_p\right)^T,\; p\geq 1$, be two continuous random vectors, with corresponding marginal distribution functions $Y_j \sim F_{Y_j},$ for $ j = 1,2$ and $X_i \sim F_{X_i},$ for $ i = 1,\ldots ,p$. The corresponding PITs vectors are $\mathbf{V} = \left( V_1,V_2\right)^T$ for the response vector, and $\mathbf{U}_{1:p} = \left(U_1,\ldots ,U_p\right)^T,\; p\geq 1$ for the predictor vector. We are interested in the stochastic behaviour of the  response/s given the predictors. The target function we want to estimate are the univariate and the bivariate conditional distribution functions $C_{V_1|\mathbf{U}_{1:p}}, C_{V_2|\mathbf{U}_{1:p}}$ and $ C_{V_1,V_2|\mathbf{U}_{1:p}},$ respectively.

\noindent For the estimation of the univariate conditional distribution functions, we use a D-vine based regression model, based on a D-vine tree structure (a tree sequence where each tree is a path). The D-vine is build by optimizing the conditional log-likelihood and adding predictors until there
is no further improvement, thus providing an automatic forward variable selection algorithm. The conditional distribution function of the response given the predictors can be obtained in a closed form, guaranteed by the fact that the response is a leaf node in all trees.
The conditional distributions $C_{V_j|\mathbf{U}_{1:p}}$  for $ j = 1,2$   are expressed with the recursion provided in \cite{joe1996families}
\begin{equation}
	C_{V_j\vert \mathbf{U}_{1:p}}\left( v_j\vert \mathbf{u}_{1:p} \right) = h_{V_j\vert U_p;\mathbf{U}_{1:p-1}}\left(C_{V_j\vert \mathbf{U}_{1:p-1}}\left(v_j\vert \mathbf{u}_{1:p-1} \right) \vert    
	C_{U_p\vert \mathbf{U}_{1:p-1}}\left(u_p\vert \mathbf{u}_{1:p-1} \right) \right),
\end{equation}
where $ h_{V_j\vert U_p;\mathbf{U}_{1:p-1}} = \frac{d}{du_p} C_{V_j, U_p; \mathbf{U}_{1:1-p}}$. By continuing this recursion, only the pair copulas defined in the vine copula are used, thus avoiding possible integration (more details can be found for example in \cite{masterMarija}).

\noindent This bivariate conditional distribution function can be derived from the $p+2$ dimensional copula $C_{V_1,V_2,\mathbf{U}_{1:p}},$ which describes the joint distribution of $(V_1,V_2,\mathbf{U}_{1:p})$. As proposed by \cite{tepegjozova2022bivariate}, we estimate this copula using a special type of R-vines, called Y-vines. Figure \ref{fig:yvine} shows a 7-dimensional Y-vine tree sequence for 2 response and 5 predictor variables. Y-vines are specifically designed so that the conditional distribution function $C_{V_1,V_2|\mathbf{U}_{1:p}}$ can be obtained from the Y-vine copula in a numerically inexpensive way. This is guaranteed since the two responses are leaf nodes in all trees of the Y-vine.  Y-vines allow for a fully symmetric treatment of the two response variables. Also, considering one response only, a D-vine regression is a submodel of the Y-vine model, so due to the advantages  of this bivariate regression model, we opt for the D-vine regression models for the univariate case (consider only only response and the predictors in each tree, the resulting vine sequence is a D-vine tree sequence or sequence of paths). 
Further, the Y-vine is build in a sequential manner, by optimising the conditional bivariate log-likelihood, thus it also has an automatic forward selection of predictors.
As shown in \cite{tepegjozova2022bivariate} the bivariate conditional distribution function $C_{V_1,V_2\vert \mathbf{U}_{1:p}}$ can be written as	
\begin{equation}\label{cond_int}
		C_{V_1,V_2\vert \mathbf{U}_{1:p}}\left(v_1,v_2\vert \mathbf{u}_{1:p}\right) 
		=\int_{0}^{v_2} c_{V_2\vert \mathbf{U}_{1:p}}\left(v_2'\vert \mathbf{u}_{1:p}\right)\cdot C_{V_1 \vert V_2, \mathbf{U}_{1:p}}\left(v_1\vert v_2', \mathbf{u}_{1:p}\right) dv_2' .		
\end{equation}

The conditional densities of the two terms on the right-hand side of Eq.\ref{cond_int} can be derived from the Y-vine in a closed form, using only the pair copulas defined in the Y-cine tree sequence. This is computationally convenient, as otherwise more multiple integration procedures would be required. 
The univariate conditional density $c_{V_2 \vert \mathbf{U}_{1:p}}\left(v_2| \mathbf{u}_{1:p}\right)$ is derived as
\begin{equation}\label{eq:densitycond5}
	c_{V_2 \vert \mathbf{U}_{1:p}}\left(v_2| \mathbf{u}_{1:p}\right)  = 
	\prod_{i=1}^{p} 
	c_{V_2,U_i; \mathbf{U}_{1:i-1} } \left( C_{V_2|\mathbf{U}_{1:i-1} } \left( v_2|\mathbf{u}_{1:i-1}  \right),  C_{U_i|\mathbf{U}_{1:i-1} } \left( u_i|\mathbf{u}_{1:i-1}  \right)    \right),        
\end{equation} 
where $c_{V_2,U_i; \mathbf{U}_{1:i-1}} \left( \cdot, \cdot   \right)$ are pair copula densities contained in the Y-vine.
The univariate conditional density $c_{V_1 \vert V_2, \mathbf{U}_{1:p}}\left(v_1\vert v_2', \mathbf{u}_{1:p}\right)$ can be expressed as

\begin{equation}\label{eq:densitycond4}
	\begin{aligned}
		c_{V_1 \vert \mathbf{U},V_2}\left( v_1| \mathbf{u},v_2\right) = &  
		\prod_{i=1}^{p} \Big[
		c_{V_1,U_i; \mathbf{U}_{1:i-1} } \left( C_{V_1|\mathbf{U}_{1:i-1} } \left( v_1|\mathbf{u}_{1:i-1}  \right),  C_{V_i|\mathbf{U}_{1:i-1} } \left( u_i|\mathbf{u}_{1:i-1}  \right)    \right)  \Big] \\
		& 
		\cdot c_{V_1,V_2; \mathbf{U}_{1:p} } \left( C_{V_1|\mathbf{U}_{1 :p} } \left( v_1|\mathbf{u}_{1:p}  \right), C_{V_2|\mathbf{U}_{1:p} } \left( v_2|\mathbf{u}_{1:p} \right)    \right),
	\end{aligned}
\end{equation}	
where $c_{V_1,U_i; \mathbf{U}_{1:i-1} } \left( \cdot, \cdot \right)$ and $c_{V_1,V_2; \mathbf{U}_{1:p} } \left( \cdot, \cdot \right)$ are also pair copula densities contained in the Y-vine. The  corresponding conditional distribution $C_{V_1 \vert V_2, \mathbf{U}_{1:p}}\left(\cdot \vert \cdot, \cdot\right)$ is obtained by univariate integration of the corresponding conditional density in Eq.\eqref{eq:densitycond4}.
\subsection{Our approach}
In order to examine what is the effect of the possible predictors on the late-frost and drought
indices and how it changes over time,  we fit a D-vine regression model for each of the two responses  and a Y-vine regression model for their joint behavior. The models are fitted for each year separately, using parametric bivariate copula families with a single parameter,  an AIC-penalized log
likelihood selection criteria on the choice of the copula family  (both implemented in the \texttt{R} package \texttt{rvinecopulib}) and the marginal distributions are fitted in a nonparametric manner, using kernel density smoothing (implemented in the \texttt{R} package \texttt{kde1d}). The use of the parametric copula families is due to the need for quantifying and analyzing the tail dependence in the models. Non-parametric marginal distribution estimation is used in order to decrease possible model misspecification bias.  Each model is set to find the 5 most influential predictors for each year. This is done because of computational limitations, due to  the large size of the data set and the number of models to be fitted (in total $69\cdot 3 = 207$ vine copula regression models are fitted). Thus, we have:

\begin{itemize}
	\item \textbf{Data periods:} 69 years, $ t \in \left[ 1952-2020\right]$.
	\item \textbf{Locations:} Per year there are 2867 gridcells where the climatological variables are evaluated, $l \in \left[  1, \ldots, 2867 \right]$. 
	\item \textbf{Spatial effects:} We model  spatial effects in the data by including as possible predictors the gridcell spatial coordinates, i.e. the \texttt{latitude} and \texttt{longitude}. Further, the inclusion of spatially varying predictors at each grid cell also capture spatial effects present in the data.
	\item \textbf{Univariate D-vine models:} For each year, two D-vine regression models are fitted on all 2867 grid points. The ones where the response is the frost index we denote as $\hat{\mathcal{D}}_{{frost}_{1952}}, \ldots, \hat{\mathcal{D}}_{{frost}_{2020}}$. The drought index is a response variable in the models $\hat{\mathcal{D}}_{{drought}_{1952}}, \ldots, \hat{\mathcal{D}}_{{drought}_{2020}}$.
	\item \textbf{Bivariate Y-vine models:} For each year, we fit a joint response Y-vine model over all grid points. The fitted $Y$-vine models we denote as $\hat{\mathcal{Y}}_{1952}, \ldots, \hat{\mathcal{Y}}_{2020}$.
	%
\end{itemize}

\subsection{Model  analysis}

\begin{figure}
	\centering
	\begin{tikzpicture}

		\node[inner sep=0pt] (Col1) at (-25, 8)  {\includegraphics[width=0.8\textwidth]{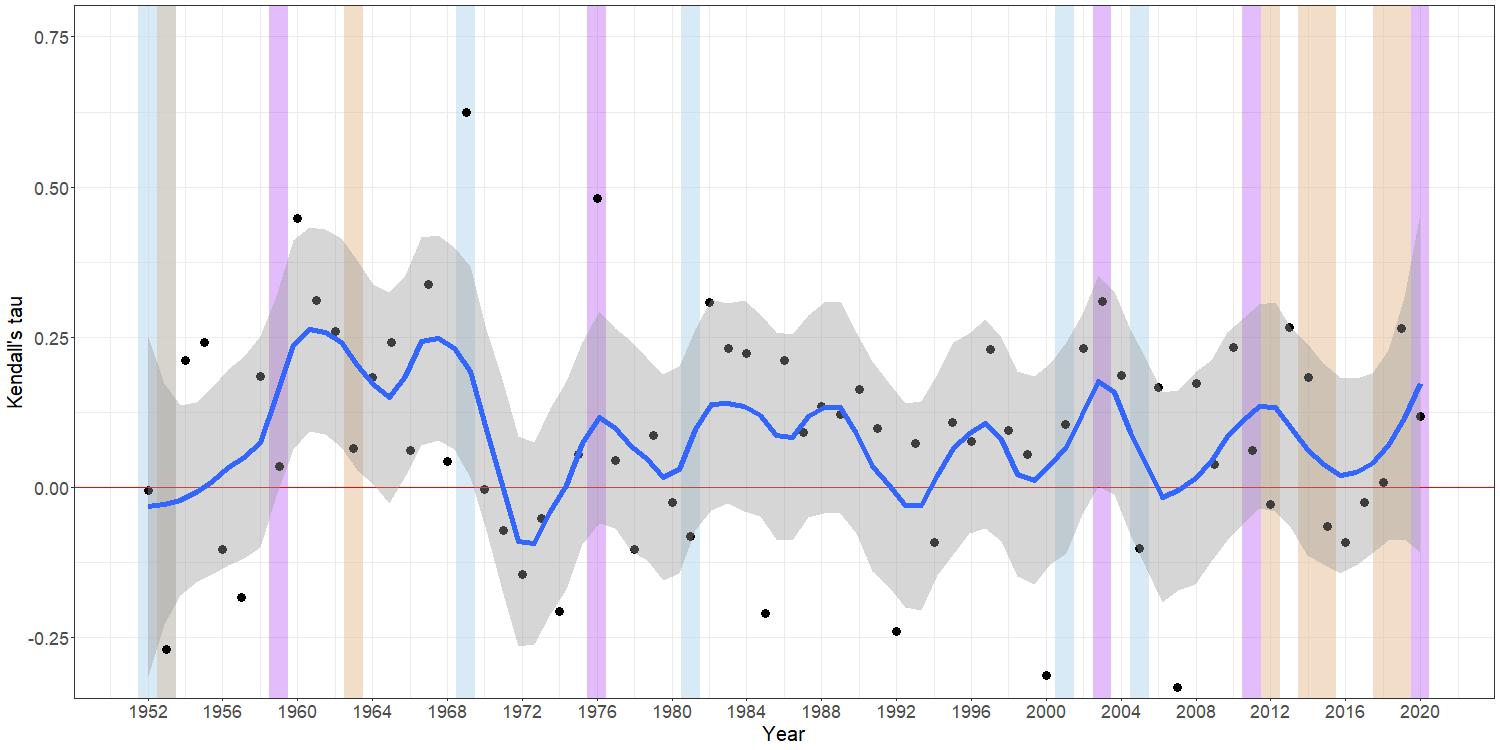}};
		
		\node[inner sep=0pt] (Col1) at (-25, 2)  {\includegraphics[width=0.8\textwidth]{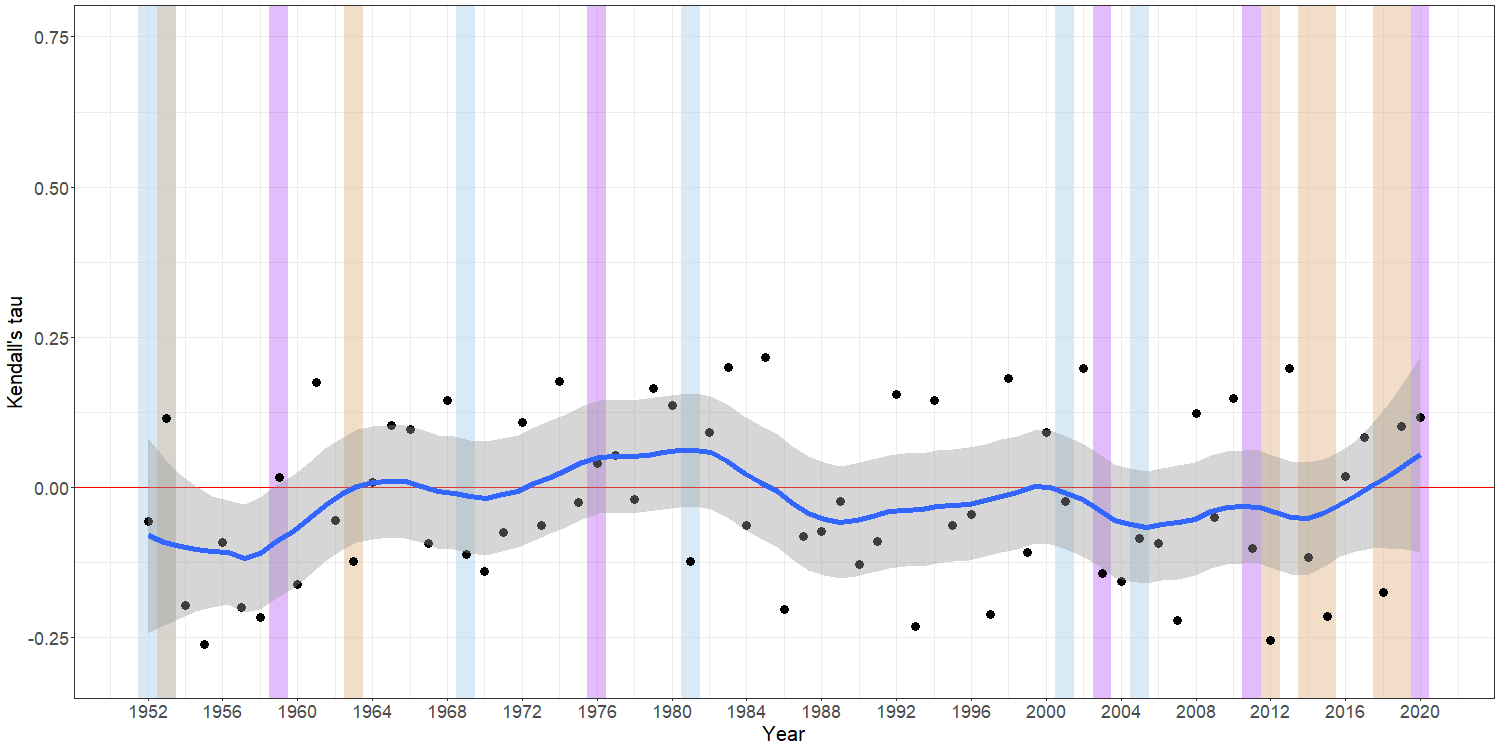}};
	\end{tikzpicture}
	\caption{x-axis: Years 1952-2020. y-axis: Top row: estimated unconditional Kendall's $\hat{\tau}_{f,d}$ value, bottom row:  conditional $\hat{\tau}_{f,d;\mathbf{u} }$. (The black points denote the estimated values at each year, the red horizontal line denotes when Kendall's $\hat{\tau}=0$, which indicates  independence. The blue line is the smoothed regression line and it's $95\%$ confidence interval. The vertical ribbons denote extreme years identified for frost risk (blue), drought risk (apricot), joint frost and drought risk (purple), and marginal drought and frost risk, but not joint risk identified(light gray).) More details on identifying these years follows in Section \ref{sc:climvineresultss}.}
	\label{fig:unconkendall}
\end{figure}

First, we analyze the unconditional and conditional dependence between the late-frost index (\textit{f}) and the drought index (\textit{d}). The top panel of Figure \ref{fig:unconkendall} shows the estimated unconditional Kendall's $\hat{\tau}$ measure of dependence, denoted as $\hat{\tau}_{f,d}$, for each year in the period 1952-2020. The smoothed line represents a fitted moving average model, and the shaded area indicates the corresponding 95\% confidence interval.
The analysis reveals the evolution of this dependence measure over the years. Initially, from approximately 1952-1969, there is an increasing trend in the dependence between the late-frost and drought indices, peaking at Kendall's $\hat{\tau}_{f,d}=0.62$ in 1969. This value represents both the maximum dependence and the maximum absolute Kendall's $\hat{\tau}_{f,d}$. Subsequently, there is a brief period of decreasing trend, followed by stabilization from around 1980-2000. During this period, there are smaller dependencies in opposite directions, with some being positive and others negative. The dependence then exhibits a decreasing trend, reaching its minimum value of Kendall's $\hat{\tau}_{f,d}=-0.33$ in 2007. The minimum absolute Kendall's $\hat{\tau}$, approximately zero ($|\hat{\tau}_{f,d}|=0.0004$), is observed in 1970. Finally, in the last five years, there is an increasing trend in the dependence.

Next, we analyse the conditional dependence between the frost and drought indices in the bottom panel of Figure \ref{fig:unconkendall}. After fitting the Y-vine models, for each year there is a pair copula fitted between the frost and drought indices, conditioned on the chosen 5 predictors. This corresponds to the last pair copula fitted in the Y-vine model. For each of  these fitted pair copulas, we extract the estimated value of the Kendall's $\tau$, denoted as $\hat{\tau}_{f,d;\mathbf{u} }$ and we plot it for each year. Here we see a different trend in the conditional dependence, than in the unconditional dependence between the frost and drought indices. In the period 1952-1985 there is an overall increasing trend in the conditional dependence, reaching a maximal value in year 1985 of $\hat{\tau}_{f,d;\mathbf{u} }=0.22$. Afterwards, there is a decreasing trend in their dependence until around the 2000s, after which an increasing trend follows again.  The maximal  absolute value is reached in year 1955 of $|\hat{\tau}_{f,d;\mathbf{u} }|=0.26$, which is the minimal overall value as well. The minimal absolute value is reached in year 1964 and it is $|\hat{\tau}_{f,d;\mathbf{u} }|=0.008$. More details on how we identify the extreme risk years, represented by the vertical ribbons in Figure \ref{fig:unconkendall} is explained later in Section \ref{climvine:riskmeasures}.


Next, to amplify the benefits of the usage of vine copula models on the data, whose main advantage is modeling non-Gaussian relationships with tail and asymmetric dependencies, we analyse how many of the selected pair copula families are Gaussian  pair copulas and how many are non-Gaussian. In the fitted Y-vine models $\hat{\mathcal{Y}}_{t}$ there are in total 21 fitted pair copulas, and in the fitted univariate D-vine models, $\hat{\mathcal{D}}_{frost_t}$ and $\hat{\mathcal{D}}_{drought_t}$ there are 15 fitted pair copulas for $ t \in \left[ 1952-2020\right]$. On average for the frost D-vine models $\hat{\mathcal{D}}_{{frost}_{t}}$ for all $ t \in \left[ 1952-2020\right],$   there are 11\%  of Gaussian pair copulas, and 89\% of non-Gaussian pair copulas fitted. For the drought D-vine models  $\hat{\mathcal{D}}_{{drought}_{t}}$ for all $ t \in \left[ 1952-2020\right],$ there are 15\%  of Gaussian pair copulas, and 85\%  of non-Gaussian pair copulas fitted. For the joint model of frost and drought, the Y-vine model $\hat{\mathcal{Y}}_{t}$ for $ t \in \left[ 1952-2020\right],$ there are 12\% of Gaussian pair copulas, and 88\%  of non-Gaussian pair copulas fitted. The exact numbers for each model for every year are given in the supplementary material. The fact that the majority of the fitted pair copulas are non-Gaussian,  indicates that the overall dependence structure in the data is non-Gaussian and can be captured by the vine copula models.

\begin{figure}[]
	\centering
	\begin{tikzpicture}[scale=0.9] 
		
		\node[rotate=90] (a) at (-33, 10)  {Frost};
		
		\node[inner sep=0pt] (Col1) at (-25, 10)  {\includegraphics[width=0.85\textwidth]{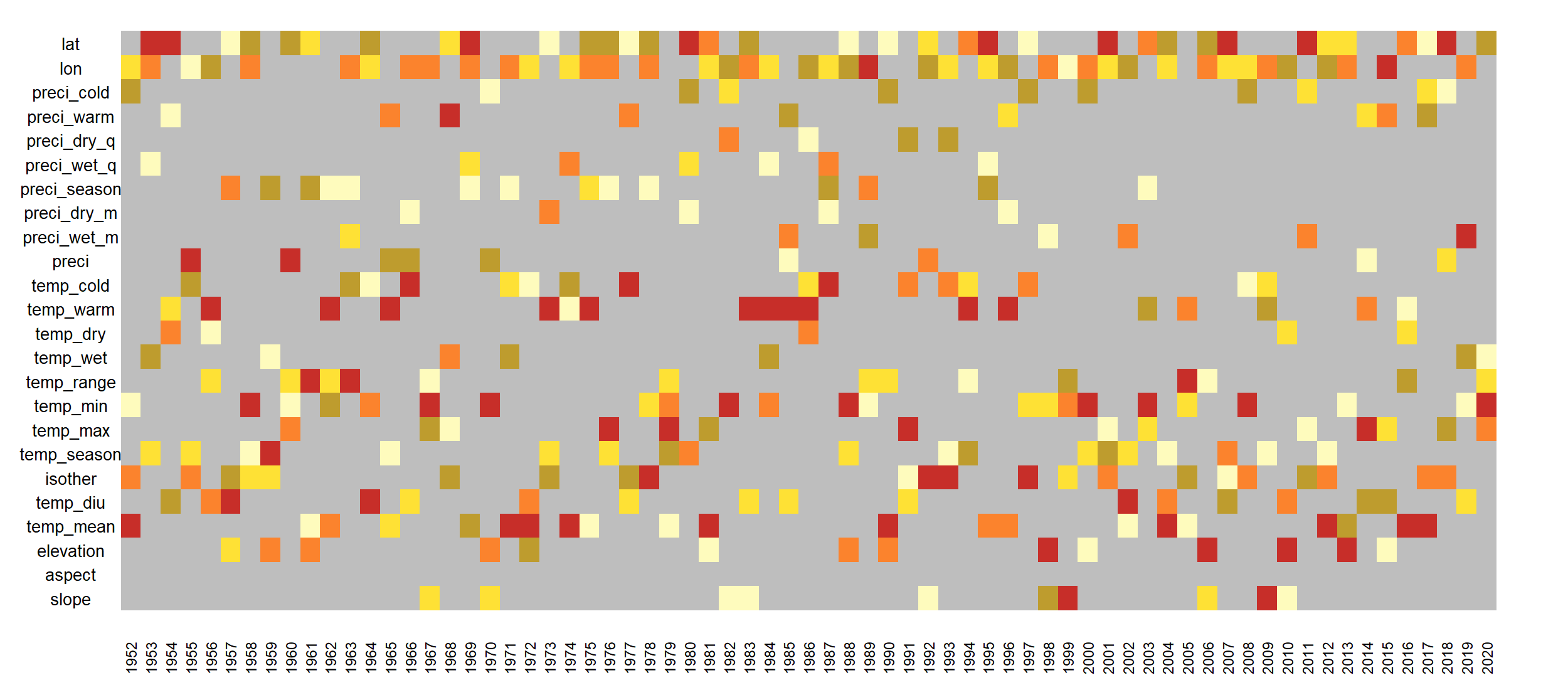}};
		\node[rotate=90] (a) at (-33, 3)  {Drought};
		
		\node[inner sep=0pt] (Col1) at (-25, 3)  {\includegraphics[width=0.85\textwidth]{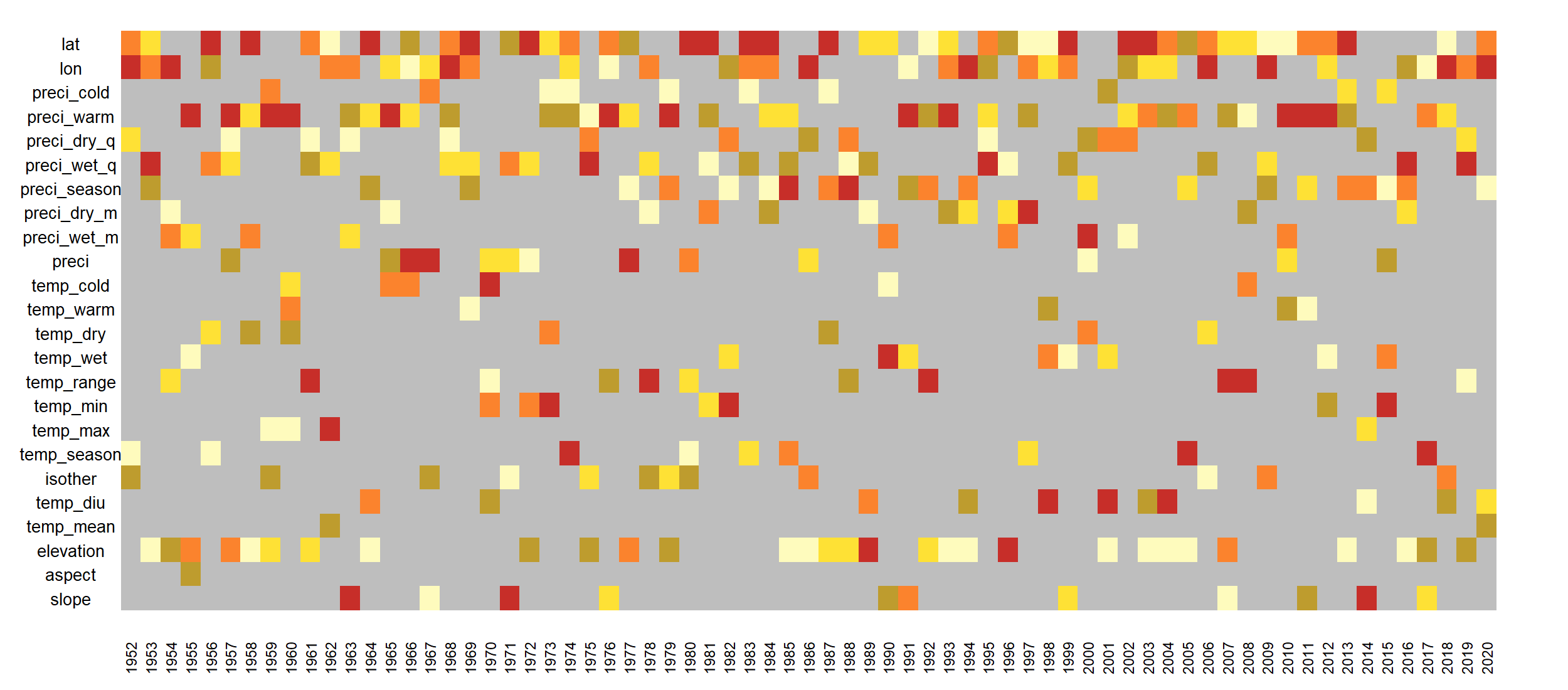}};
		
		\node[rotate=90] (a) at (-33, -4)  {Frost+Drought};
		
		\node[inner sep=0pt] (Col1) at (-25, -4)  {\includegraphics[width=0.85\textwidth]{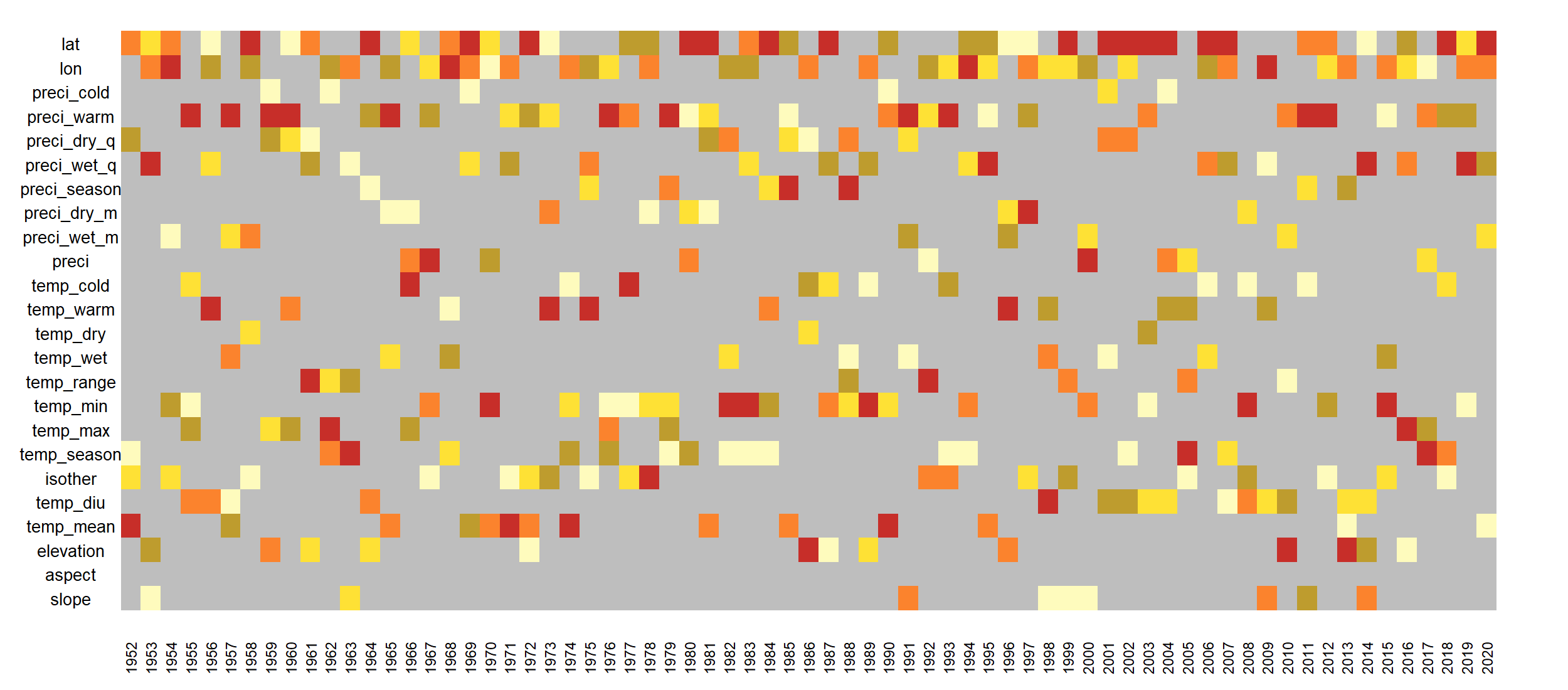}};
		
		\node[inner sep=0pt] (Col1) at (-25, -8)  {\includegraphics[width=0.9\textwidth]{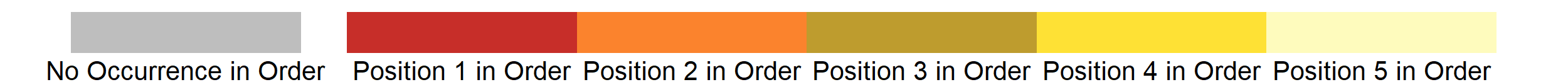}};
	\end{tikzpicture}
	\caption{Orders of the chosen predictors in the fitted annual models.}   
	\label{climvine:orders}
\end{figure}
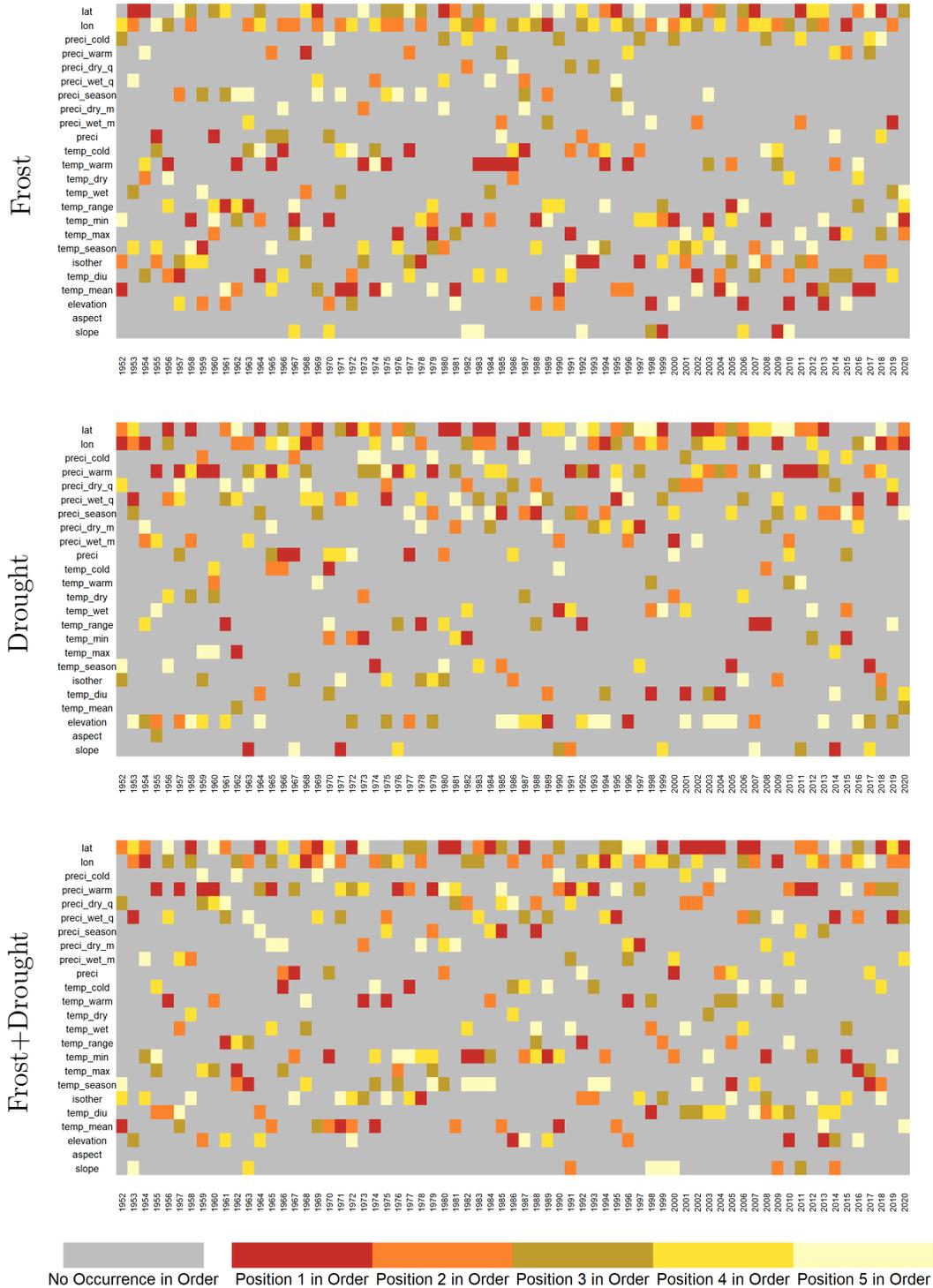

\begin{figure}[]
	\centering
	\includegraphics[width=\textwidth]{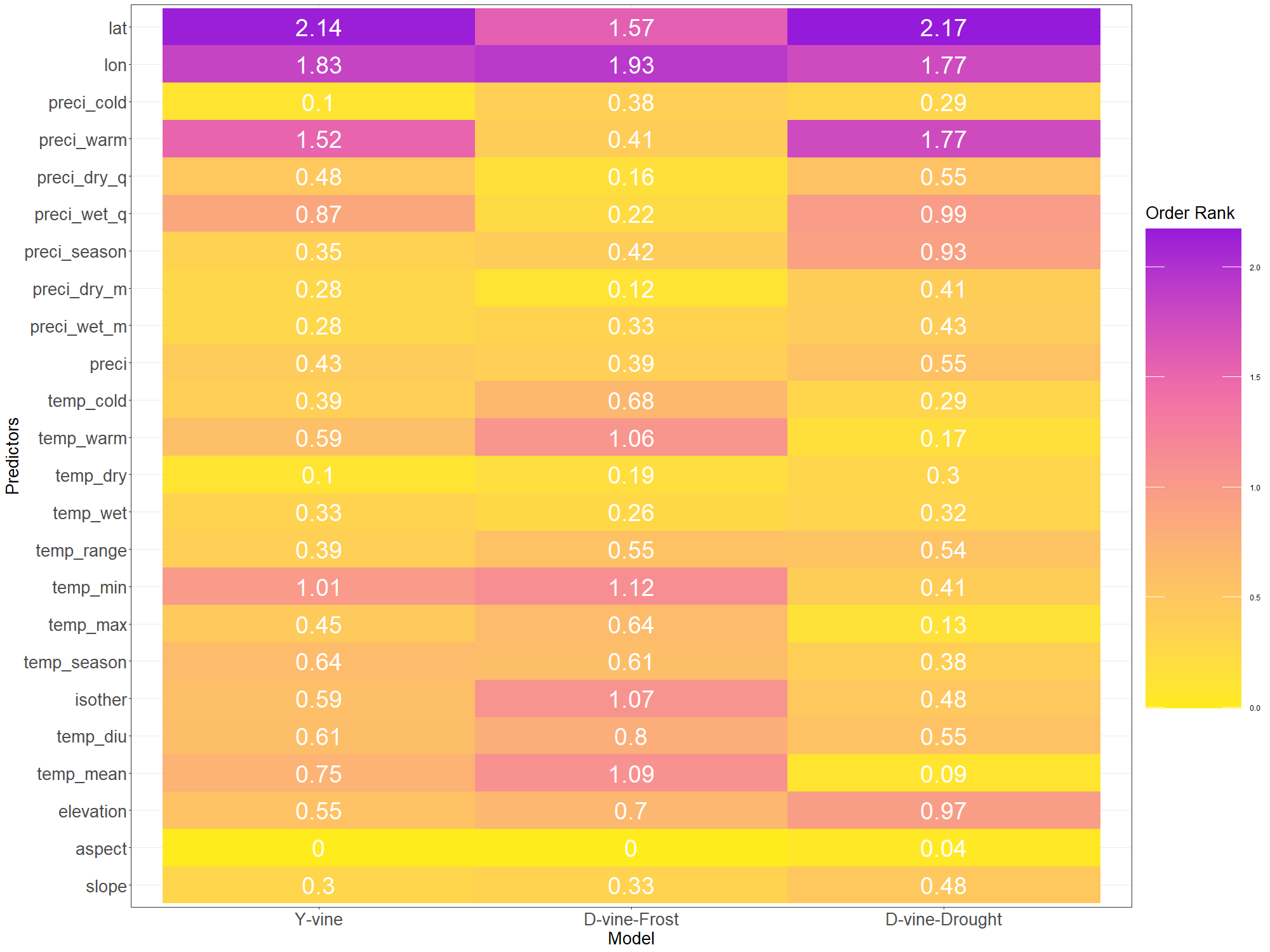}
	\caption{Order ranks for each chosen predictor in the fitted annual models.}
	\label{fig:orderranking}
\end{figure}

Each of the fitted models selects the 5 most influential predictors for the frost and drought indices, and the  5 most influential predictors for the joint modeling of them. To analyse how the influence of the predictors vary over the years and which of the possible predictors are chosen  over the historical period from 1952 to 2020, we show in Figure \ref{climvine:orders} for each year which 5 predictors are chosen by each model. Their influence  depends on the position in the order. The first predictor in the order is the most influential one on the response/s, the second in the second most influential one and so on. The 5 predictors that are chosen the most by each of the 69 models, not taking into account the position they are chosen in, are the following: 
for $\hat{\mathcal{D}}_{{frost}_{t}} $ \texttt{longitude} (43 times), \texttt{latitude} (35 times),  \texttt{temp\_min} (23 times), \texttt{isotherm} (each 22 times) and \texttt{temp\_mean} (21 times);
for $ \hat{\mathcal{D}}_{{drought}_{t}} $  \texttt{latitude} (44 times), \texttt{longitude}  and  \texttt{preci\_warm} (36 times each), \texttt{elevation }(30 times), \texttt{preci\_wet\_q} (23 times), \texttt{preci\_season} (22 times); 
and for $\hat{\mathcal{Y}}_{t}$  \texttt{latitude} (41 times), \texttt{longitude }(39 times),  \texttt{preci\_warm} (30 times), \texttt{temp\_min} (23 times), \texttt{preci\_wet\_q} and \texttt{isotherm} (each 19 times).
Further, we provide a ranking on the predictors by their position in the order, calculated as 
$rank( X_i) = \frac{\sum_{k=1}^{5}
	n_k^i (6-k)}{69}$, where $i=1,\ldots, p$ and $n_k^i$ denotes the count how many time the predictor $i$ appeared in the k-th position in the order, with $k= 1, \ldots, 5$. Figure \ref{fig:orderranking} shows the order ranks for each predictor and for each model, where a higher value implies a greater influence on the response/s. 
Out of the chosen predictors for all 3 models, we can conclude that the spatial effects have  very influential role, as \texttt{latitude} and \texttt{longitude} are chosen by all 3 models in the  orders. Further, for the $\hat{\mathcal{D}}_{{frost}_{t}}$ influential predictors are also the temperature based predictors, as \texttt{temp\_min}, \texttt{temp\_warm}, \texttt{temp\_mean}. 
For the $  \hat{\mathcal{D}}_{{drought}_{t}} $  influential predictors are the precipitation based predictors, as \texttt{preci\_warm}, \texttt{preci\_wet\_q}, and also the \texttt{elevation } spatial predictor. For the joint Y-vine regression models $\hat{\mathcal{Y}}_{t}$ influential predictors are both temperature based predictors, such as \texttt{temp\_min}, \texttt{isotherm}, but also precipitation based predictors, as \texttt{preci\_wet\_q}, \texttt{preci\_warm}. Additional investigations regarding the order of the chosen predictors are given in the supplementary material.

\section{Univariate and bivariate conditional probabilistic risk measures of extreme events}\label{climvine:riskmeasures}

For the fitted vine models, we propose a probabilistic risk measure, which is defined as the conditional probability of the random variable/s to be less that a specified threshold given the predictors. 
Denote the frost index random variable as $Y_{f,t,l}$ and the drought index random variable as $Y_{d,t,l}$ for  year $ t \in \left[ 1952, \ldots, 2020\right]$ at gridcell (location) $l \in \left[  1, \ldots, 2867 \right]$. The corresponding marginal distribution functions are denoted as  $F_{Y_{f,t,l}}$ and $F_{Y_{d,t,l}}$ respectively. Denote the 5 ordered predictors chosen by each model as the vector $\mathbf{X}_{t,l}= \left( X_{1,t,l}, \ldots, X_{5,t,l}\right)^T$  with corresponding marginal distribution functions 
$ F_{1,t,l}, \ldots, F_{5,t,l}$ for $ t \in \left[ 1952, \ldots ,2020\right]$ and $l \in \left[  1, \ldots, 2867 \right]$. 

\subsection{General framework}
Given a threshold vector $\mathbf{p}=\left(y_{f},y_{d}\right)^T $, we define the conditional probability of the  occurrence of the random variable $Y_{f,t,l} \leq y_f$ given a set of predictors $\mathbf{X}_{t,l} = \mathbf{x}_{t,l}$, i.e. $ P\left( Y_{f,t,l} \leq y_f \vert \mathbf{X}_{t,l}=\mathbf{x}_{t,l}\right),$ within a time period $t$ for all $l$, as a risk measure for the occurrence of frost. This conditional probability is estimated as 
\begin{equation}\label{event_prob_frost}
	\hat{P}\left( Y_{f,t,l} \leq y_f \vert \mathbf{X}_{t,l} = \mathbf{x}_{t,l}\right) \coloneqq C_{\hat{\mathcal{D}}_{{frost}_{t}}} \left( V_{f,t,l} \leq F_{Y_{f,t,l}}\left(y_{f}\right) \vert \mathbf{U}_{t,l}= \mathbf{u}_{t,l}  \right) ,
\end{equation}
where $V_{f,t,l} = F_{Y_{f,t,l}}\left(Y_{f,t,l}\right)$, and $\mathbf{U}_{t,l}= \left( F_{1,t,l}\left(X_{1,t,l}\right), \ldots, F_{5,t,l}\left(X_{5,t,l}\right) \right)^T$ using the estimated $\hat{\mathcal{D}}_{{frost}_{t}}$ model with its chosen order of predictors.
We denote the right hand side of Equation \eqref{event_prob_frost} as $\hat{P}_{\hat{\mathcal{D}}_{{frost}_{t}}}\left(y_f | \mathbf{x}_{t,l}\right)$.
Following the same analogy, a risk measure for the occurrence of drought, $\hat{P}_{\hat{\mathcal{D}}_{{drought}_{t}}}\left(y_d| \mathbf{x}_{t,l}\right)$ is defined, where  $\mathbf{x}_{t,l}$ contains  the observations of the 5 chosen predictors in the order of the $\hat{\mathcal{D}}_{{drought}_{t}}$ model for each time $t$ and location $l$. 
The joint risk measure for the joint occurrence of frost and drought, given a threshold vector $\mathbf{p}=\left(y_{f},y_{d}\right)^T $, is  the conditional probability of a random  $Y_{f,t,l} \leq y_f$ and $Y_{d,t,l}\leq y_d$ given a set of predictors, within a time period $t$. We estimate it as
\begin{equation}\label{event_prob_joint}
	\small
	\hat{P}\left( Y_{f,t,l} \leq y_f, Y_{d,t,l} \leq y_d \vert \mathbf{X}_{t,l} =\mathbf{x}_{t,l}\right) \coloneqq C_{\hat{\mathcal{Y}}_{t}} \left( V_{f,t,l} \leq F_{Y_{f,t,l}}\left(y_{f}\right),V_{d,t,l} \leq F_{Y_{d,t,l}}\left(y_{d}\right) \vert \mathbf{U}_{t,l}= \mathbf{u}_{t,l}  \right),
\end{equation}
where $V_{d,t,l} = F_{Y_{d,t,l}}\left(Y_{d,t,l}\right)$ and  $\mathbf{u}_{t,l}$ contains  the observations of the 5 chosen predictors in the order of the $\hat{\mathcal{Y}}$ model for each time $t$ at location $l$.  We denote the right hand side of Equation \eqref{event_prob_joint} as $\hat{P}_{\hat{\mathcal{Y}}_{t}}\left( \mathbf{p} |\mathbf{x}_{t,l}\right)$.

In order to estimate the proposed conditional probabilities, we chose the threshold to be $\mathbf{p}=\left(y_{f},y_{d}\right) = \left(-2,-1.5\right)$. Thus, we estimate the  conditional risk probability of extreme frost $\hat{P}_{\hat{\mathcal{D}}_{{frost}_{t}}}\left(-2| \mathbf{x}_{t,l}\right)$, the conditional risk probability of extreme drought $\hat{P}_{\hat{\mathcal{D}}_{{drought}_{t}}}\left(-1.5|\mathbf{x}_{t,l}\right)$ and the conditional risk probability of joint extreme frost and drought $\hat{P}_{\hat{\mathcal{Y}}_{t}}\left(-2, -1.5|\mathbf{x}_{t,l}\right)$, where for each model the conditioning values $\mathbf{u}_{t,l}$ differ. 
For the drought index, we set the threshold at -1.5 which represents a commonly accepted threshold beyond which drought conditions are classified as severely to extremely dry \citep{sletteHowEcologistsDefine2019}. Since such a commonly accepted threshold is not available for the frost index, we set the threshold at -2, signifying two standard deviations below the mean of the frost index. The greater these conditional probabilities are, the greater the chances are of 'extreme' drought, frost or joint frost and drought risk.  

\subsection{Results}\label{sc:climvineresultss}

\begin{figure}[]
	\centering
	\begin{tikzpicture}[scale=0.9] 
		
		
		\node[inner sep=0pt] (Col1) at (-25, 10)  {\includegraphics[width=\textwidth]{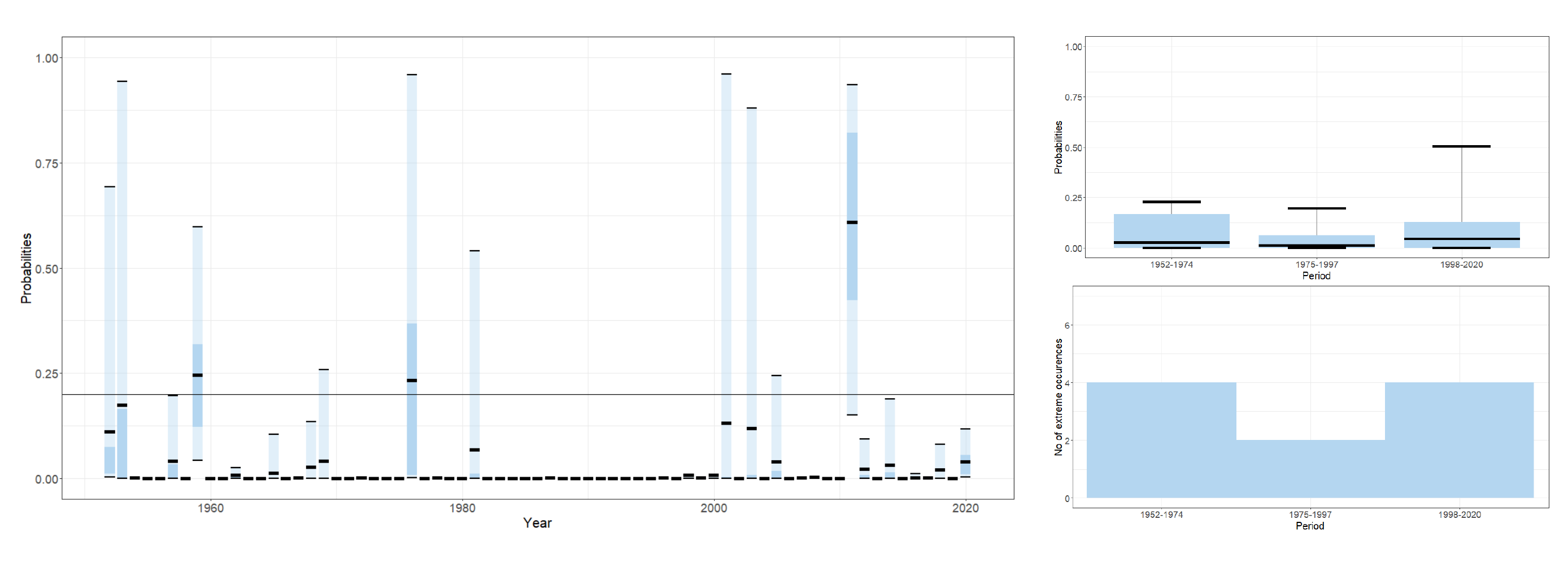}};
		
		\node[inner sep=0pt] (Col1) at (-25, 3)  {\includegraphics[width=\textwidth]{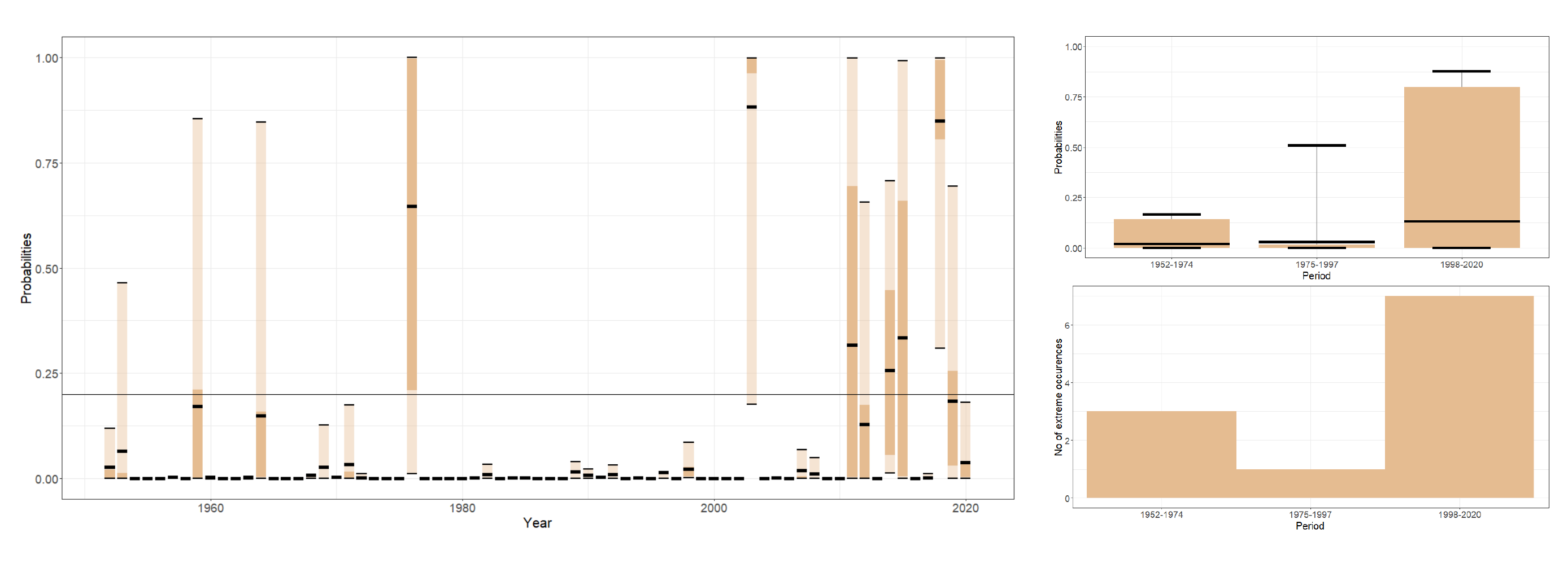}};
		
		
		\node[inner sep=0pt] (Col1) at (-25, -4)  {\includegraphics[width=\textwidth]{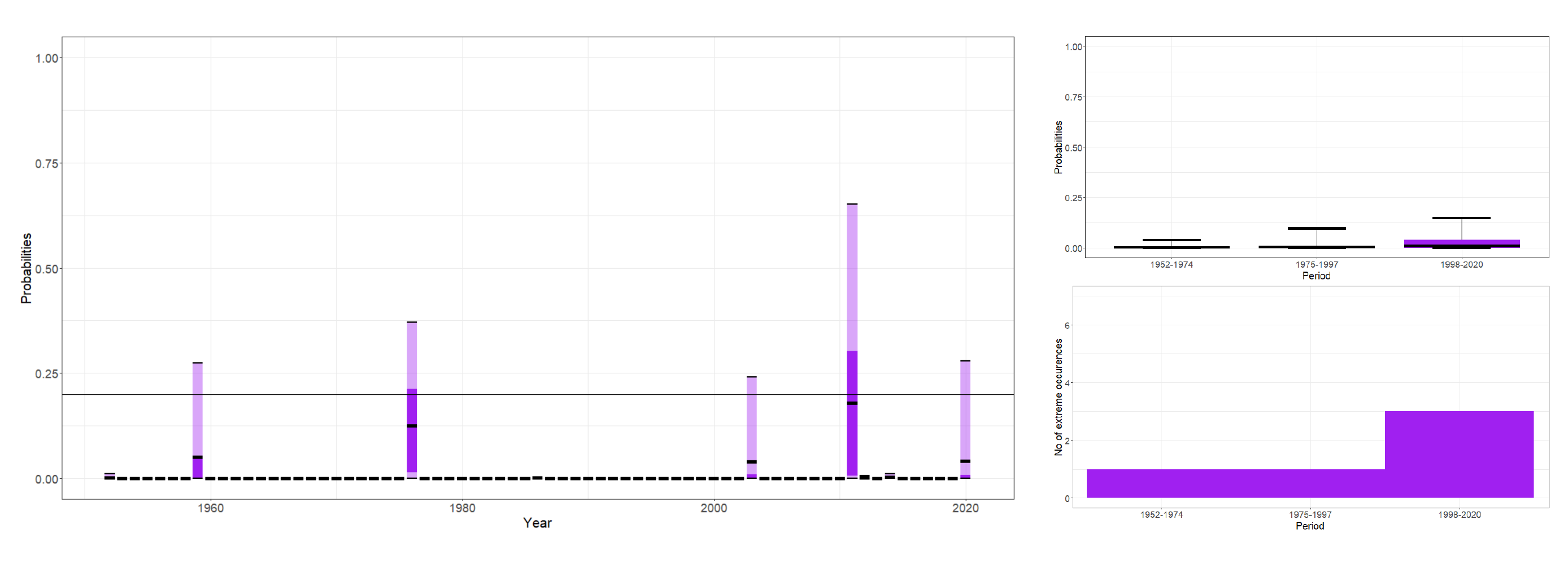}};
		
	\end{tikzpicture}
	\caption{Top row: conditional probabilities of \textbf{annual extreme frost}  $\hat{P}_{\hat{\mathcal{D}}_{{frost}_{t}}}\left(-2 | \mathbf{x}_{t,l}\right)$,  Middle row: conditional probabilities of \textbf{annual extreme drought} $\hat{P}_{\hat{\mathcal{D}}_{{drought}_{t}}}\left(-1.5 | \mathbf{x}_{t,l}\right)$, Bottom row: conditional probabilities of \textbf{annual jointly extreme frost and drought} $\hat{P}_{\hat{\mathcal{Y}}_{t}}\left(-2, -1.5|\mathbf{x}_{t,l}\right)$.}   
	\label{climvine:condpro3}
\end{figure}

The estimated conditional probabilities from the 3 fitted vine regression models are given in Figure \ref{climvine:condpro3}. 
Per year the conditional probabilities are estimated for each location, and subsequently their 90\% (light coloured) and 50\% (dark coloured) confidence intervals based on empirical quantiles over grid cells and empirical means are shown in the left panel of each figure. Additionally, the 98\%  (shown as whiskers) and 90\%  (as boxes) confidence intervals and means (top-right panels) are shown for three selected time periods, i.e. years 1952-1974, 1975-1997 and 1998-2020. Further, a year is considered to have an \textbf{"annual extreme event occurrence" if the 0.95 empirical quantile of the estimated conditional risk probability for that year over all gridcells is greater than 0.2} (indicated by the  horizontal line), in which case at least 5\% of the locations  have a 20\% or higher chance of an extreme event occurring.  The number of years which exhibit such   extreme occurrences, within each of the considered time periods, are shown  in the bottom-right panels for each row of Figure \ref{climvine:condpro3}. 


\begin{figure}[]
	\centering
	\begin{tikzpicture}[scale=0.9] 
		
		\node[rotate=90] (a) at (-22.2, 10)  {1959};
		
		\node[inner sep=0pt] (Col1) at (-20, 10)  {\includegraphics[width=0.24\textwidth]{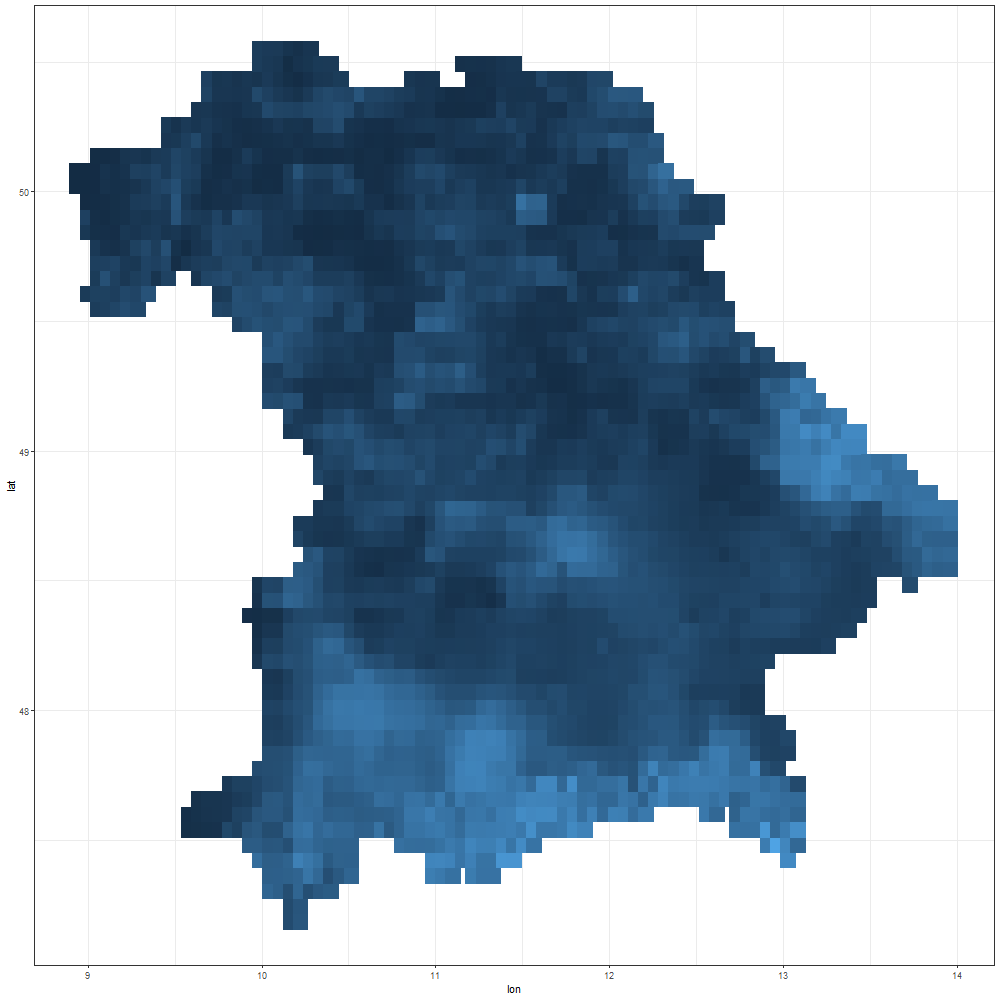}};
		
		\node (Col111) at (-20, 12.3)  {Frost};
		
		\node[inner sep=0pt] (Col1) at (-15, 10)  {\includegraphics[width=0.24\textwidth]{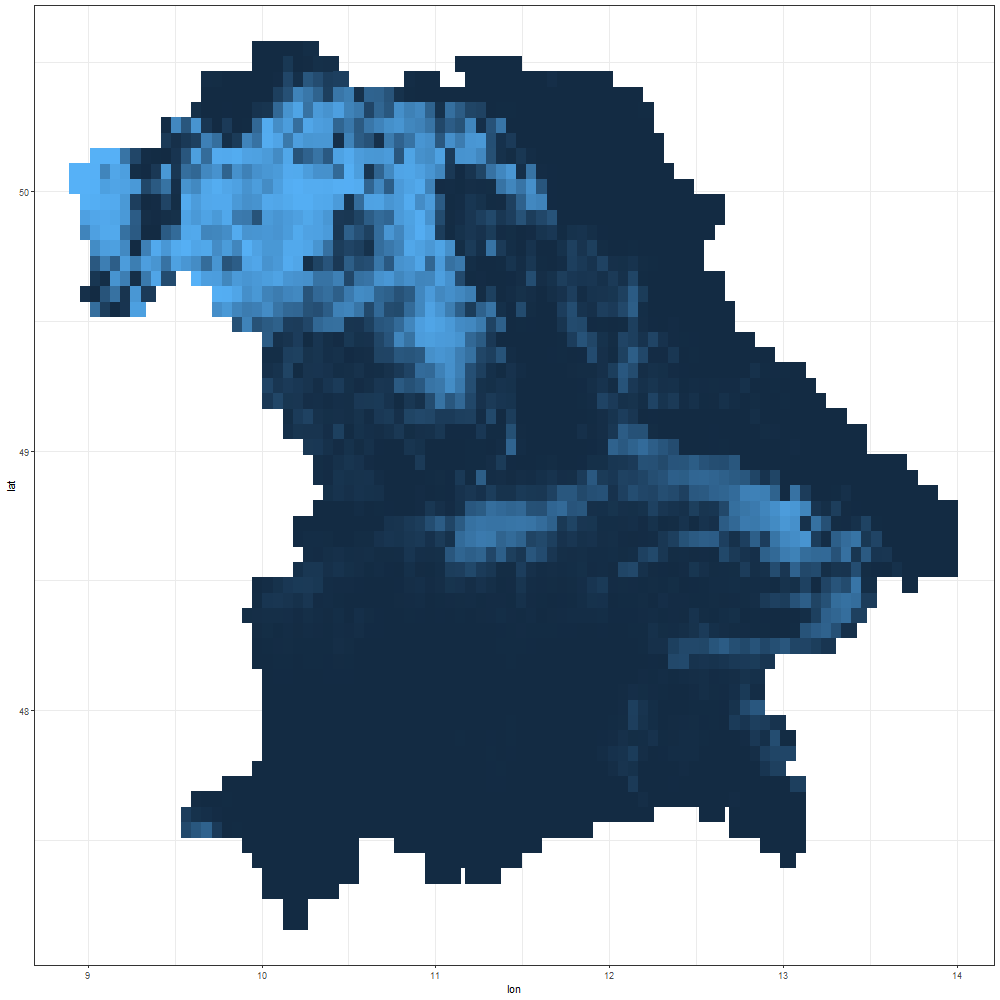}};
		
		\node (Col111) at (-15, 12.3)  {Drought};
		\node[inner sep=0pt] (Col1) at (-10, 10)  {\includegraphics[width=0.24\textwidth]{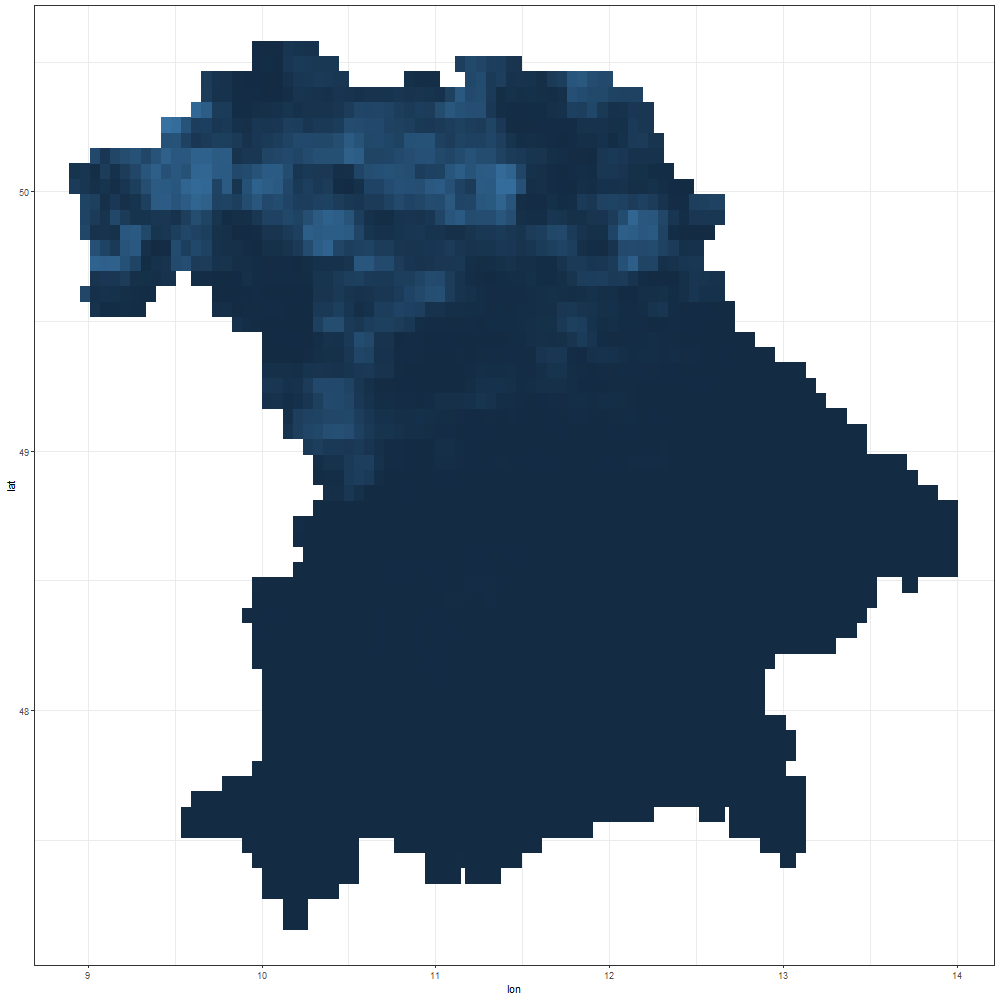}};
		\node (Col111) at (-10, 12.3)  {Frost+Drought};
		\node[rotate=90] (a) at (-22.2, 6)  {1976};
		
		\node[inner sep=0pt] (Col1) at (-20, 6)  {\includegraphics[width=0.24\textwidth]{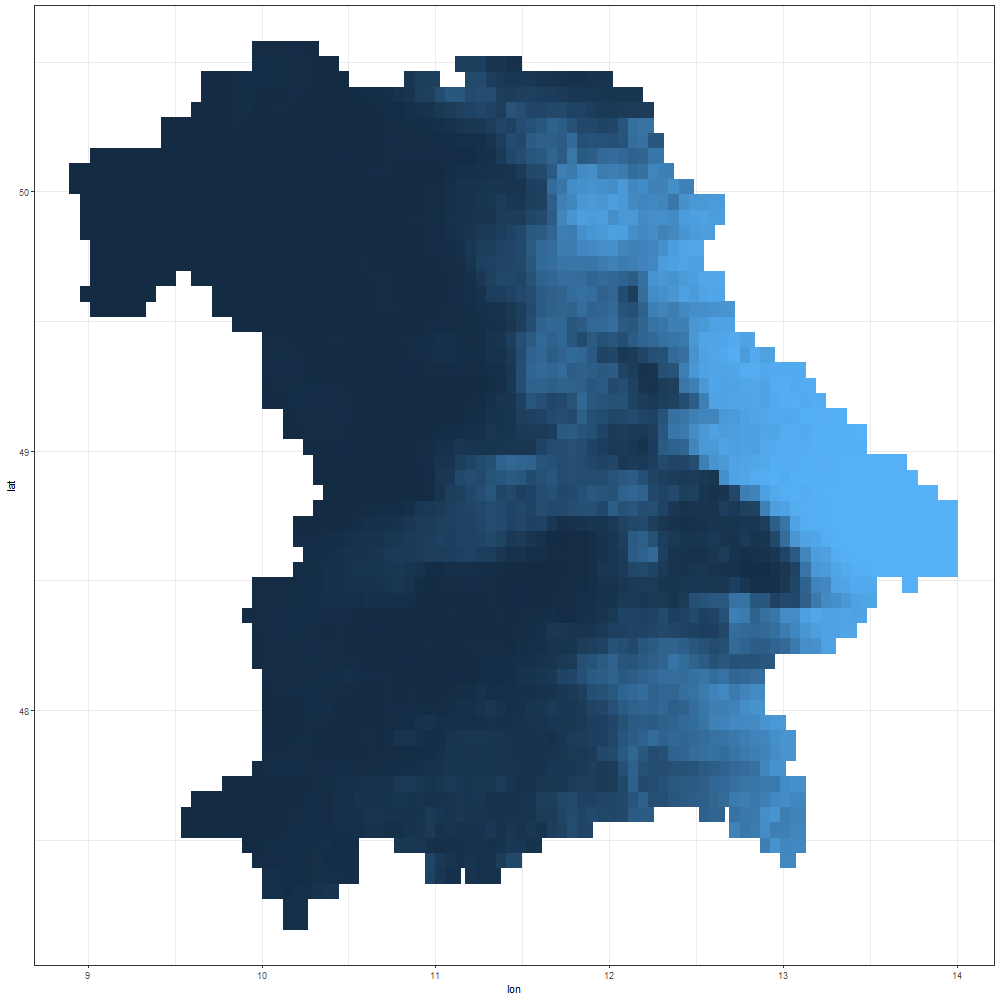}};
		
		\node[inner sep=0pt] (Col1) at (-15, 6)  {\includegraphics[width=0.24\textwidth]{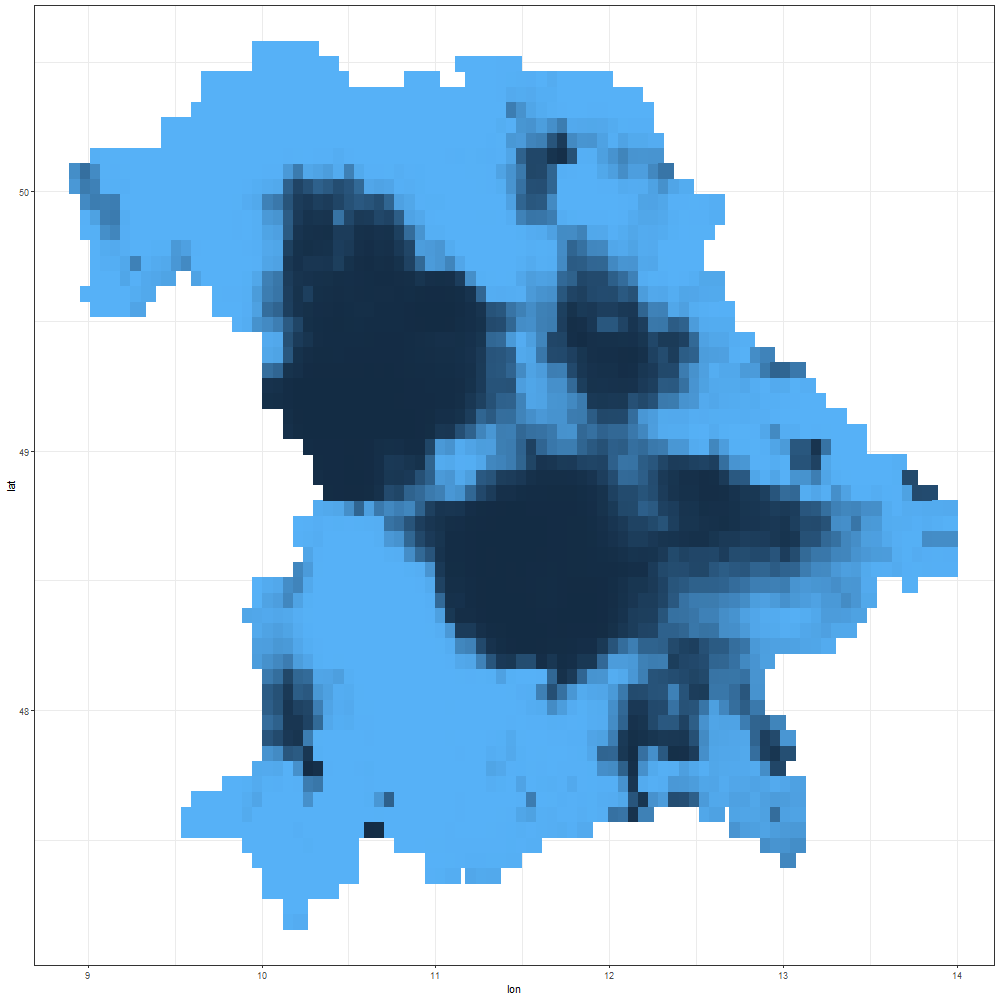}};
		
		\node[inner sep=0pt] (Col1) at (-10, 6)  {\includegraphics[width=0.24\textwidth]{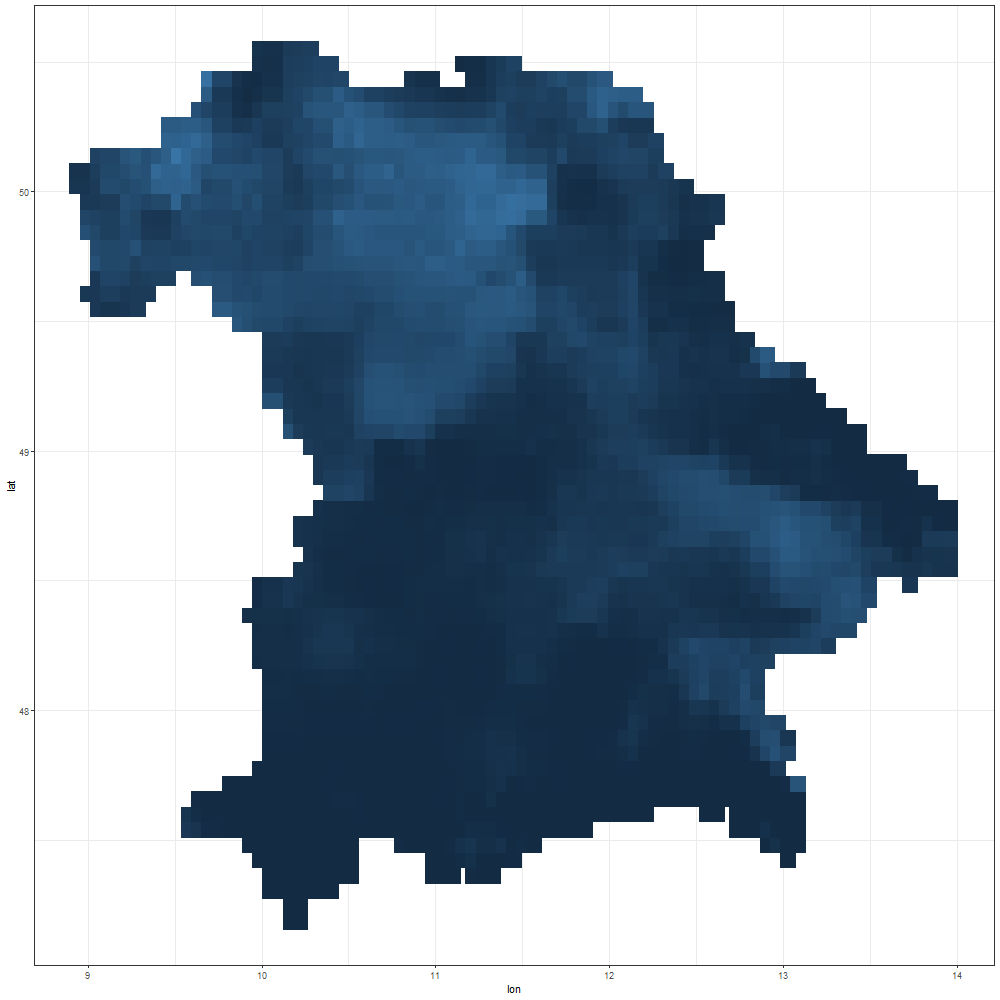}};
		\node[rotate=90] (a) at (-22.2, 2)  {2003};
		
		\node[inner sep=0pt] (Col1) at (-20, 2)  {\includegraphics[width=0.24\textwidth]{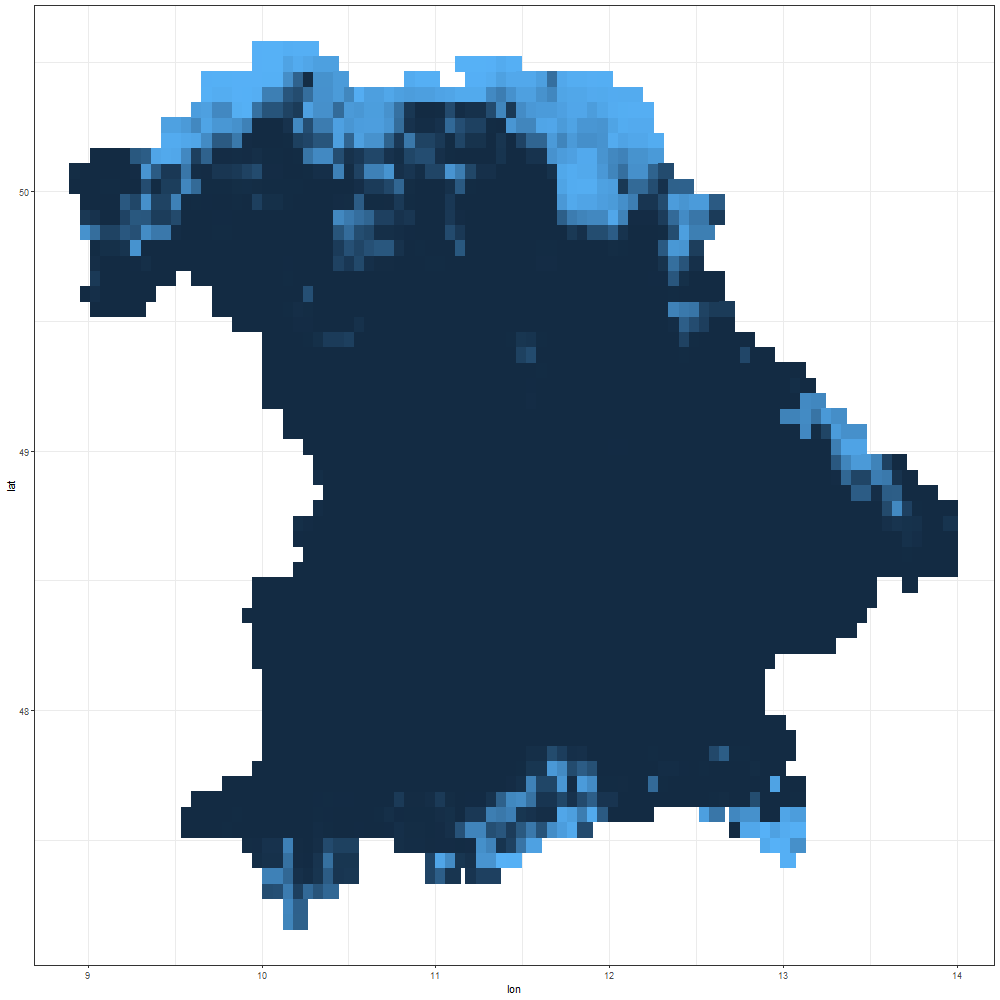}};
		
		\node[inner sep=0pt] (Col1) at (-15, 2)  {\includegraphics[width=0.24\textwidth]{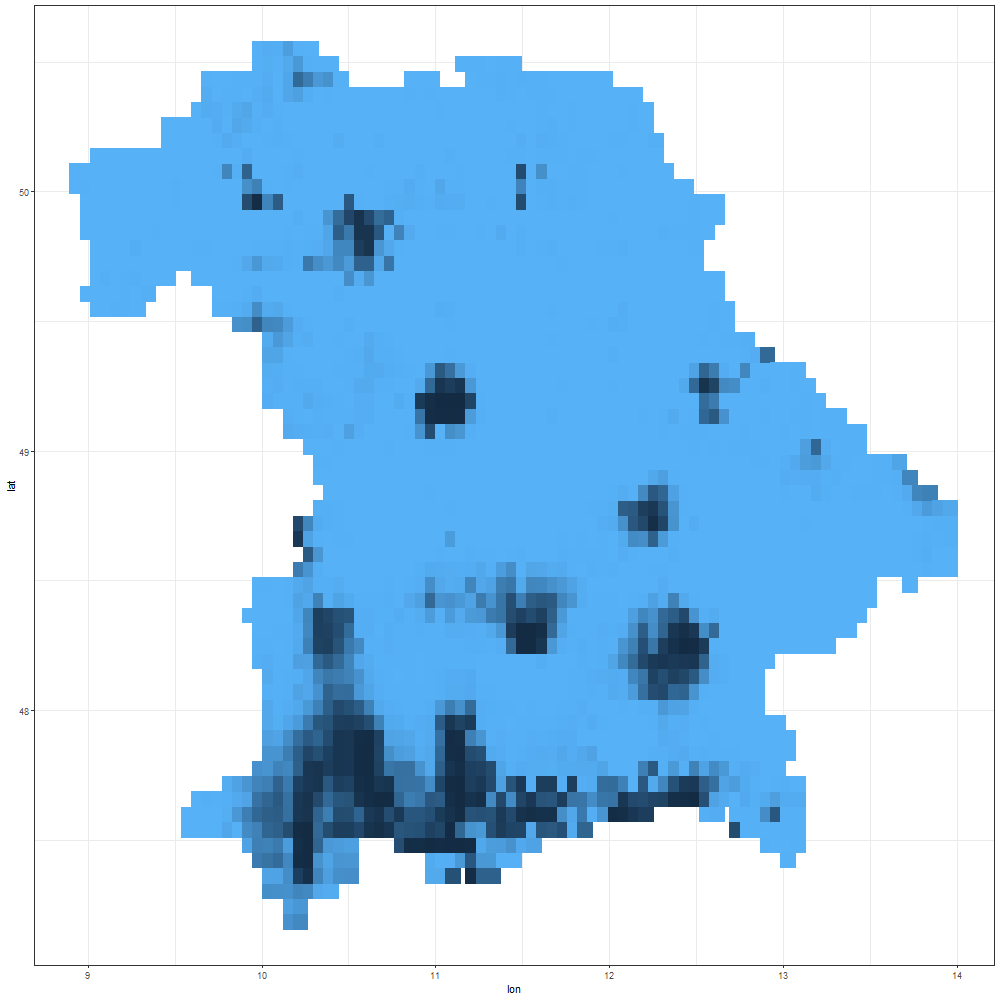}};
		
		\node[inner sep=0pt] (Col1) at (-10, 2)  {\includegraphics[width=0.24\textwidth]{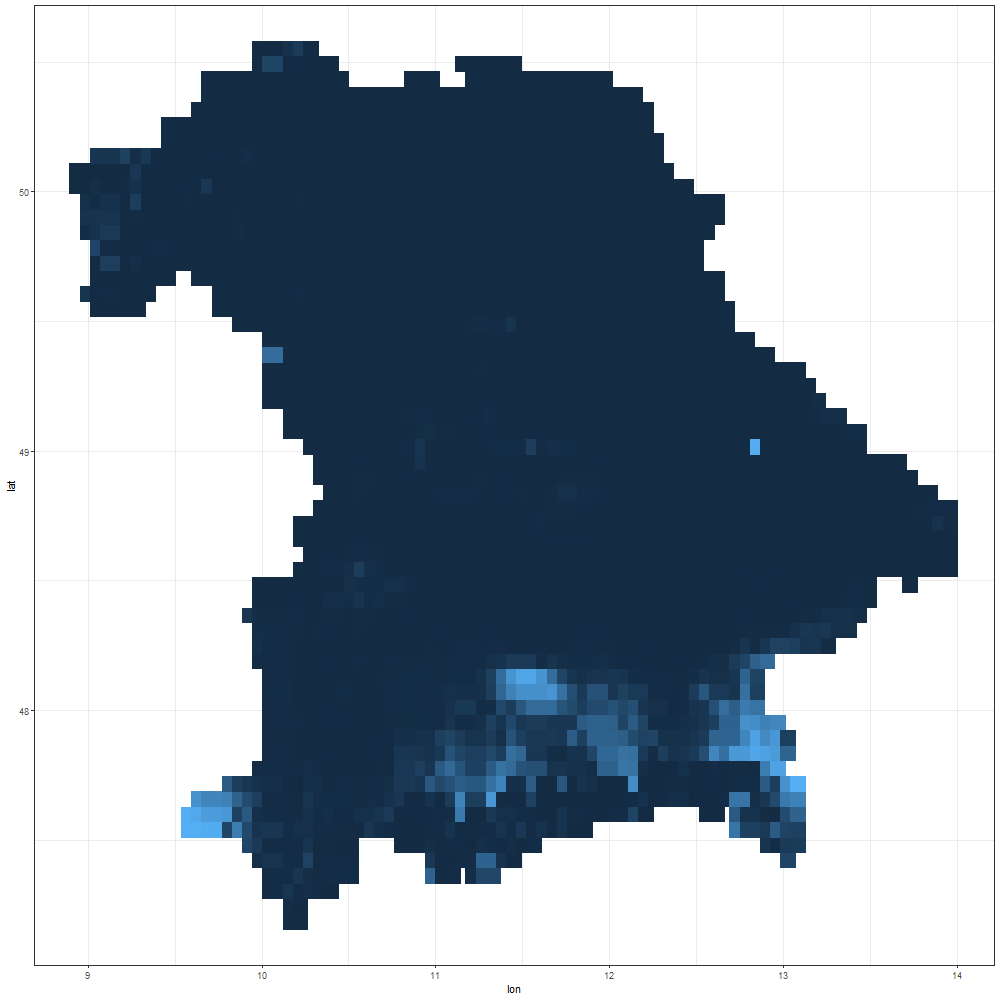}};
		\node[rotate=90] (a) at (-22.2, -2)  {2011};
		
		\node[inner sep=0pt] (Col1) at (-20, -2)  {\includegraphics[width=0.24\textwidth]{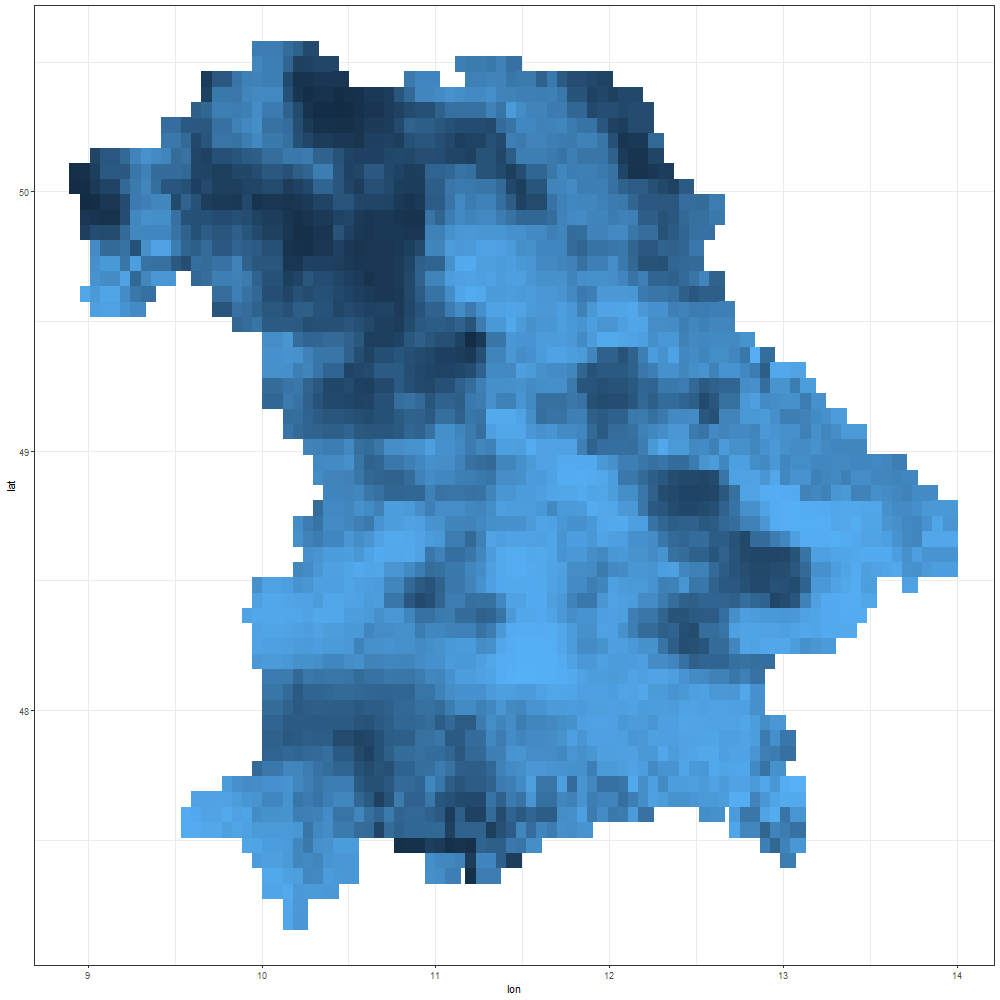}};
		
		\node[inner sep=0pt] (Col1) at (-15, -2)  {\includegraphics[width=0.24\textwidth]{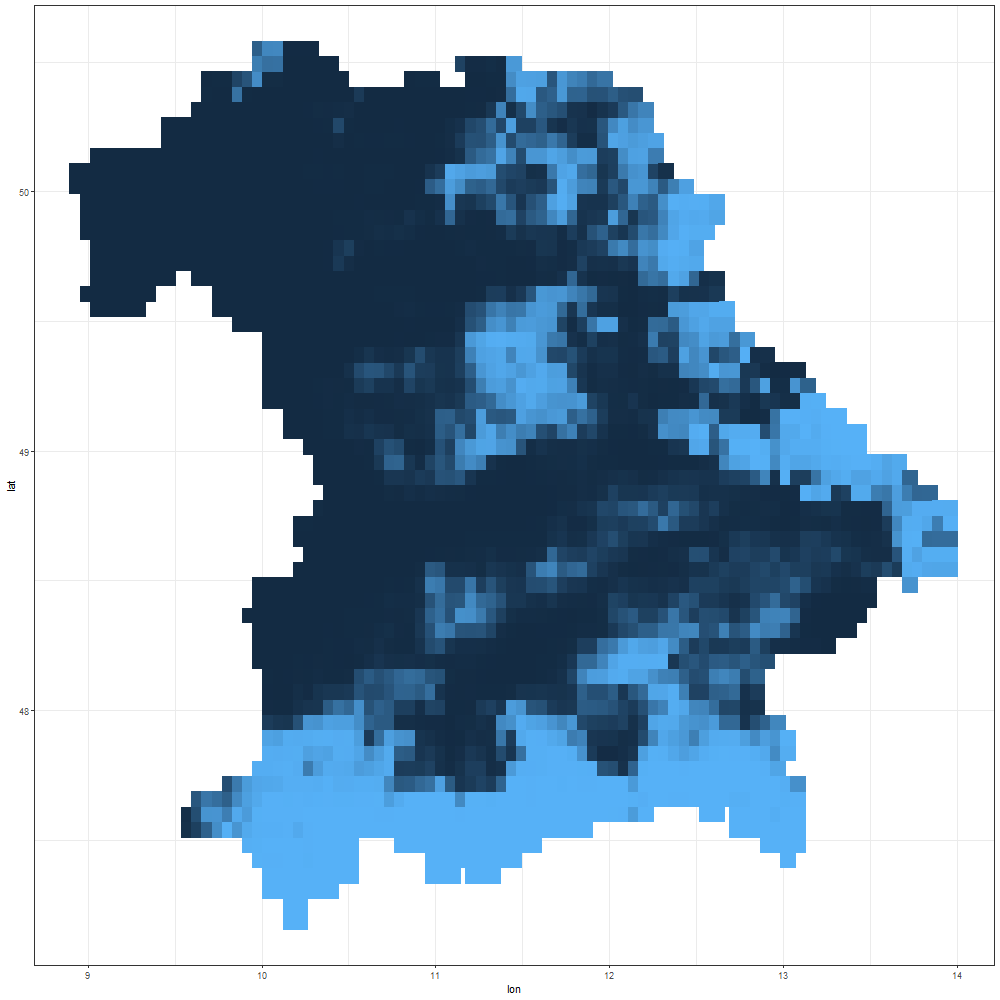}};
		
		\node[inner sep=0pt] (Col1) at (-10, -2)  {\includegraphics[width=0.24\textwidth]{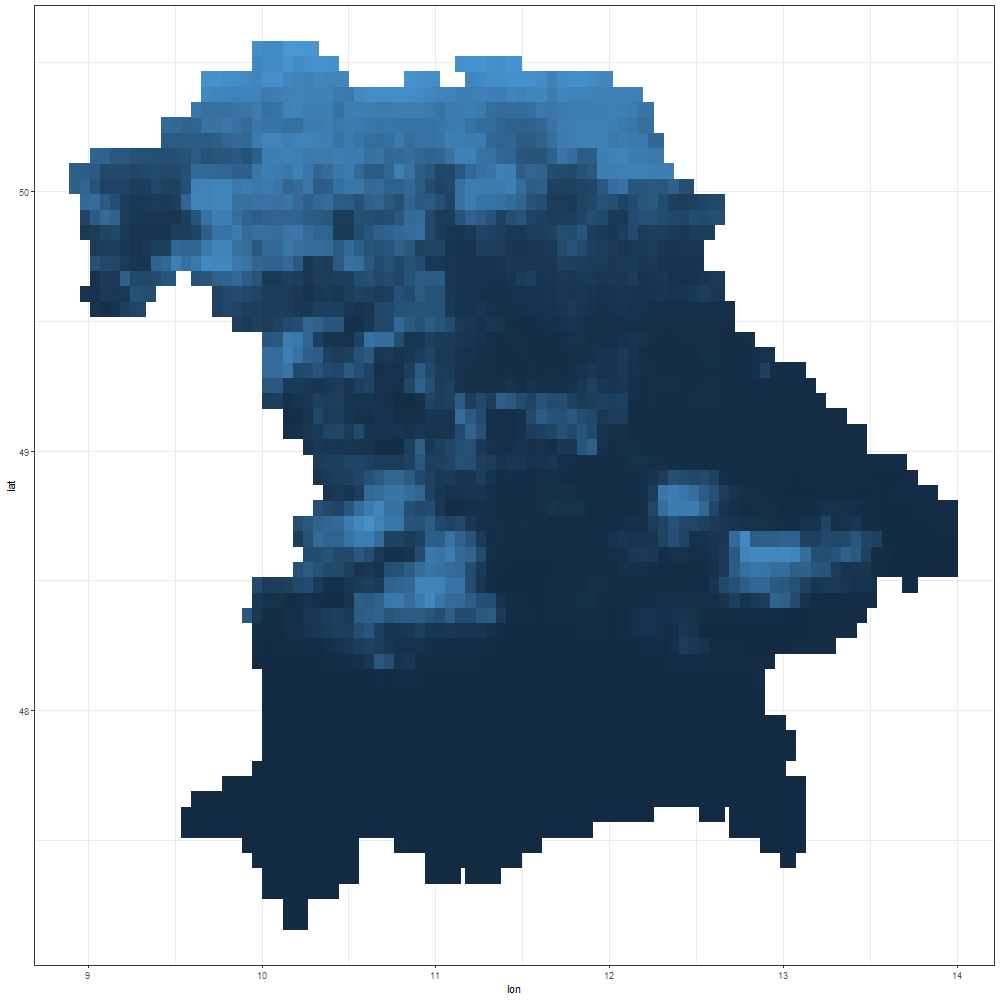}};
		\node[rotate=90] (a) at (-22.2, -6)  {2020};
		
		\node[inner sep=0pt] (Col1) at (-20, -6)  {\includegraphics[width=0.24\textwidth]{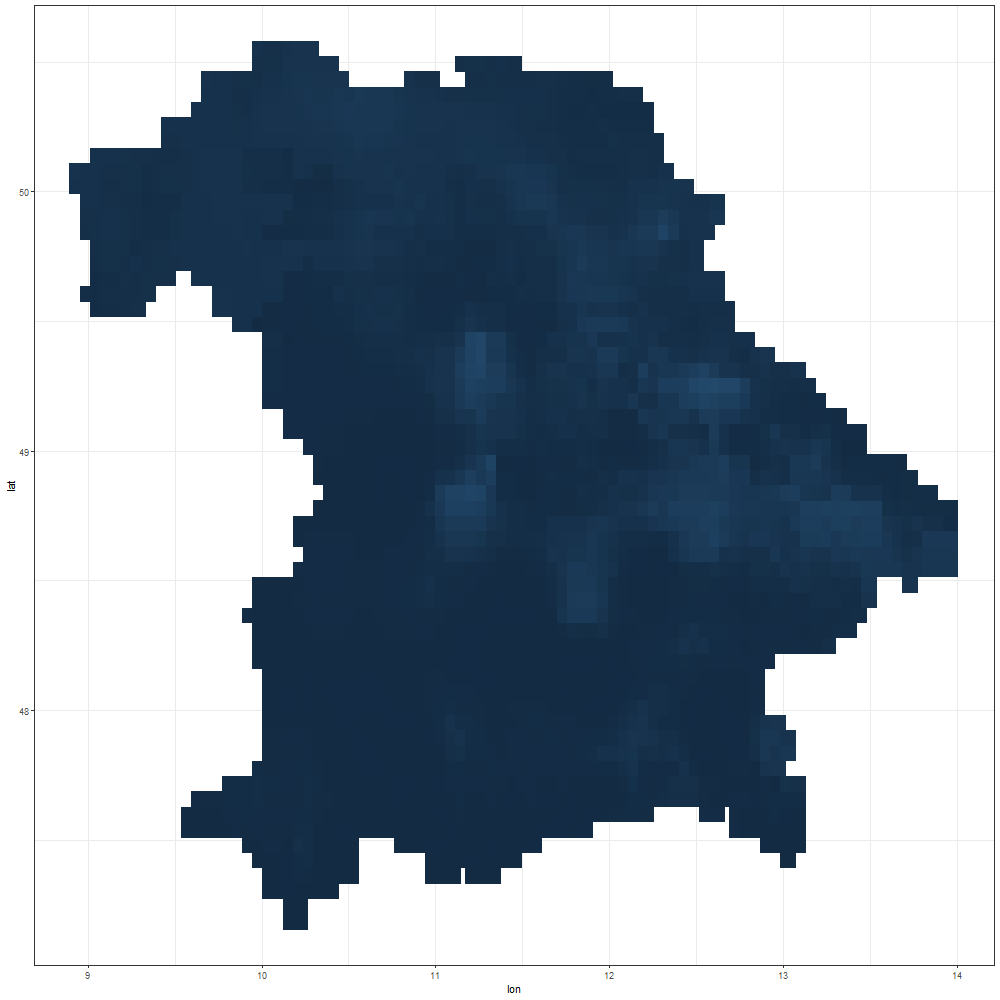}};
		
		\node[inner sep=0pt] (Col1) at (-15, -6)  {\includegraphics[width=0.24\textwidth]{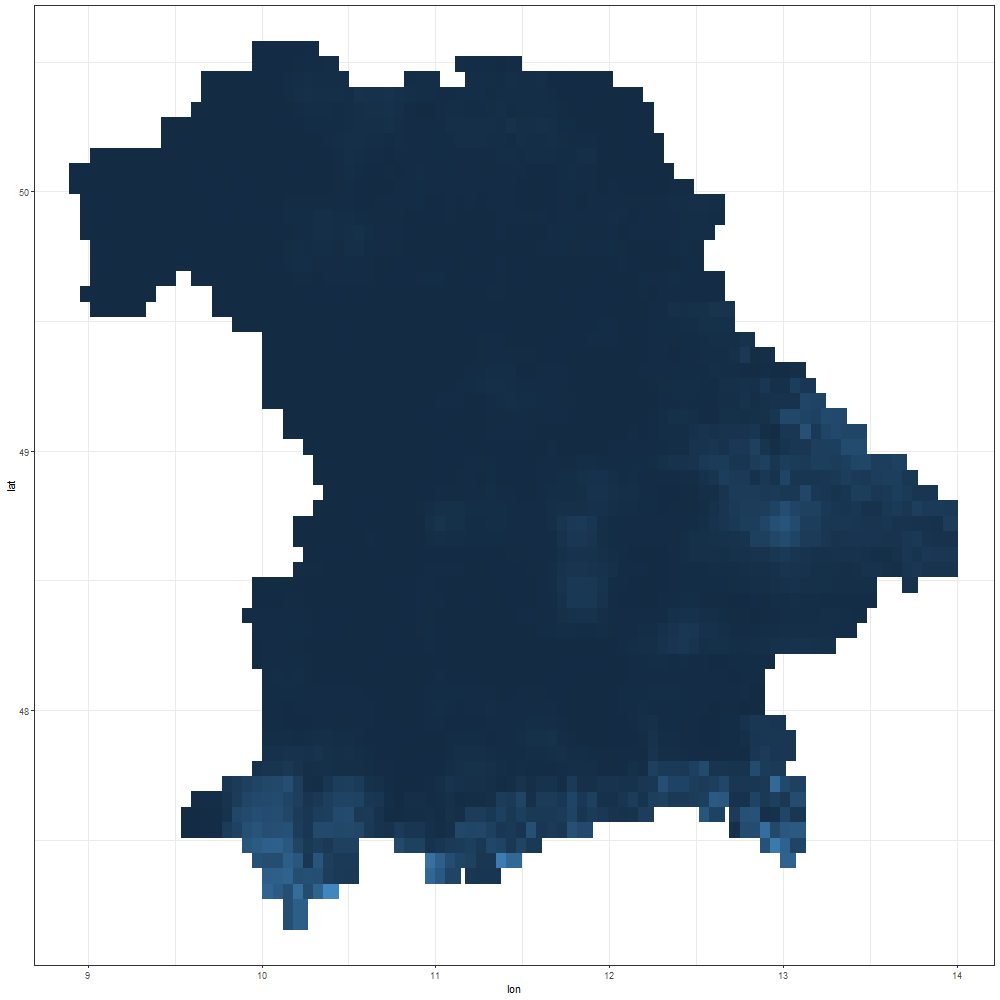}};
		
		\node[inner sep=0pt] (Col1) at (-10, -6)  {\includegraphics[width=0.24\textwidth]{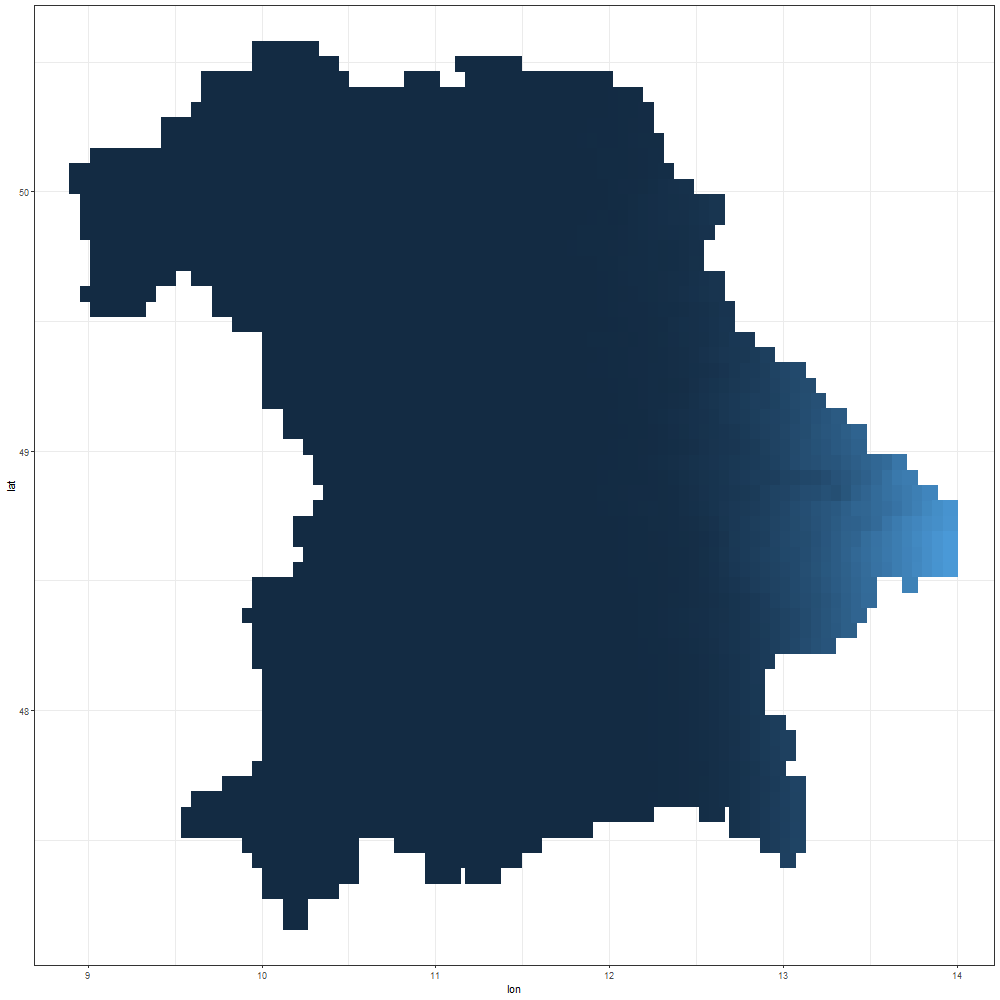}};
		\node[inner sep=0pt] (Col1) at (-15, -8.4)  {\includegraphics[width=0.6\textwidth]{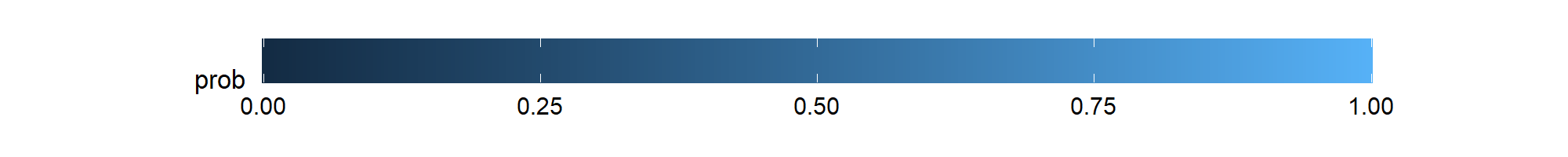}};
	\end{tikzpicture}
	\caption{ $\hat{P}_{\hat{\mathcal{D}}_{{frost}_{t}}}\left(-2| \mathbf{x}_{t,l}\right)$,    $\hat{P}_{\hat{\mathcal{D}}_{{drought}_{t}}}\left(-1.5|\mathbf{x}_{t,l}\right)$ and  $\hat{P}_{\hat{\mathcal{Y}}_{t}}\left(-2, -1.5|\mathbf{x}_{t,l}\right)$, for the years  identified as extreme by the joint Y-vine model for all gridcells. }
	\label{fig:mapsextremeyeaars}
\end{figure}
\subsubsection*{Extreme annual frost events}
The top row of Figure \ref{climvine:condpro3} shows a summary of the estimated conditional probabilities associated with an extreme frost event. It indicates that in approximately 23 years out of 69 there is a non zero probability of extreme events happening in at least 5\% of the locations in Bavaria. As the means of extreme events are fairly constant between the three 23 year periods, there seems not to be a significantly increased risk of extreme frost between the first and third period (top-right panel). Also, there is a total of 4  annual extreme events in the first period, 2 such in the second and 4 in the third period, resulting in a total of 10 annual extreme events for the frost risk (bottom-right panel). These identified risky years for the frost  are: 1952, 1953, 1959, 1969, 1976, 1981, 2001, 2003, 2005, 2011. 

\subsubsection*{Extreme annual drought events}
The middle row in Figure \ref{climvine:condpro3} summarizes the estimated conditional probabilities associated with an extreme drought event. Again, quite often there is a non-zero probability of extreme drought occurs in at least 5\% of the locations in Bavaria, in approximately 26 years out of 69. However, in contrast to frost, there is a clear increase in the average estimated conditional probability of an extreme drought event occurring, as well as the frequency of occurrence (top-right panel). In the period between 1998 and 2020 there is almost every year a high probability of an extreme drought event occurring over all locations. Further, the frequency of the annual extreme events increases as well, with 3  extreme events in the period 1952-1974 and a single  extreme event in the period 1975-1997, to a total of 7  extreme events in 1998-2020 (bottom-right panel).
These identified risky years for the drought index are: 1953, 1959, 1963, 1976, 2003, 2011, 2012, 2014, 2015, 2018, 2019.

\subsubsection*{Extreme annual frost and drought events}
The bottom row in Figure \ref{climvine:condpro3}
gives a summary of the estimated joint conditional probabilities of both, extreme frost and extreme drought events, occurring. There is  a clear increase in the average of the joint conditional probabilities, approximately by a factor of three (top-right panel), as well as a significant increase in occurrence frequencies between the last period (1998-2020) and the first two (1952-1974 and 1975-1997) (bottom-right panel). A significantly high risk of annual jointly extreme events is present in the years 1959, 1976, 2003, 2011 and 2020. Furthermore, in the years that there is an increased joint risk, there are also increased marginal risks of frost and drought. Basically, the joint Y-vine model identified 5 very extreme joint events, which were also identified by the univariate D-vine models for both the frost and drought risks, except for the year 2020, where  both the conditional frost  and drought risks are just below the threshold for an extreme event, but it is a quite high non-zero value.

\subsubsection*{Spatial distribution of identified extreme joint annual events}
Figure \ref{fig:mapsextremeyeaars} shows the estimated conditional probabilities of  frost $\hat{P}_{\hat{\mathcal{D}}_{{frost}_{t}}}\left(-2| \mathbf{x}_{t,l}\right)$,  drought  $\hat{P}_{\hat{\mathcal{D}}_{{drought}_{t}}}\left(-1.5|\mathbf{x}_{t,l}\right)$ and joint events $\hat{P}_{\hat{\mathcal{Y}}_{t}}\left(-2, -1.5|\mathbf{x}_{t,l}\right)$, 
for the years 1959, 1976, 2003, 2011 and 2020, for each of the considered locations or gridcells in Bavaria. At a given location, if there are high univariate conditional probabilities of  frost and  drought, there is not necessarily a high bivariate conditional probability of a joint  event. An example is the north of Bavaria in the year 2003. Despite having high chances of  frost and  drought individually, there is almost no chance of a joint  event occurring at those locations. 
Even the existence of small  conditional dependence in the responses given the predictors, indicates that the separate univariate analysis is not sufficient  for the joint analysis \citep{tepegjozova2022bivariate}. The estimated pair copula family between the responses conditioned on the 5 chosen predictors by the joint Y-vine model and the corresponding estimated Kendall's $\hat{\tau}$ for the 5 selected years are the following: 1959 is a Joe copula with $\hat{\tau}_{f,d;\mathbf{u} }= 0.02$ (lower tail dependence), 1976 is a Gumbel copula with $\hat{\tau}_{f,d;\mathbf{u} }= 0.04$ (lower tail dependence), 2003 is a Joe copula with $\hat{\tau}_{f,d;\mathbf{u} }= -0.14$,  2011 is a Clayton copula with $\hat{\tau}_{f,d;\mathbf{u} }= -0.10$ and 2020 is a Frank copula with $\hat{\tau}_{f,d;\mathbf{u} }= 0.12$ (no tail dependence). This implies that the responses are not conditionally independent given the predictors, as that case would imply estimated $\hat{\tau}=0.$ This indicates that separate assessment of risks, interpreted together will likely overestimate the joint risk and fail to detect true regions of interest.  The joint assessment is preferred since it does not assume conditional independence, i.e. it \textbf{does not} hold $\hat{P}_{\hat{\mathcal{Y}}_{t}}\left(\cdot, \cdot|\mathbf{x}_{t,l}\right) = \hat{P}_{\hat{\mathcal{D}}_{{frost}_{t}}}\left(\cdot| \mathbf{x}_{t,l}\right) \cdot \hat{P}_{\hat{\mathcal{D}}_{{drought}_{t}}}\left(\cdot|\mathbf{x}_{t,l}\right).$

\section{Survival probability and return periods}\label{climvine:returntimes}
\subsection{Survival probabilities}
The survival function is the complement of the cumulative distribution function, and it gives the probability that an event will not occur. We estimate the survival function of the  event starting from time $s$ until the  time period $T$ by subtracting from 1 the sum of the estimated conditional probabilities occurring in each year between time $s$ and $T \leq 2020$ (more in \cite[Chapter 2]{klein1997censoring}).
Then, the survival probability for the frost index is estimated as 
\begin{equation}\label{eq:survival}
	\hat{S}_{\hat{\mathcal{D}}_{frost}}(s,T) \coloneqq 1 - \sum_{t=s}^{T}\hat{P}_{\hat{\mathcal{D}}_{frost}}\left(-2| \mathbf{x}_{t,l}\right).
\end{equation}
\noindent In the same manner, the estimated survival probability for the drought index $\hat{S}_{\hat{\mathcal{D}}_{drought}}(s,T)$ is defined.

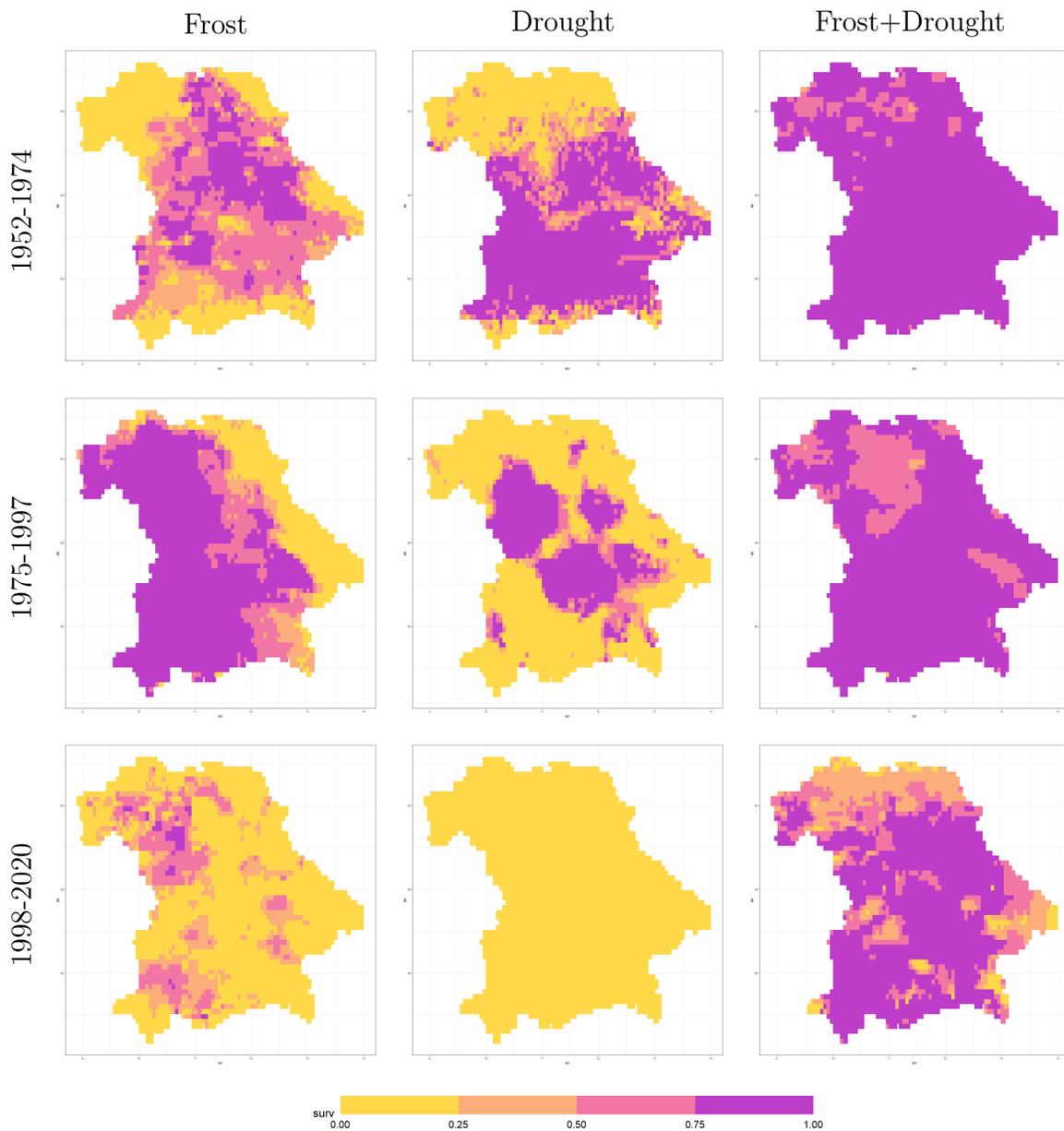
\begin{figure}
	\centering
	\begin{tikzpicture}
		
		\node[rotate=90] (a) at (-22.8, 10)  {1952-1974};
		
		\node[inner sep=0pt] (Col1) at (-20, 10)  {\includegraphics[width=0.29\textwidth]{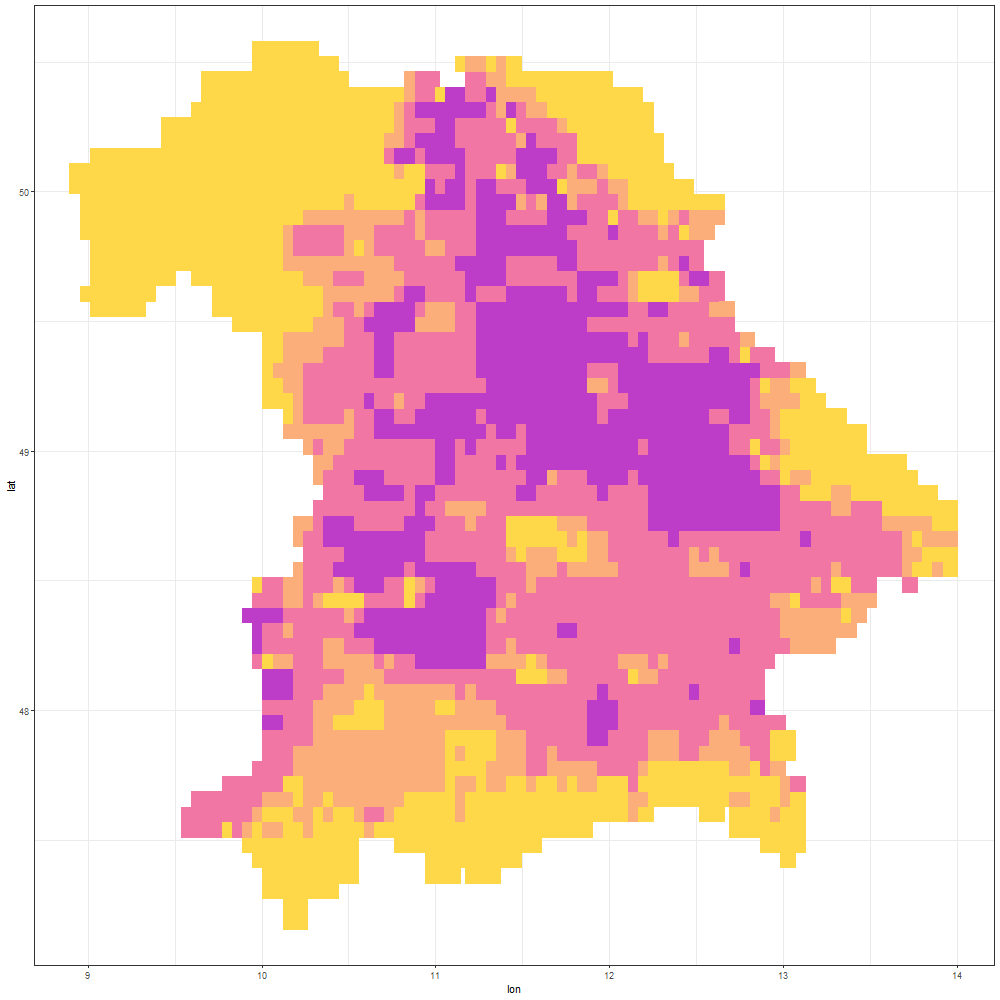}};
		
		\node (Col111) at (-20, 12.7)  {Frost};
		
		\node[inner sep=0pt] (Col1) at (-15, 10)  {\includegraphics[width=0.29\textwidth]{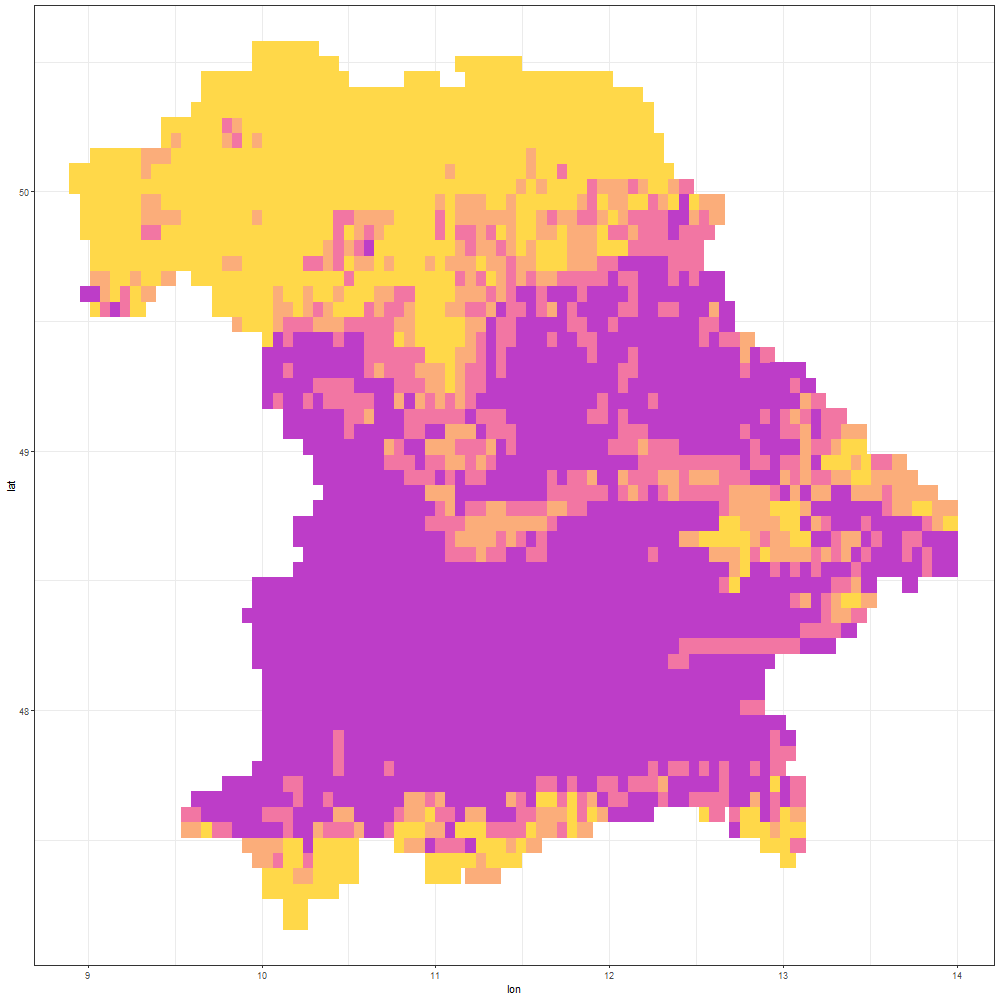}};
		
		\node (Col111) at (-15, 12.7)  {Drought};
		\node[inner sep=0pt] (Col1) at (-10, 10)  {\includegraphics[width=0.29\textwidth]{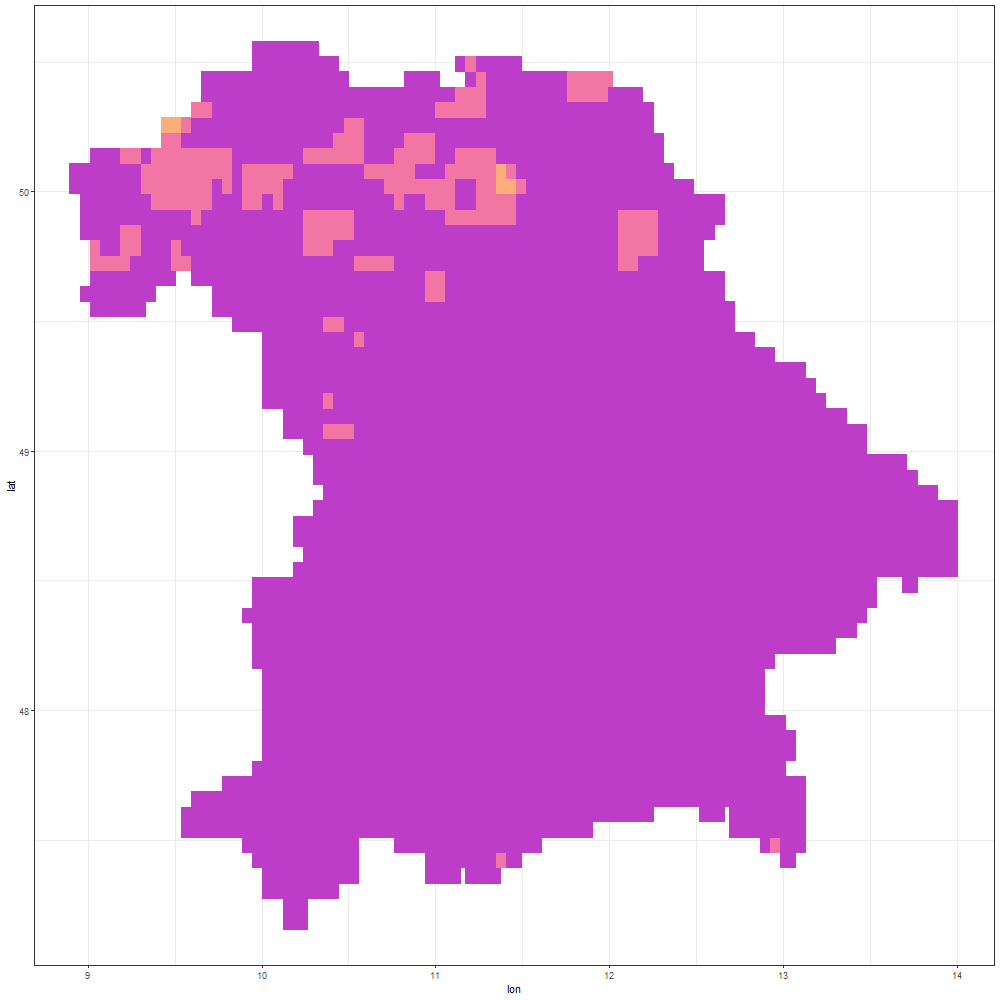}};
		\node (Col111) at (-10, 12.7)  {Frost+Drought};
		\node[rotate=90] (a) at (-22.8, 5)  {1975-1997};
		
		\node[inner sep=0pt] (Col1) at (-20, 5)  {\includegraphics[width=0.29\textwidth]{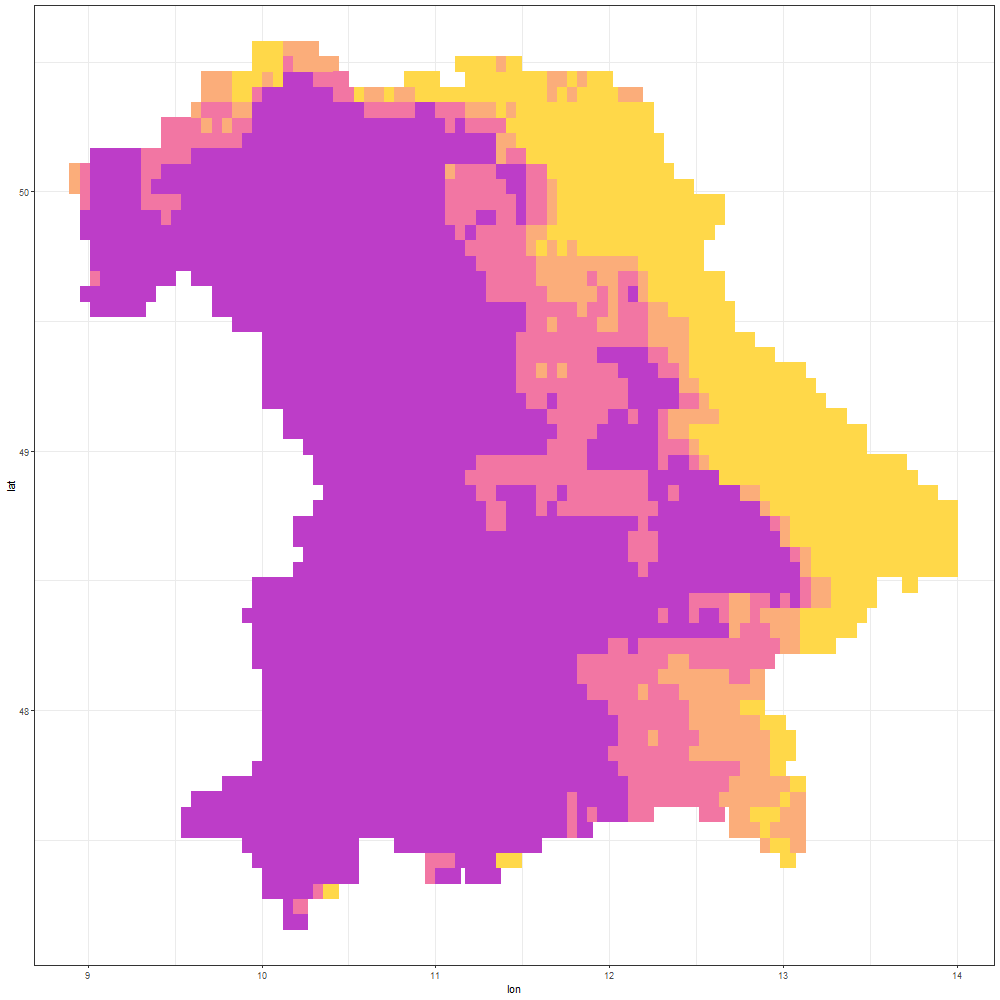}};

		\node[inner sep=0pt] (Col1) at (-15, 5)  {\includegraphics[width=0.29\textwidth]{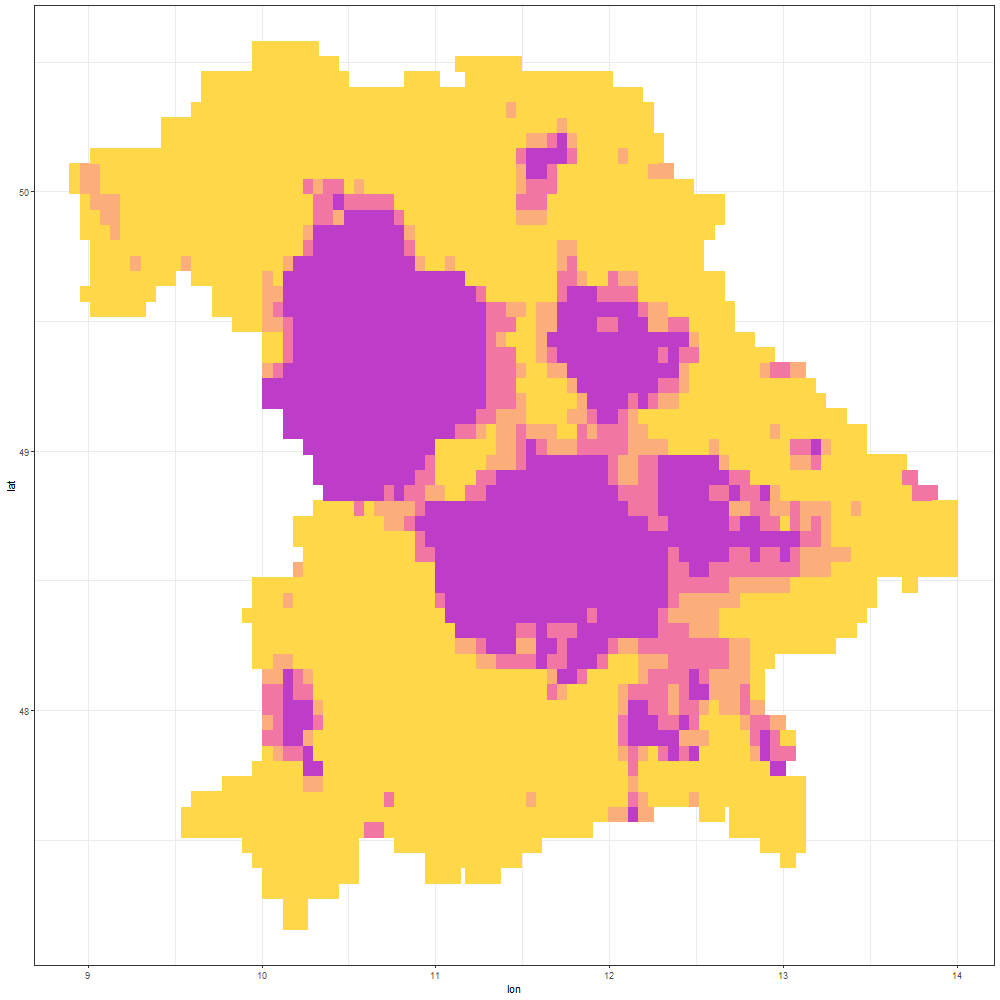}};

		\node[inner sep=0pt] (Col1) at (-10, 5)  {\includegraphics[width=0.29\textwidth]{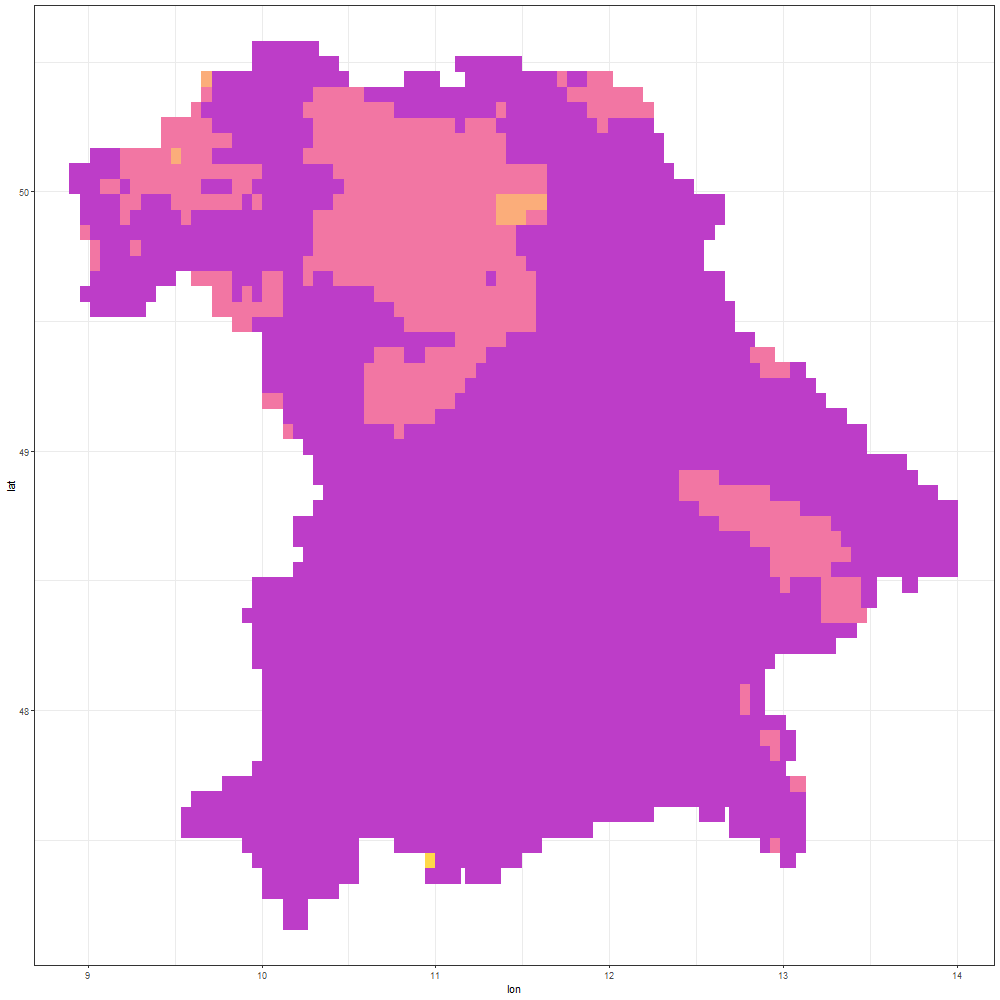}};
		\node[rotate=90] (a) at (-22.8, 0)  {1998-2020};
		
		\node[inner sep=0pt] (Col1) at (-20, 0)  {\includegraphics[width=0.29\textwidth]{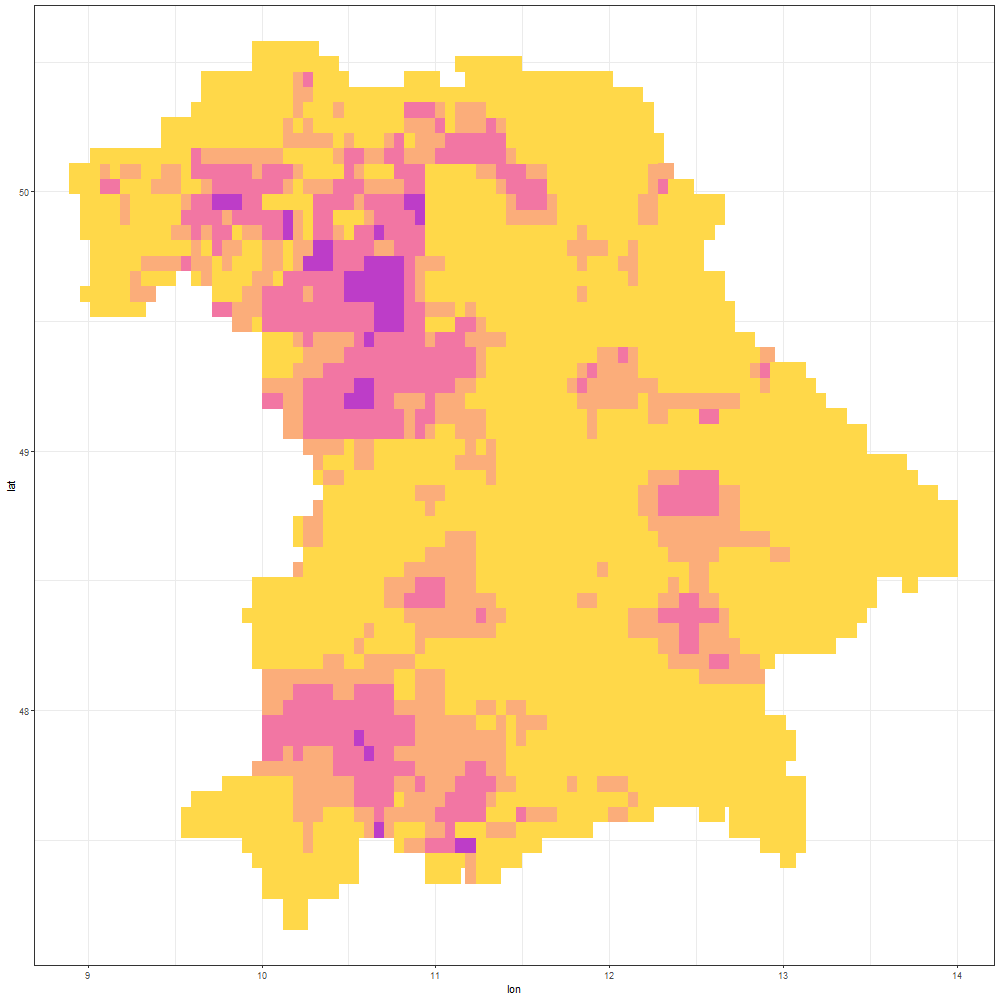}};

		\node[inner sep=0pt] (Col1) at (-15, 0)  {\includegraphics[width=0.29\textwidth]{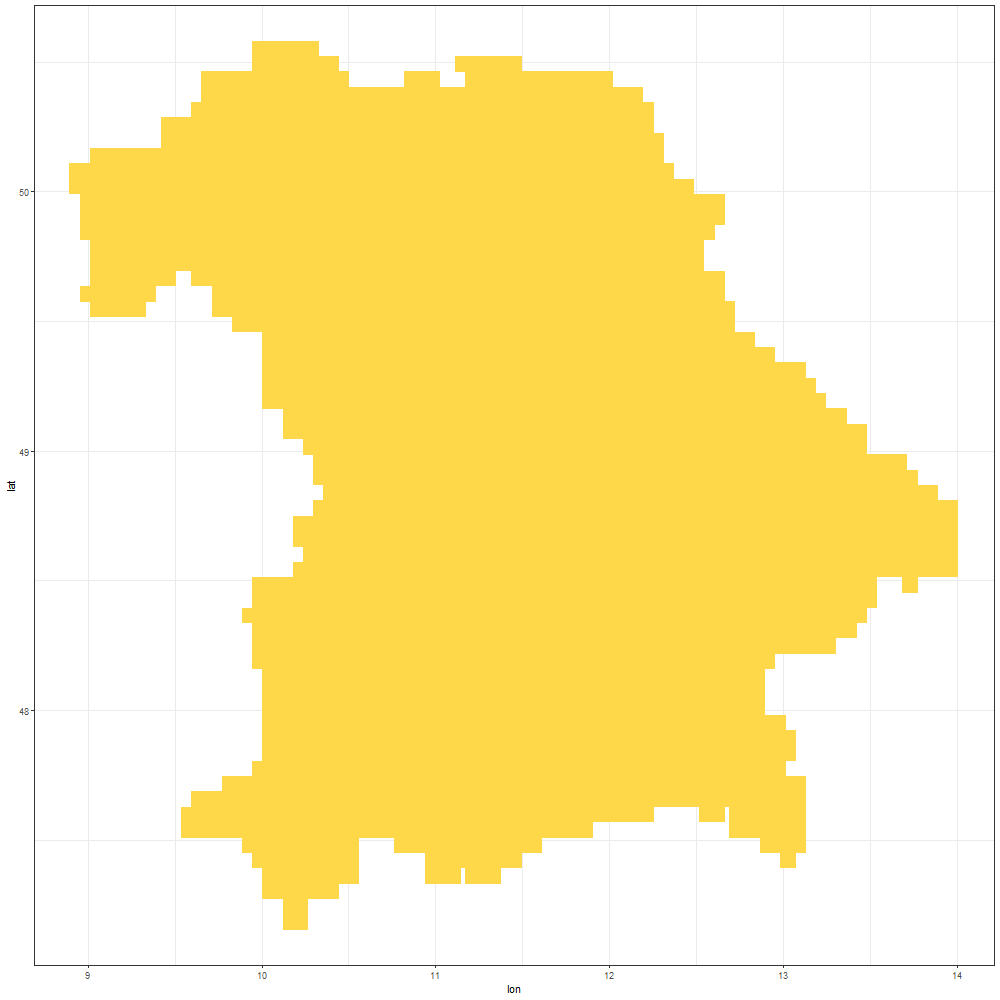}};

		\node[inner sep=0pt] (Col1) at (-10, 0)  {\includegraphics[width=0.29\textwidth]{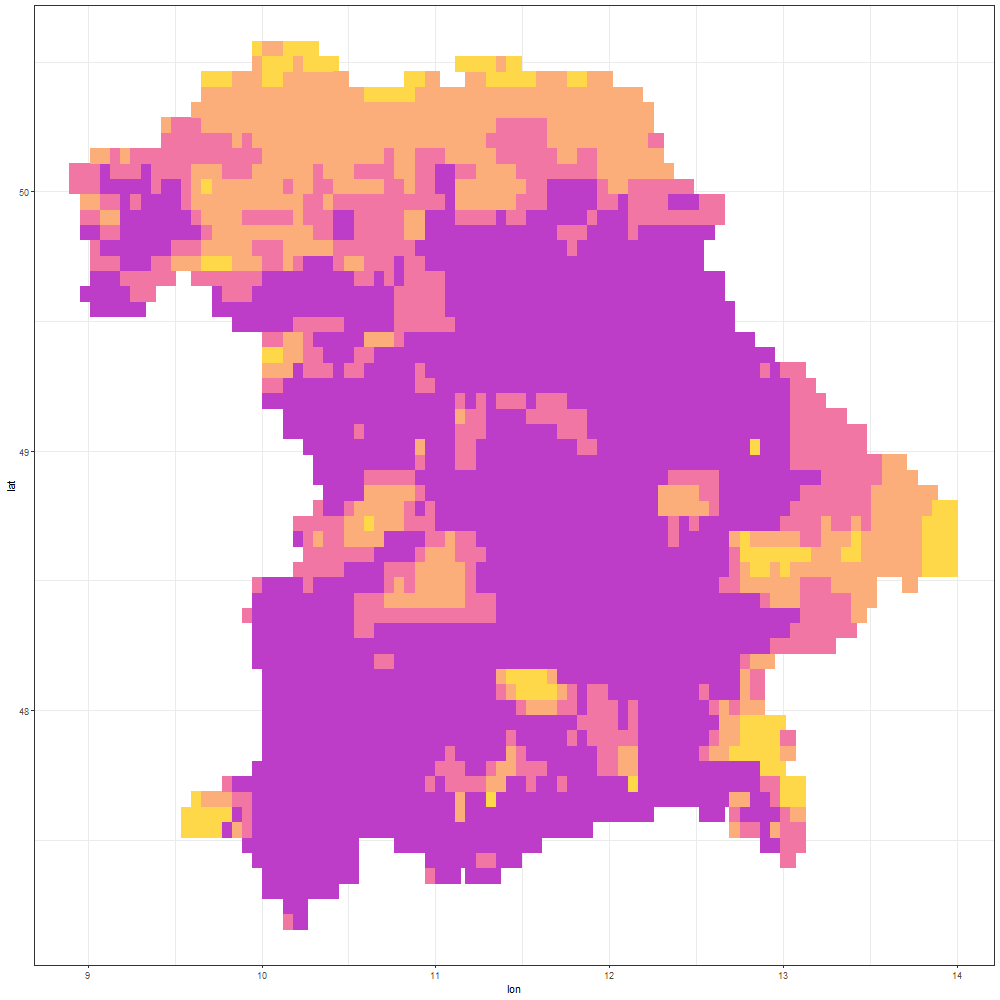}};
		
		\node[inner sep=0pt] (Col1) at (-15, -3)  {\includegraphics[width=0.6\textwidth]{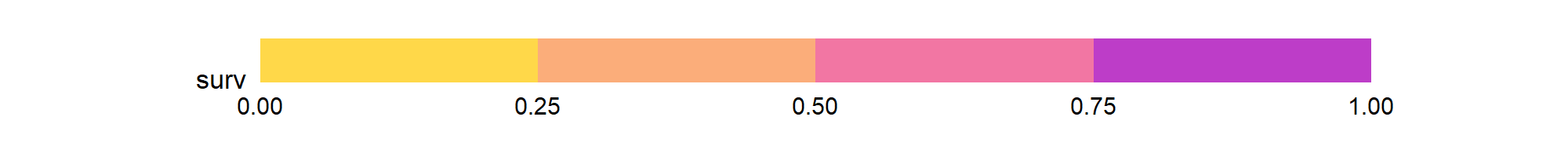}};
	\end{tikzpicture}
	\caption{Estimated survival probabilities   $\hat{S}_{\hat{\mathcal{D}}_{frost}}(s,T),$ $\hat{S}_{\hat{\mathcal{D}}_{drought}}(s,T),$  and  $\hat{S}_{\hat{\mathcal{Y}}}(s,T)$   for periods $(s,T) = \{ (1952,1974), (1975,1997), (1998,2020) \}$ (from top to bottom row).} 
	\label{fig:survival_periods}
\end{figure}
For the joint bivariate case, the survival probability is estimated as
\begin{equation}\label{eq:survivaljoint}
	\hat{S}_{\hat{\mathcal{Y}}}(s,T) \coloneqq 1 - \sum_{t=s}^{T}\hat{P}_{\hat{\mathcal{Y}}_{t}}\left(-2, -1.5| \mathbf{x}_{t,l}\right).
\end{equation}
Figure \ref{fig:survival_periods} shows the estimated survival probabilities for all considered locations for pairs $(s, T)$  evaluated at $(1952,1974), (1975,1997),$ and $ (1998,2020)$, going from top to bottom. The survival probabilities can be interpreted as the probability of a location not experiencing an extreme event in the time periods 1952-1974 (top row), 1975-1997 (middle row) and 1998-2020 (bottom row). Yellow regions have estimated survival probabilities close to 0, thus those regions have a high risk of an extreme event happening beyond the year $T$. Purple regions have estimated survival probabilities close to 1, meaning that those regions have a low risk of an extreme event happening beyond a given year $T$.
For example, purple coloured locations (i.e survival probability $\geq$ 0.75) indicate that the estimated probability of an extreme drought event not occurring in a given time period is above 75\%.
The period 1952-1974 has low estimated individual survival probabilities, i.e. high risks of extreme  occurrences, in approximately half of the considered locations, while the other half has  quite high estimated survival probabilities for both frost and drought. There are almost no chances of a joint occurrence in the majority of locations and the extreme joint events seem to be located in the north of Bavaria. In the next period, between 1975-1997, there is a small risk for a frost event, except in  the east regions of Bavaria. The estimated survival probabilities for  drought are quite low in the majority of the locations considered,  indicating that this period had higher drought risks associated than the previous period considered. The joint estimated survival probabilities are quite high for almost all locations, indicating low risk of a joint event. Also, in this period there is only one joint extreme event in 1976, in the same region as identified in the second row, third column of Figure \ref{fig:mapsextremeyeaars}. 
On the other hand, the last time period considered, 1998-2020, has almost exclusively zero estimated univariate survival probabilities for both frost and drought throughout all locations considered in Bavaria. Joint extreme events are also very likely in  for example, the north and east regions. The estimated survival probabilities derived in all three cases, both univariate and the joint, indicate a significant increase in individual risk for frost, drought and their joint occurrence in the last 20 years compared to the other two considered periods. 
\subsection{Return periods}
The return period of an event is the expected time until the event reoccurs. Depending on the usage goal and data at hand, it is defined as the expected time interval at which an event of a given magnitude is exceeded for the first time or the average of the time intervals between two exceedances of a given threshold \citep{volpi2015one}. Motivated by the goal of our study, we use the first definition 
based on the waiting time until an event happens. We define the  
\textbf{event happened if the estimated survival probability hits a threshold of 0.5}, 
which indicates the first time there will be a greater chance of the event to happen than not to, i.e.
the return period is the number of years  at which the probability of surviving is equal to or greater than $0.5$, or $50\%$. The return period for the frost index is then defined as 
$$R_{\hat{\mathcal{D}}_{frost}} \coloneqq \inf_{t \in \left[1952,2020 \right]  } \left\lbrace t| \hat{S}_{\hat{\mathcal{D}}_{frost}}(1952, t) \leq 0.5 \right\rbrace .$$
Similarly, the return periods for the univariate D-vine model for the drought index are defined and denoted as $R_{\hat{\mathcal{D}}_{drought}}$. In the bivariate case, we define the return period  as
$$R_{\hat{\mathcal{Y}}} \coloneqq \inf_{t \in \left[1952,2020 \right]  } \left\lbrace t| \hat{S}_{\hat{\mathcal{Y}}}(1952, t) \leq 0.5 \right\rbrace .$$

\begin{figure}
	\centering
	\begin{tikzpicture}
		
		\node[inner sep=0pt] (Col1) at (-20, 10)  {\includegraphics[width=0.8\textwidth]{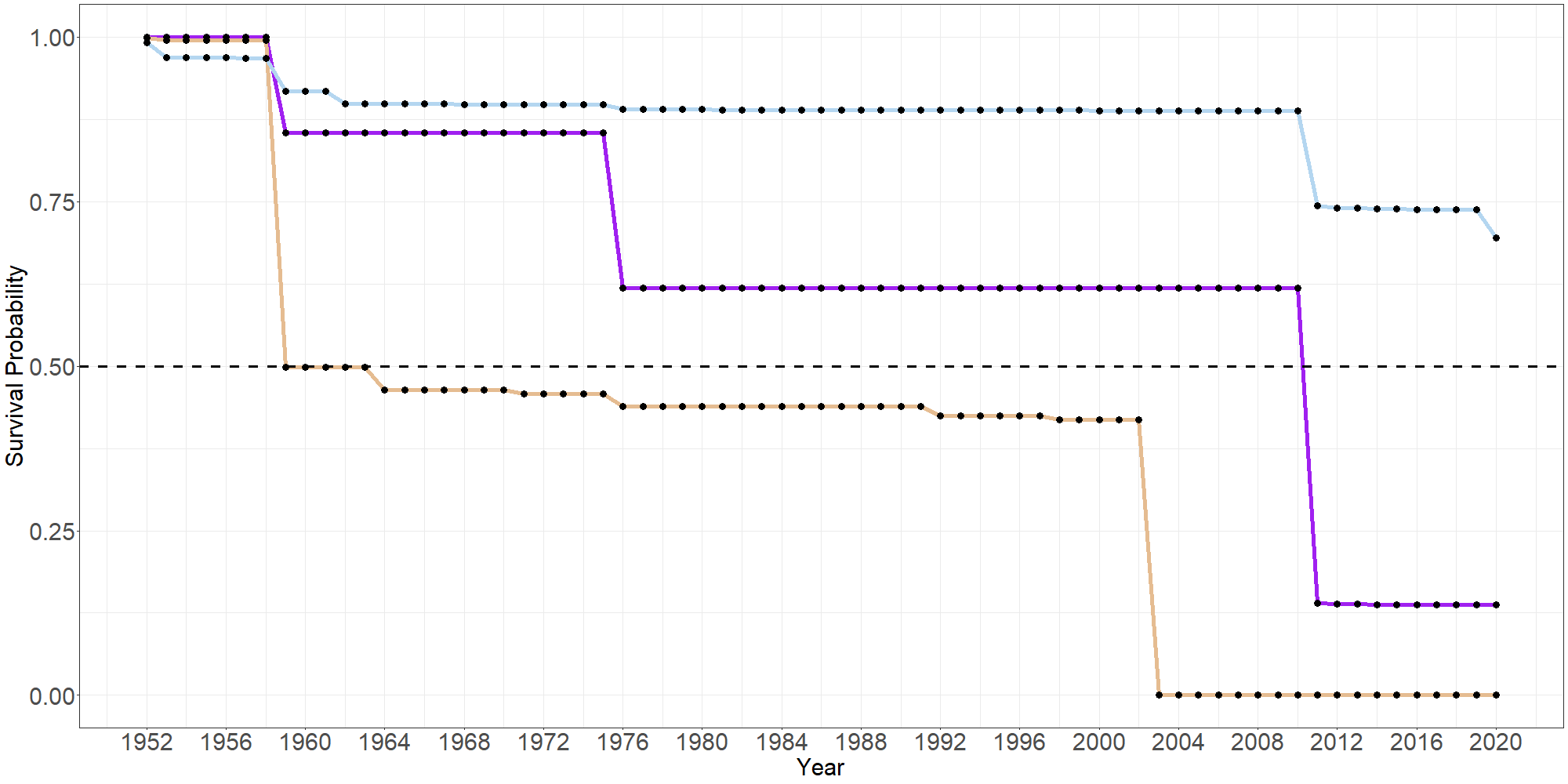}};
		
		\node[inner sep=0pt] (Col1) at (-12.55, 10)  {\includegraphics[width=0.15\textwidth]{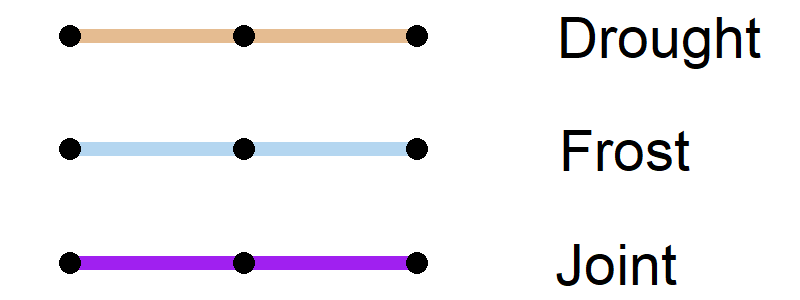}};
	\end{tikzpicture}
	\caption{Plotted are $\hat{S}_{\hat{\mathcal{D}}_{frost}}(1952,2020)$, $	\hat{S}_{\hat{\mathcal{D}}_{drought}}(1952,2020)$ and $	\hat{S}_{\hat{\mathcal{Y}}}(1952,2020)$ for the randomly chosen location with latitude coordinate 49.4874 and longitude coordinate 10.809.}
	\label{fig:climvinejointreturns}
\end{figure}

\noindent We obtain the waiting time by  evaluating $\hat{S}_{\hat{\mathcal{Y}}}(1952,t)$ for increasing values of $t$ and record the first time when $\hat{S}_{\hat{\mathcal{Y}}}(1952,t) \leq 0.5$ happens.
Since this approach assumes that the survival function is continuous and strictly decreasing, we interpolated between the discrete values of our estimated survival function $\hat{S}_{\hat{\mathcal{Y}}}(1952,t)$.  In Figure \ref{fig:climvinejointreturns} we plot the estimated survival functions for each of the 3 estimated conditional probabilites for a randomly chosen location. The dotted line represent the threshold 0.5, and its intersection with the survival functions indicates the value of the return period. Here, the return period for the drought event is   8 years, the return period for the frost risk cannot be precisely determined, as the threshold has not been reached for this location and the return period for the joint extreme event is 60 years.

\begin{figure}[]
	\centering
	\begin{tikzpicture}
		
		\node[inner sep=0pt] (Col1) at (-20, 10)  {\includegraphics[width=0.29\textwidth]{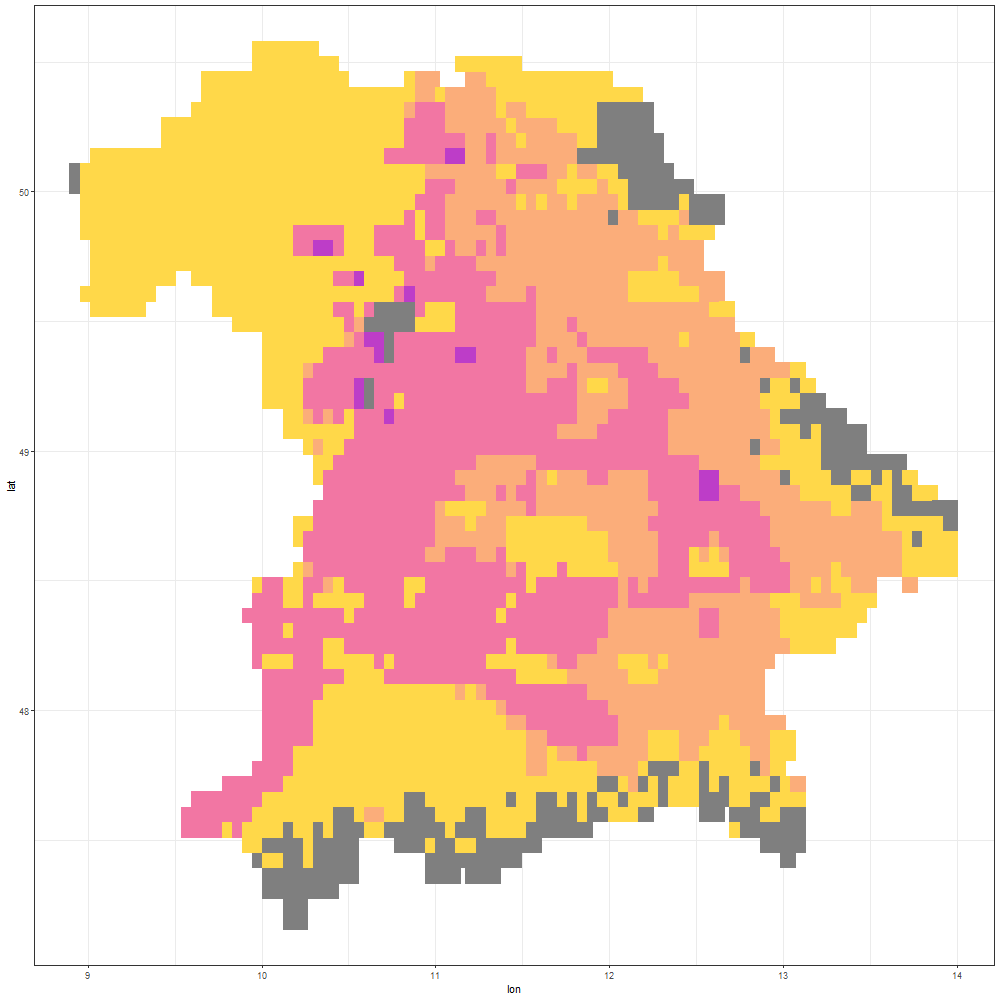}};
		
		\node (Col111) at (-20, 12.7)  {Frost};
		
		\node[inner sep=0pt] (Col1) at (-15, 10)  {\includegraphics[width=0.29\textwidth]{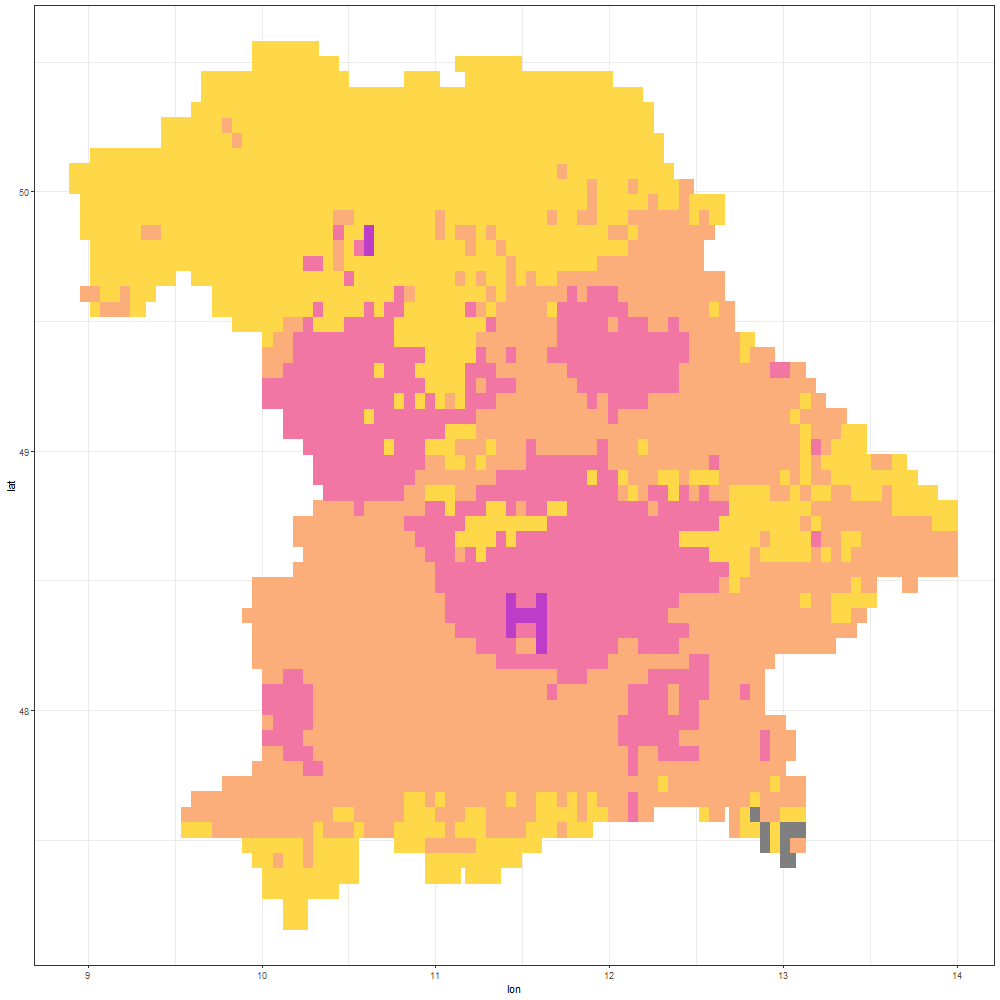}};
		
		\node (Col111) at (-15, 12.7)  {Drought};
		\node[inner sep=0pt] (Col1) at (-10, 10)  {\includegraphics[width=0.29\textwidth]{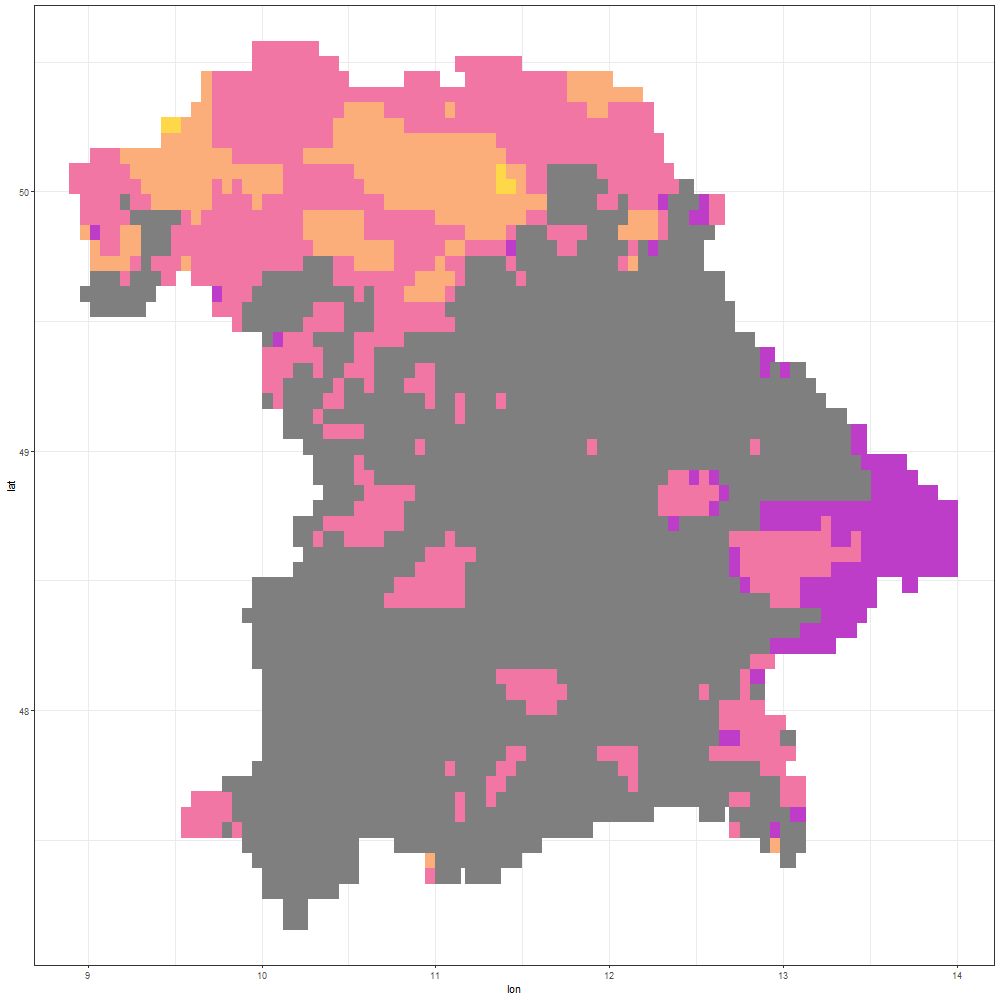}};
		\node (Col111) at (-10, 12.7)  {Frost+Drought};
		\node[inner sep=0pt] (Col1) at (-15, 7.2)  {\includegraphics[width=0.5\textwidth]{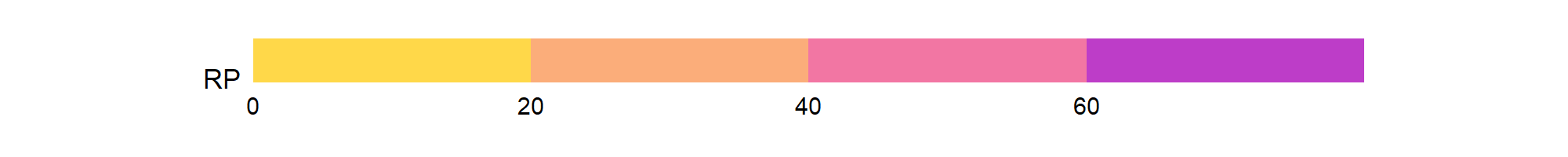}};
	\end{tikzpicture}
	\caption{Return periods for $R_{\hat{\mathcal{D}}_{frost}},  R_{\hat{\mathcal{D}}_{drought}},$  $R_{\hat{\mathcal{Y}}}.$ Regions are coloured based on the estimated return period. Distinguished are return periods of 0-20 years (yellow), 21-40 (orange), 41-60 (pink) and $>60$ (purple). Gray regions denote regions where the threshold has not been reached.}
	\label{fig:returntimes}
\end{figure}
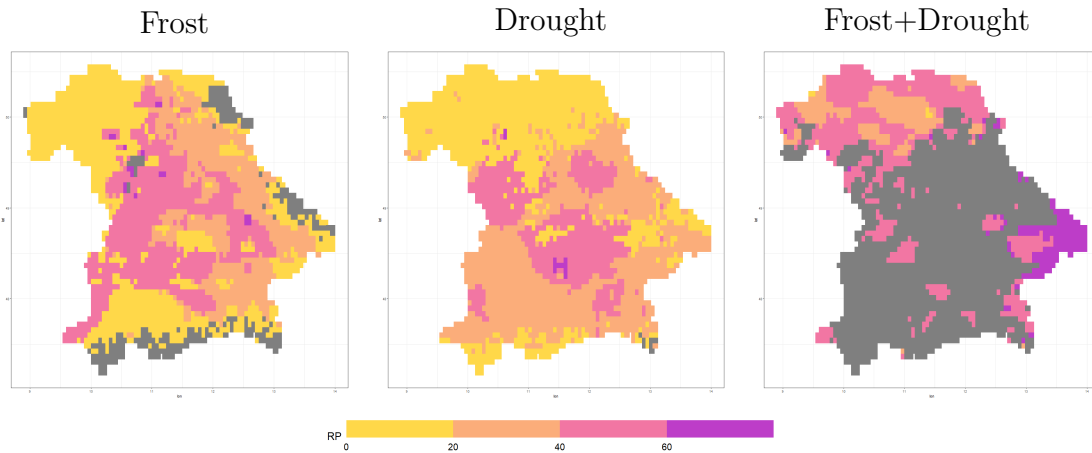

Figure \ref{fig:returntimes} shows the estimated return periods for all  locations and  the 3 different models considered, 
$R_{\hat{\mathcal{D}}_{frost}},  R_{\hat{\mathcal{D}}_{drought}},$  $R_{\hat{\mathcal{Y}}}.$ Regions are coloured based on the value of the estimated return period. We distinguish between regions with a return periods of 0-20 years (yellow), 21-40 years (orange), 41-60 years (pink) and more than $60$ years (purple). Gray regions show regions for which  the threshold of 0.50 has not been reached in the 69 studied years. Based on these plots we can identify \textbf{temporal "at-risk regions"} based on the estimated return periods. The temporal highest risk regions have the lowest return times and the lowest risk regions have the highest return times  or return times that are greater than 69 years. For  the  return periods associated with the univariate regression models, the highest  risk is shown approximately in the northern,  central-east border regions and southern regions of Bavaria. For the return period of the joint vine regression model we see lower risks and the estimated highest risks are in the northern and central-east border regions.

\section{Conclusion and outlook}\label{climvine:conclusionoutlook}

In this study, we  applied vine copula based regression models studying the effect of climate change on two important climatological variables individually and  jointly, after showing that there is a need to go beyond multivariate Gaussian approaches. 
We utilized D-vine copula models to analyze drought and frost indices separately, and the Y-vine copula model for joint modeling. By fitting annual models, we proposed conditional risk measures to quantify univariate and bivariate risks of extreme events. We identified years with extreme risks and conducted survival probability and return times analyses to identify "at-risk" spatial and temporal regions. 



Our results demonstrate the potential of joint probability modeling of drought and late spring frost to inform forest management. \textit{Fagus sylvatica}, as a species with a large geographic distributional range, has developed a high amount of genetic variability on the level of the meta-population \citep{schuldt2016adaptable}, leading to genotypes with locally specific adaptation to climatic constraints. The resulting provenances additionally display a high degree of phenotypic plasticity \citep{robson2018phenotypic}, leading to provenances with e.g. earlier or later leaf-out \citep{chmura2002variability}. As a consequence, especially provenances from the southern part of the distribution are more drought resistant \citep{cavin2017highest}, and have been discussed as potential alternatives to local provenances under climate change conditions \citep{rose2009marginal}. Our results indicate that forest adaptation to increasing frequency and intensity of drought events in the course of climate change must not only optimize for drought resistance, but simultaneously consider the adaptation to late spring frost, predominantly in the areas identified as risk regions by means of joint probability modeling. As there are clear trade-offs between drought- and frost resistance across the distributional range of the species \citep{muffler2020lowest}, provenance choice especially for the regions identified as risk regions must consider the adaptation to both extremes, which is available in some provenances \citep{seho2021internationale}.
For evaluating future risk, our approach should be expanded to incorporate future climate conditions. For adaption planning, an important information from this application will then be how the joint probability and frequency of co-occurring drought and frost events will shift under different climate change scenarios. Furthermore, the approach should be applied to other important tree species with a known sensitivity for both drought and late spring frost. Our study is currently restricted to Bavaria, and while an expansion to larger regions would be desirable, it is currently limited by the availability of high-resolution climate projections that adequately capture extreme events in primary climate variables, particularly drought and spring late-frost. Recent advancements in climate downscaling, utilizing deep convolutional neural networks with batch normalization and residual networks, offer promising prospects for generating accurate high-resolution climate datasets for the European domain.\\

\noindent \textbf{Acknowledgments}\\
This work was supported by the  Munich Data Science Institute Seed Fund: Climvine. Tepegjozova and Czado are also supported by the Deutsche Forschungsgemeinschaft[DFG CZ 86/6-1].
	
\bibliographystyle{abbrvnat}
\bibliography{bibliographynew}

\end{document}


\section{Data description and exploratory analysis}

\begin{landscape}	
	\begin{figure}
		\centering
		\begin{tikzpicture}
			
			\node[inner sep=0pt] (Col1) at (-26, 7)  {\includegraphics[width=0.67\textwidth]{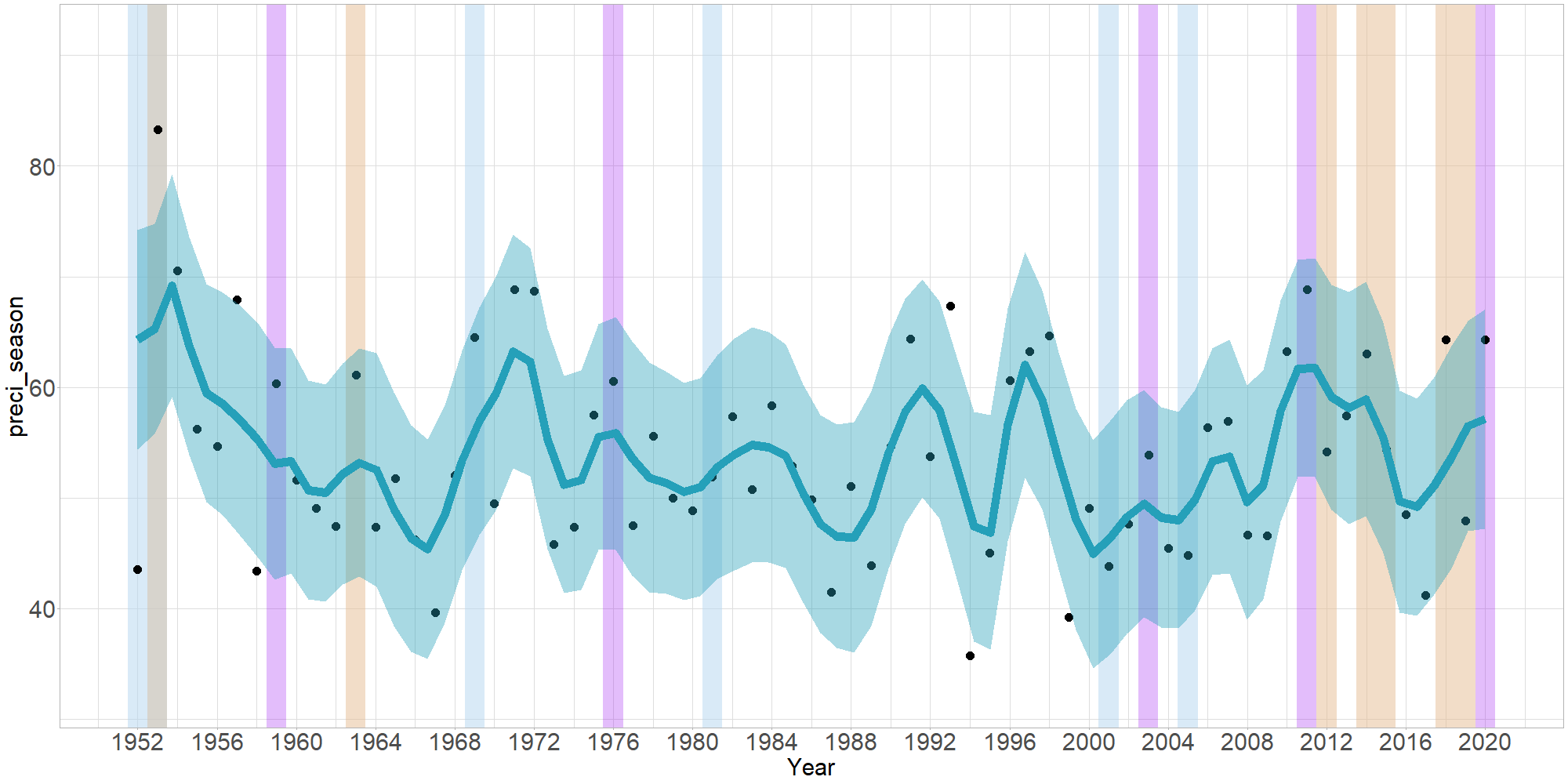}};
			
			\node[inner sep=0pt] (Col1) at (-16, 7)  {\includegraphics[width=0.67\textwidth]{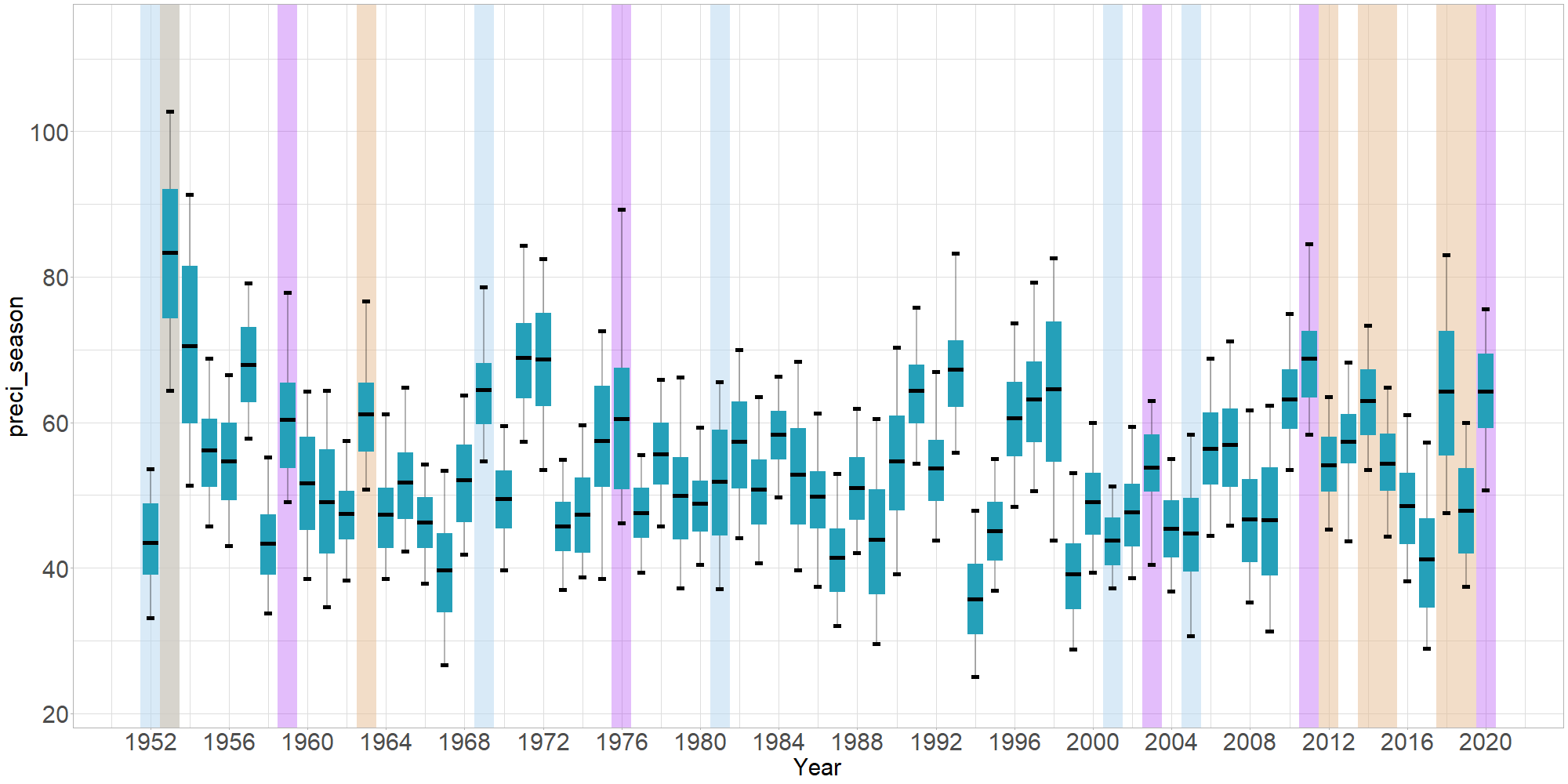}};

			
			
			
		\end{tikzpicture}		
	\caption{\textit{Left column}: the points denote the mean observations per year over all gridcells (1952-2020), the smoothed line is a fitted moving average model, the shaded area is the 95\% CI for each variable. \textit{Right column:} the horizontal line represents the 95\% CI per year, the box represents the 50\% CI, and the horizontal line is the annual mean value over all gridcells.  The vertical ribbons denote extreme years identified for frost risk (blue), drought risk (apricot), joint frost and drought risk (purple), and marginal drought and frost risk, but not joint risk identified (light gray).}
	\label{climvine:edaplotss}
	\end{figure}
\end{landscape}

 In each left panel for each variable in Figure \ref{climvine:edaplotss}, we show the annual mean observation over all gridcells per each year in the period 1952-2020. The smoothed line is the fitted moving averages model and the shaded area is the corresponding 95\% confidence interval (CI) for each variable. Using these plots we can observe how the mean values change over the historical period and what is the time trend for each variable. For example, we can observe the increasing trend for the  \texttt{temp\_mean} or   \texttt{temp\_warm}  over the years or the clearly decreasing trend of \texttt{preci} over the last 20 years (2000-2020). Also, since our main goal is understanding the drought and frost indices, we can easily see the decreasing trend  for both of these variables in the last 20 years (2000-2020), which implies worse frost and drought conditions. Also, we can  identify outliers in these variables, years which had low average values of frost and drought indices. For example, the lowest value of the annual average frost index is achieved in the year  2011, while the lowest value of the drought index is in the year 2003. 

The right panels of Figure \ref{climvine:edaplotss} for each variable, shows  the annual mean averages over all gridcells, the corresponding  95\% (interval between whiskers) and 50 \%  (interval given by the boxes) confidence interval per year over all gridcells. This plot gives more information about the variability of each variable over all the possible gridcells for each year. For example, for \texttt{temp\_wet} and \texttt{temp\_dry} we can see very big variability in the observations for some years, while for others the majority of the observations are very close to each other. This implies that there are years in which these two variables vary over the locations we have considered, while for others, almost all the locations exhibit similar conditions. 

The years that are identified as extreme by each model are also highlighted with a vertical ribbon in the background of Figure \ref{climvine:edaplotss}. The blue ribbon represents years identified only by the frost D-vine model, the apricot colored ribbon represents years identified only by the drought D-vine model,  purple ribbons represents years identified by the joint Y-vine model and light gray (only year 1953) ribbon is where both the univariate  models identifies risks, but not the joint Y-vine model.

\begin{figure}
	\includegraphics[width = \textwidth]{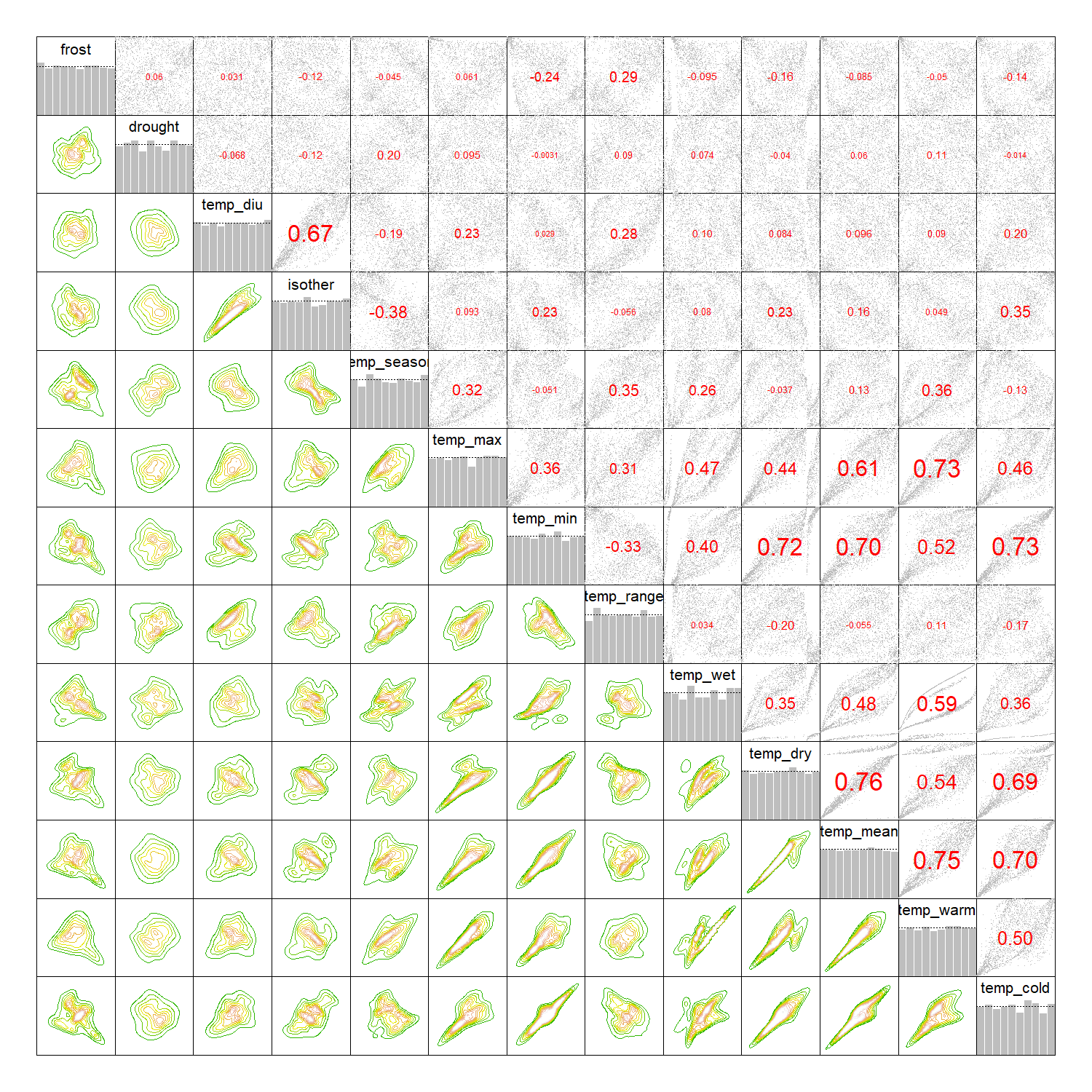}
	\centering
	\caption{Lower diagonal: marginally normalized contour plots, upper diagonal: pairwise scatter plots with the associated empirical Kendall's $\hat{\tau}$ values and on the diagonal: histograms of the u-data,  for the 2 responses and the \textbf{temperature related} predictor variables  for all 2867 locations in year \textbf{2011}.}
	\label{fig:pairsplot2011_1}
\end{figure}

\begin{figure}
	\includegraphics[width = \textwidth]{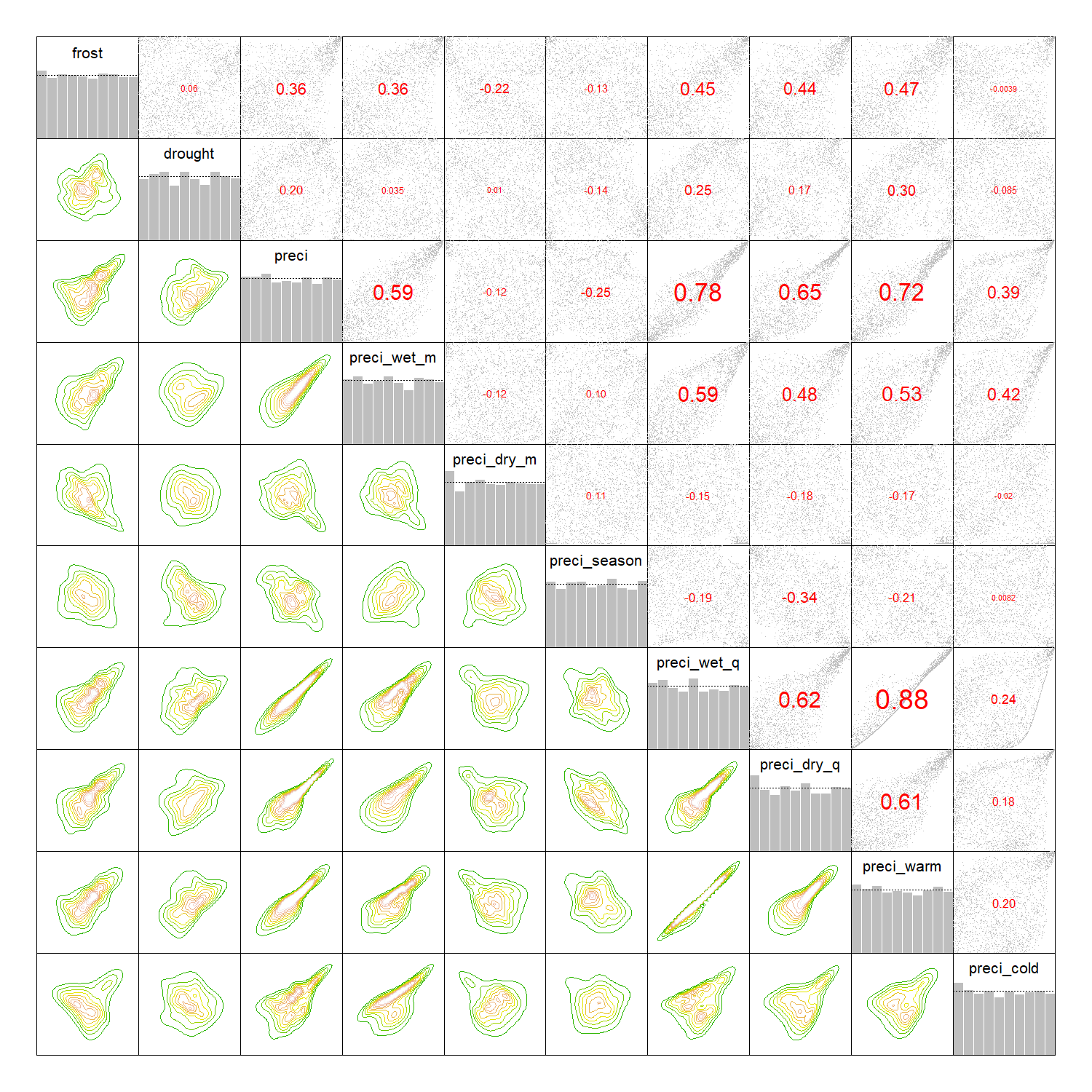}
	\centering
	\caption{Lower diagonal: marginally normalized contour plots, upper diagonal: pairwise scatter plots with the associated empirical Kendall's $\hat{\tau}$ values and on the diagonal: histograms of the u-data,  for the 2 responses and the \textbf{precipitation related} predictor variables  for all 2867 locations in year \textbf{2011}.}
	\label{fig:pairsplot2011_2}
\end{figure}

\begin{figure}
	\includegraphics[width = \textwidth]{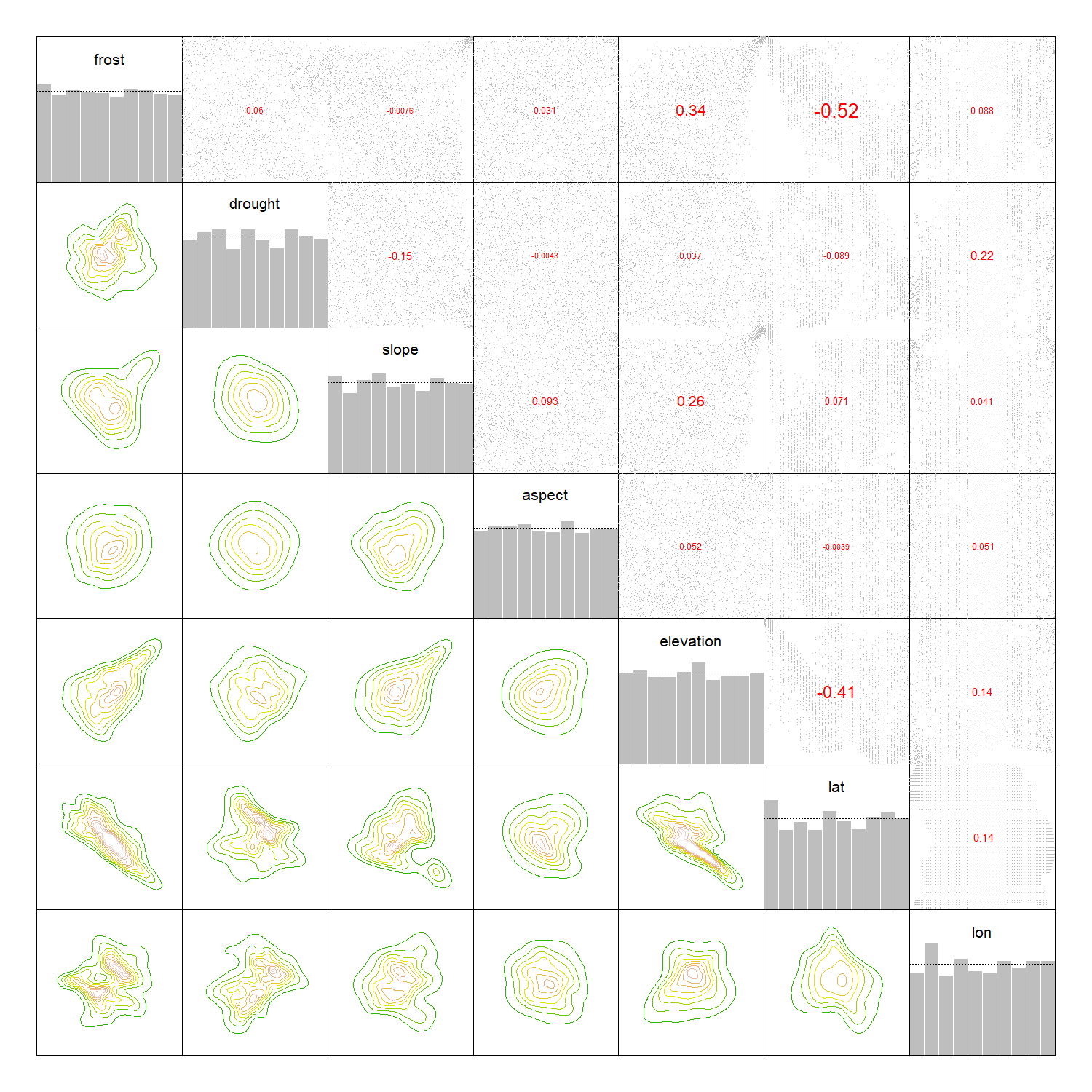}
	\centering
	\caption{Lower diagonal: marginally normalized contour plots, upper diagonal: pairwise scatter plots with the associated empirical Kendall's $\hat{\tau}$ values and on the diagonal: histograms of the u-data,  for the 2 responses and the \textbf{topographic} predictor variables   for all 2867 locations in year \textbf{2011}.}
	\label{fig:pairsplot2011_3}
\end{figure}

Figures  \ref{fig:pairsplot2011_1}, \ref{fig:pairsplot2011_2}, \ref{fig:pairsplot2011_3} shows marginally normalized contour plots, where the marginal distributions are fitted in a nonparametric manner, using kernel density smoothing, for year 2011. Each plot is based on all 2867 locations for the two responses and a subset of the predictors.

\begin{landscape}	
	\begin{figure}
	\centering
	\begin{tikzpicture}
		
		\node[inner sep=0pt] (Col1) at (-26, 7)  {\includegraphics[width=0.67\textwidth]{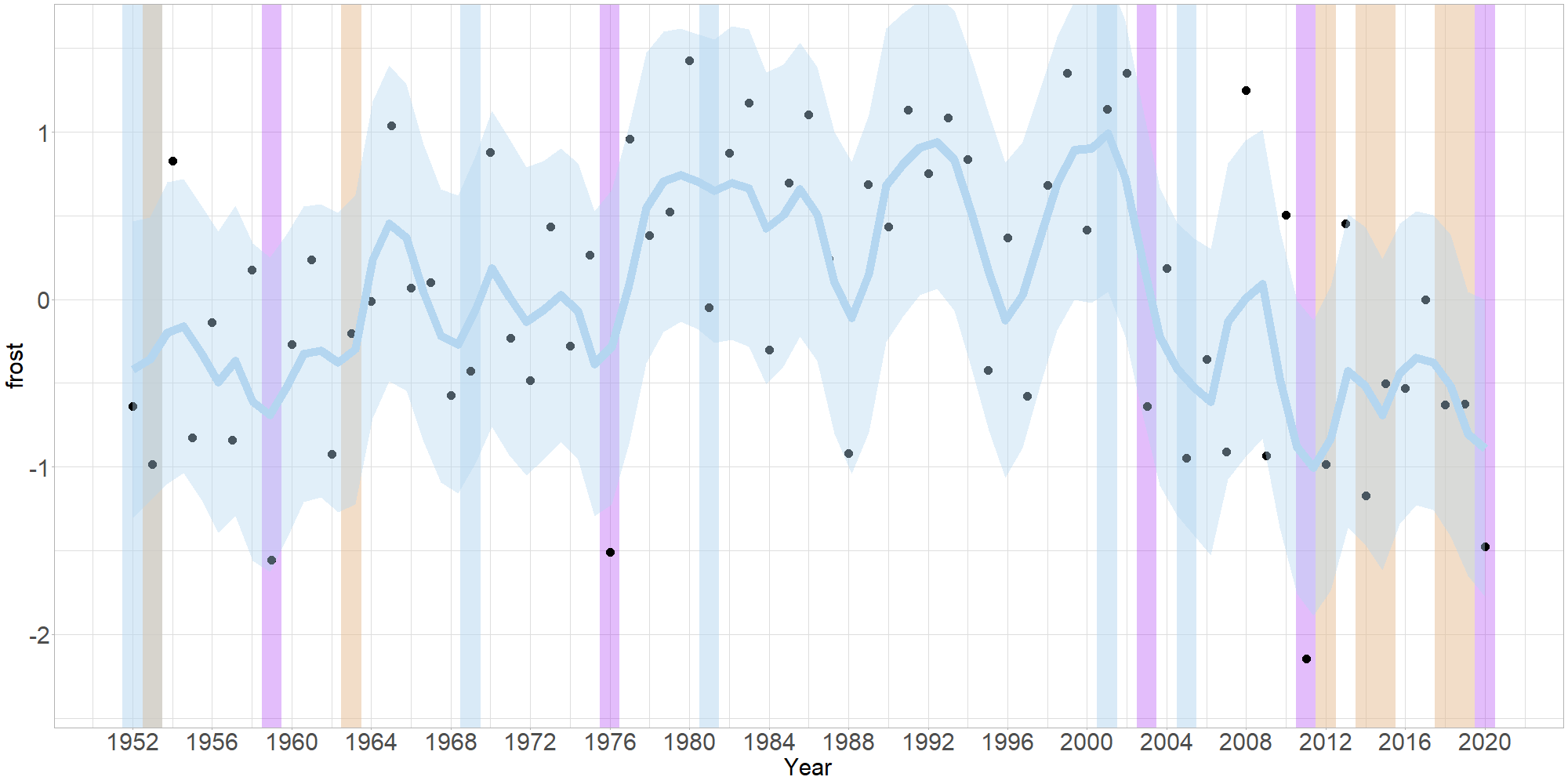}};
		
		\node[inner sep=0pt] (Col1) at (-16, 7)  {\includegraphics[width=0.67\textwidth]{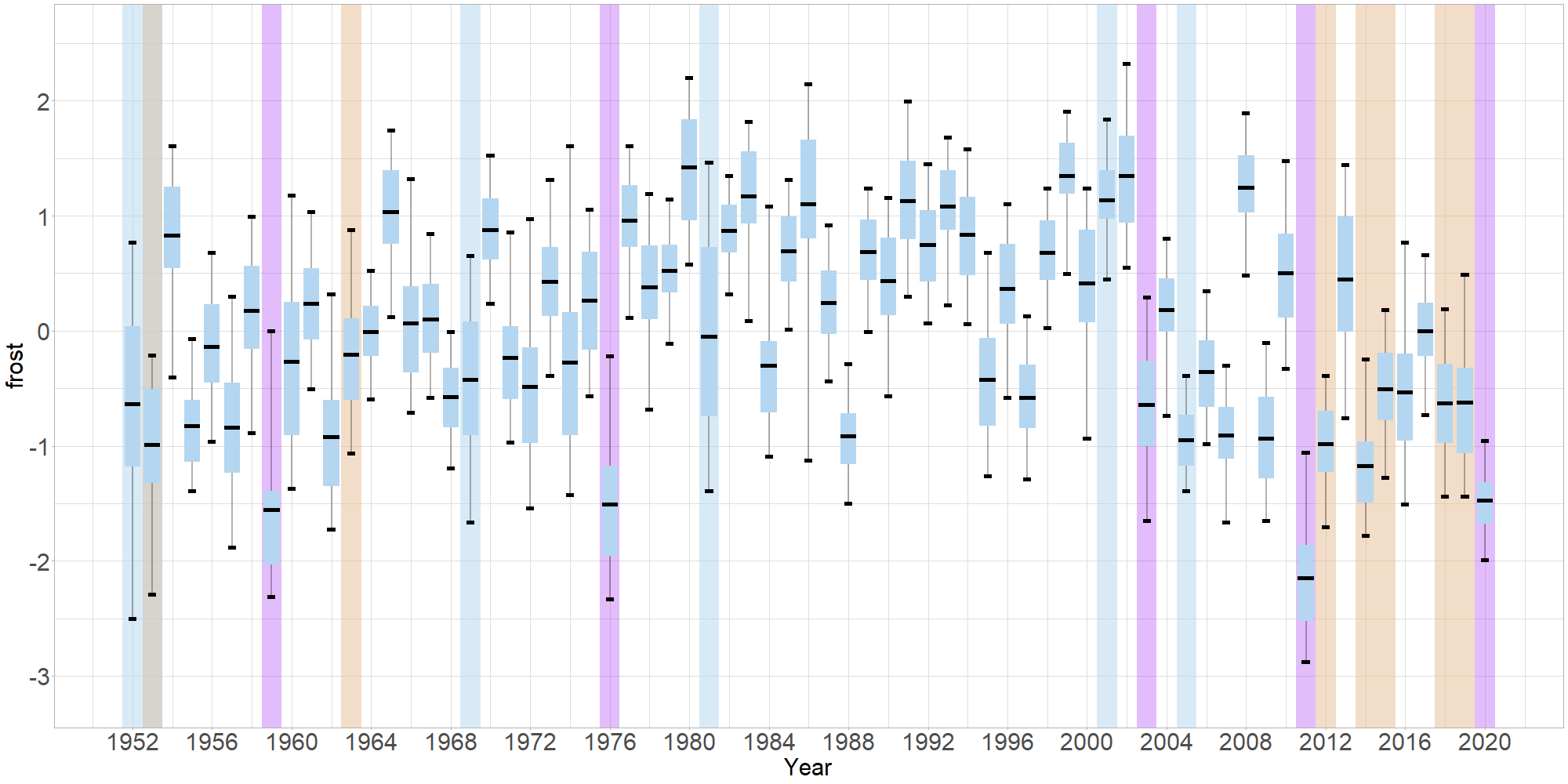}};

		
		\node[inner sep=0pt] (Col1) at (-26, 0)  {\includegraphics[width=0.67\textwidth]{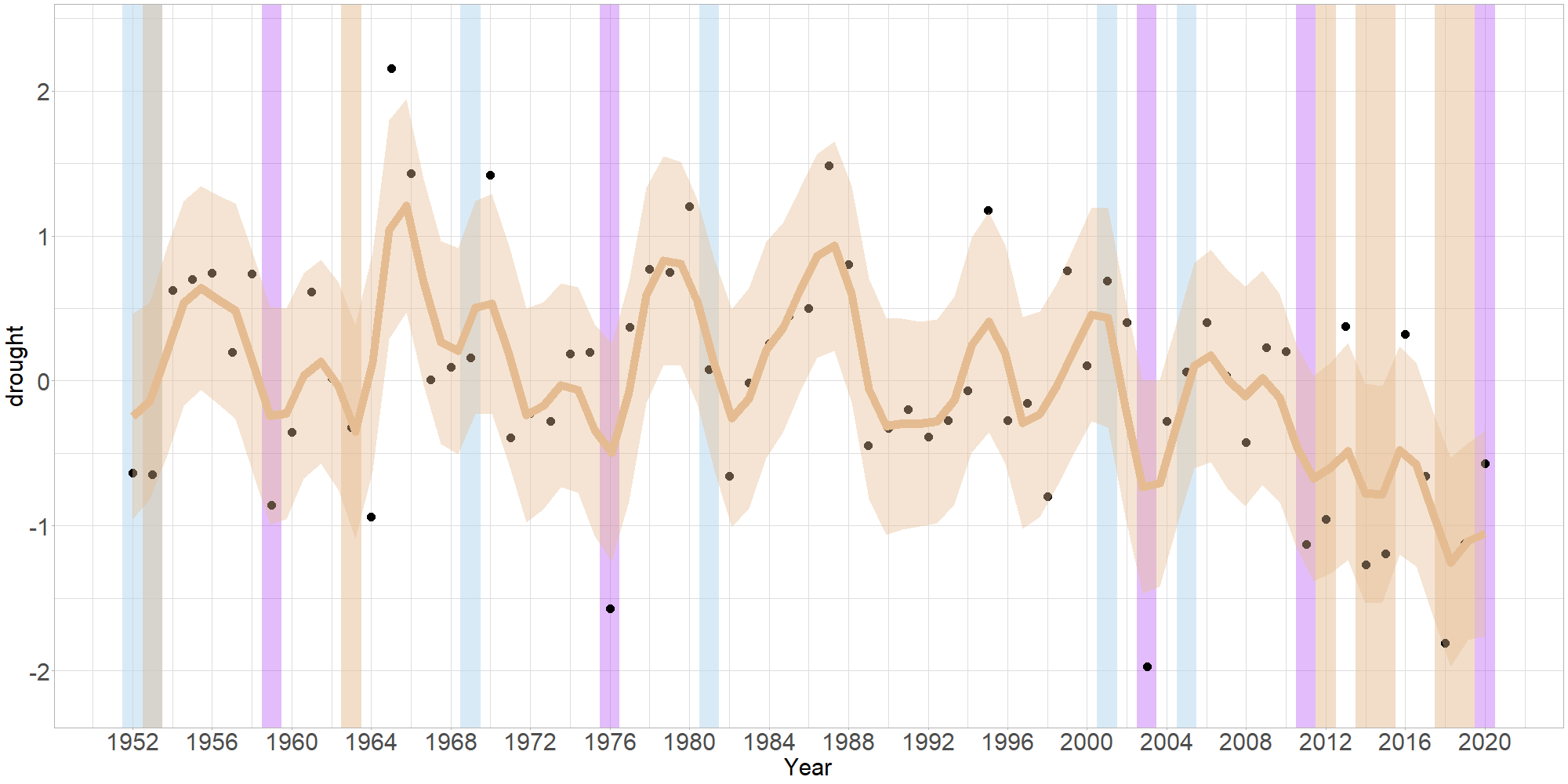}};
		
		\node[inner sep=0pt] (Col1) at (-16, 0)  {\includegraphics[width=0.67\textwidth]{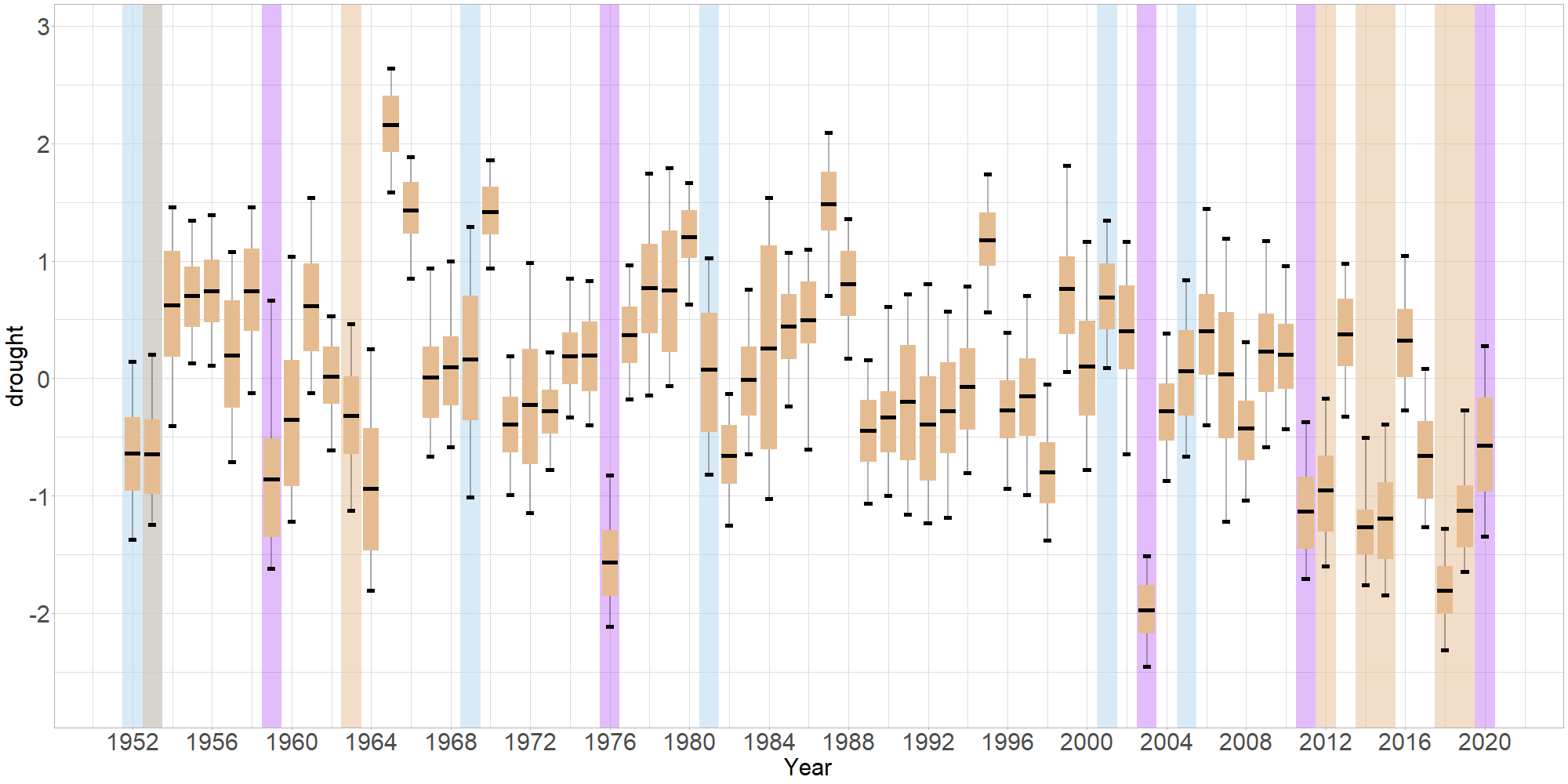}};
%
%
%
		
		
		
		\end{tikzpicture}		
	\end{figure}
\end{landscape}

\begin{landscape}	
	\begin{figure}
		\centering
		\begin{tikzpicture}
			
			\node[inner sep=0pt] (Col1) at (-26, 7)  {\includegraphics[width=0.67\textwidth]{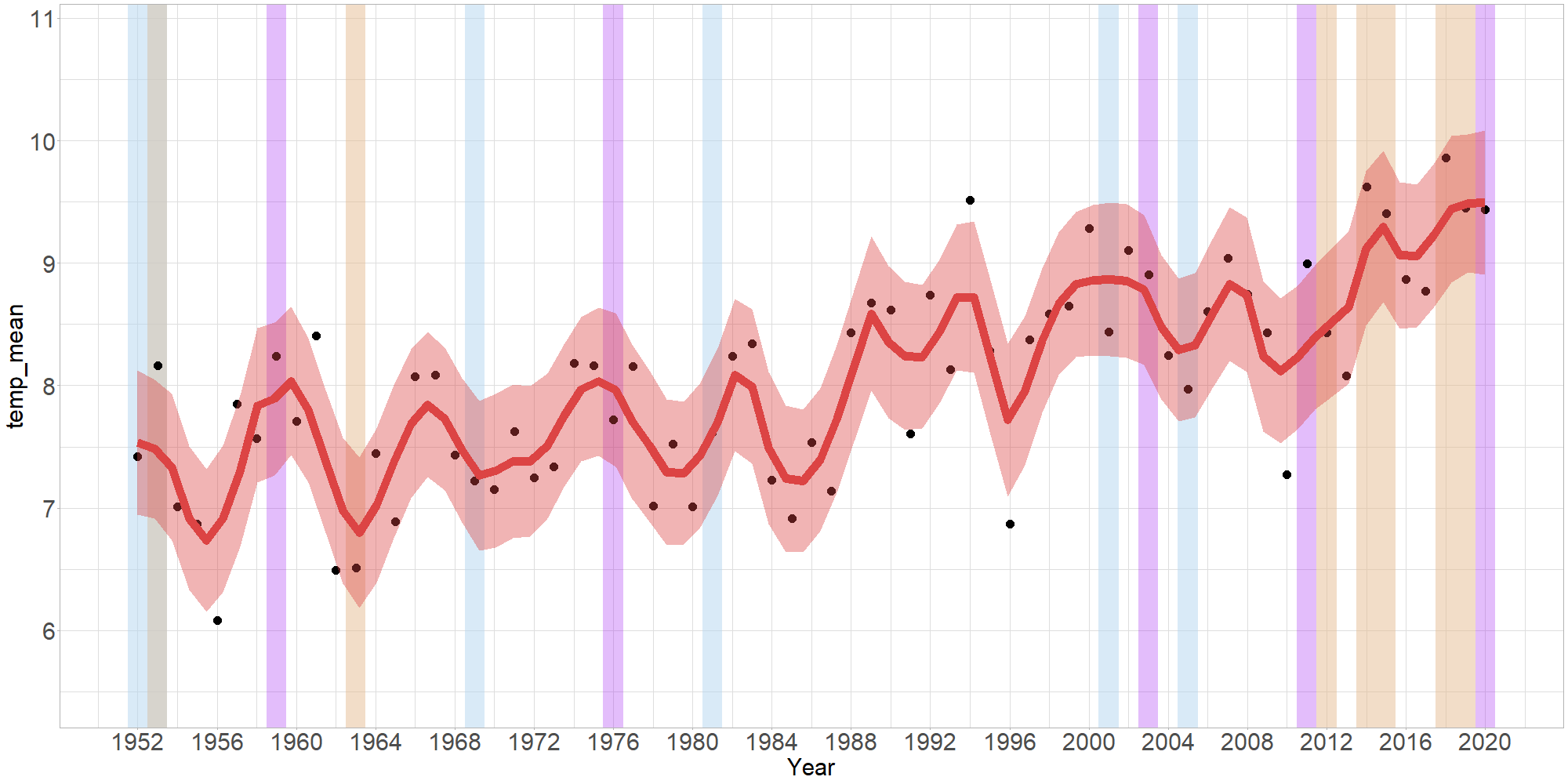}};
			
			\node[inner sep=0pt] (Col1) at (-16, 7)  {\includegraphics[width=0.67\textwidth]{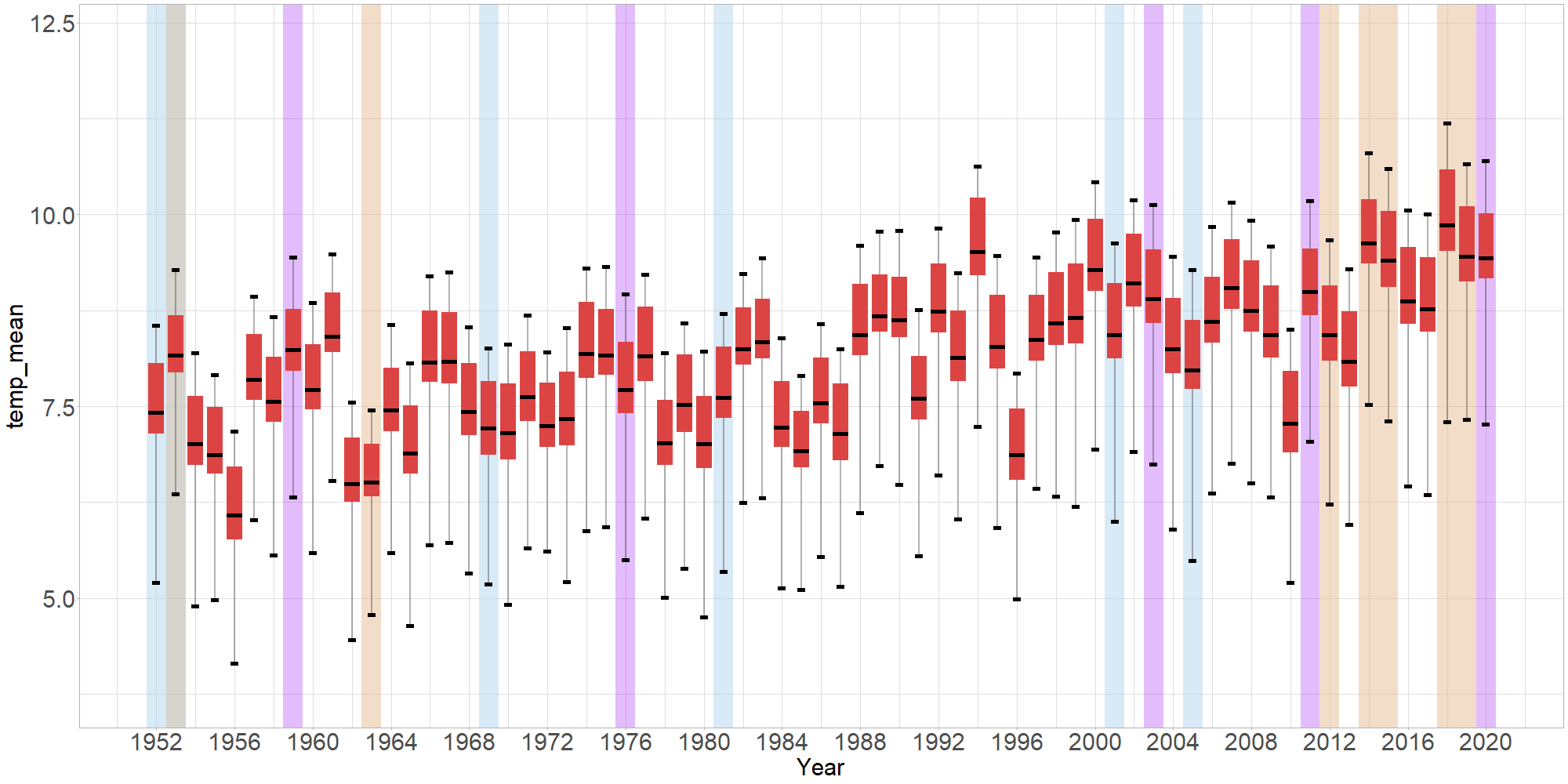}};

			
			\node[inner sep=0pt] (Col1) at (-26, 0)  {\includegraphics[width=0.67\textwidth]{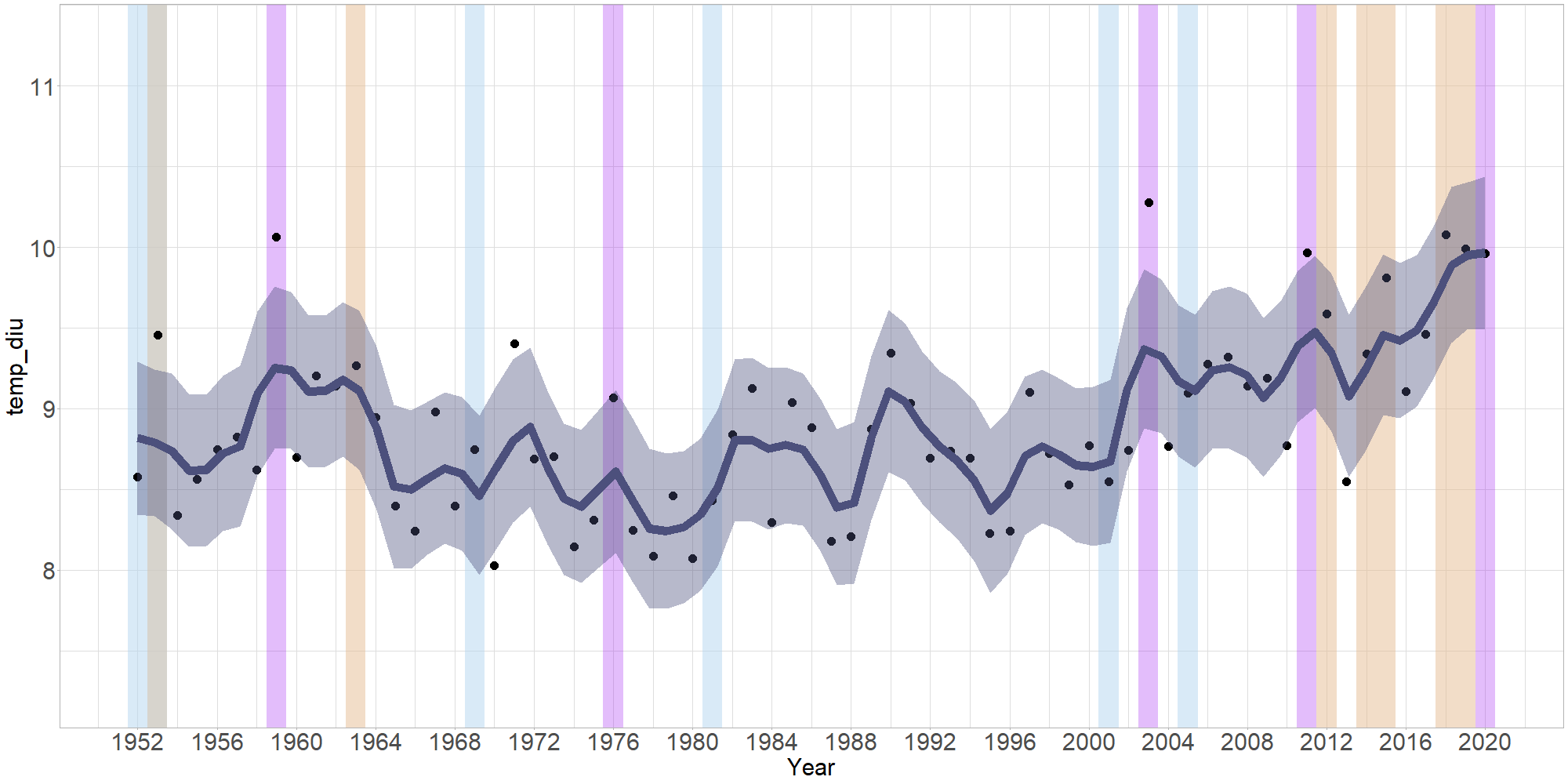}};
			
			\node[inner sep=0pt] (Col1) at (-16, 0)  {\includegraphics[width=0.67\textwidth]{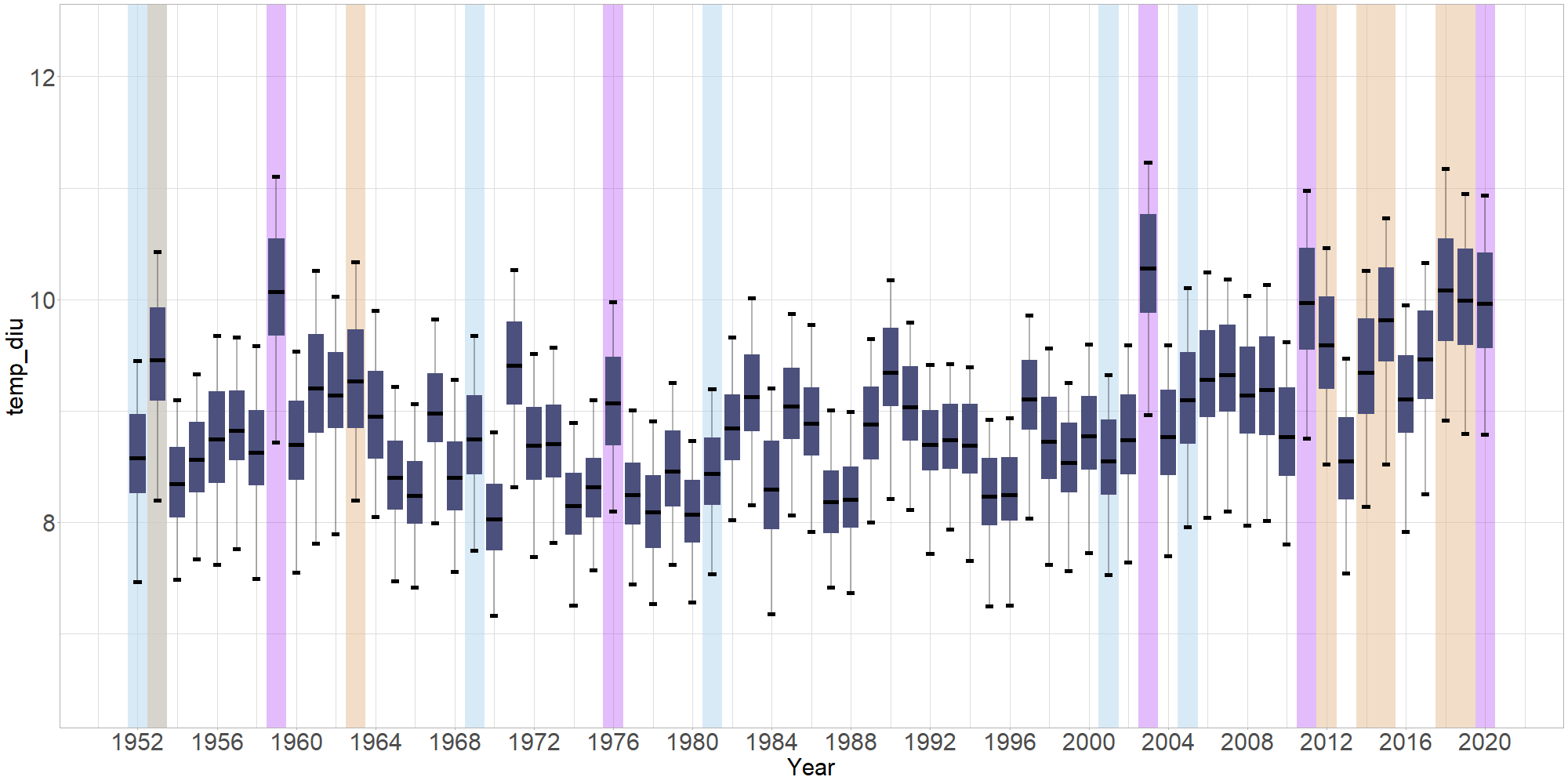}};

		\end{tikzpicture}		
\end{figure}
\end{landscape}	

\begin{landscape}	
	\begin{figure}
		\centering
		\begin{tikzpicture}
			
			\node[inner sep=0pt] (Col1) at (-26, 7)  {\includegraphics[width=0.67\textwidth]{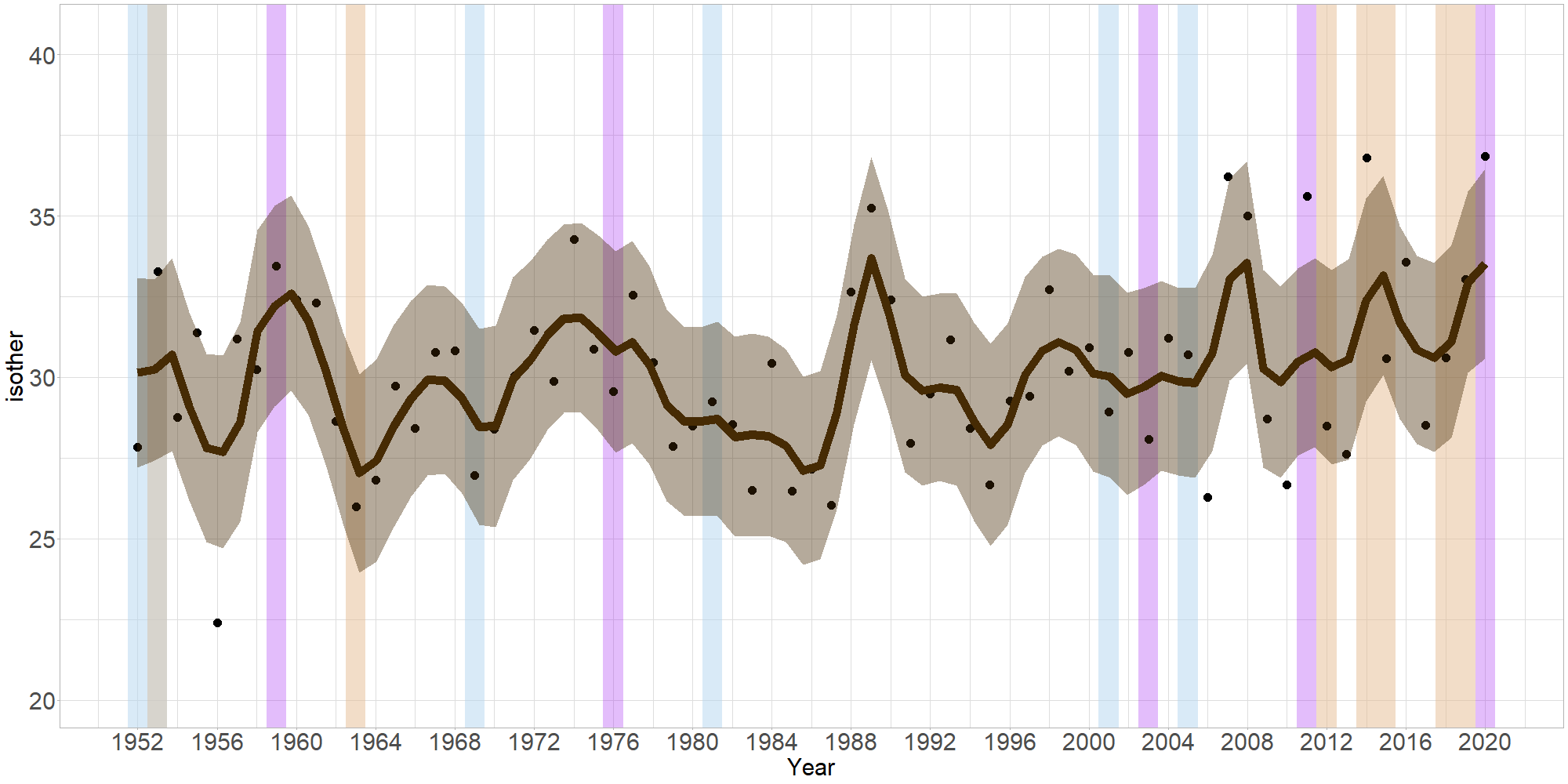}};
			
			\node[inner sep=0pt] (Col1) at (-16, 7)  {\includegraphics[width=0.67\textwidth]{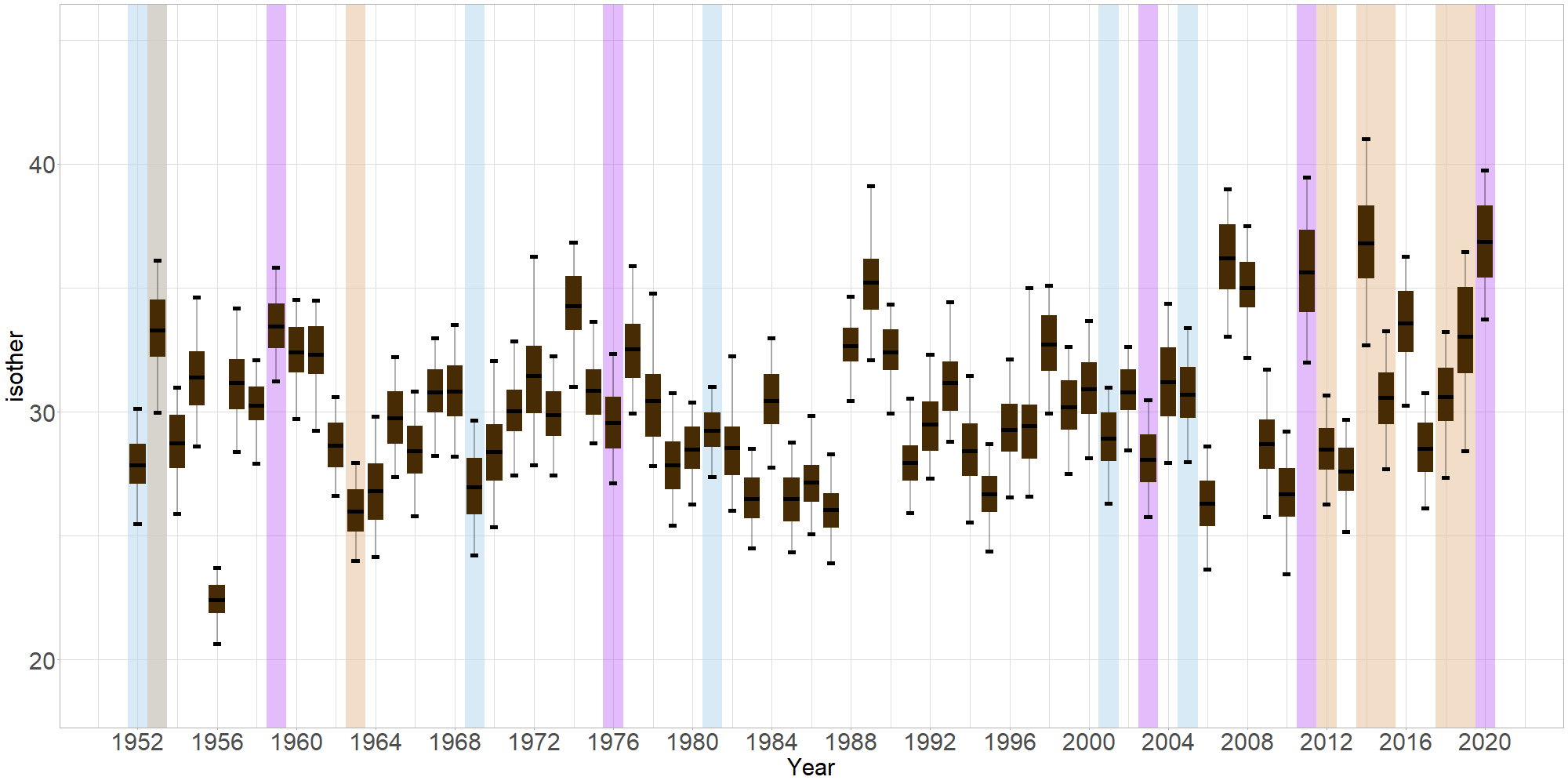}};

			
			\node[inner sep=0pt] (Col1) at (-26, 0)  {\includegraphics[width=0.67\textwidth]{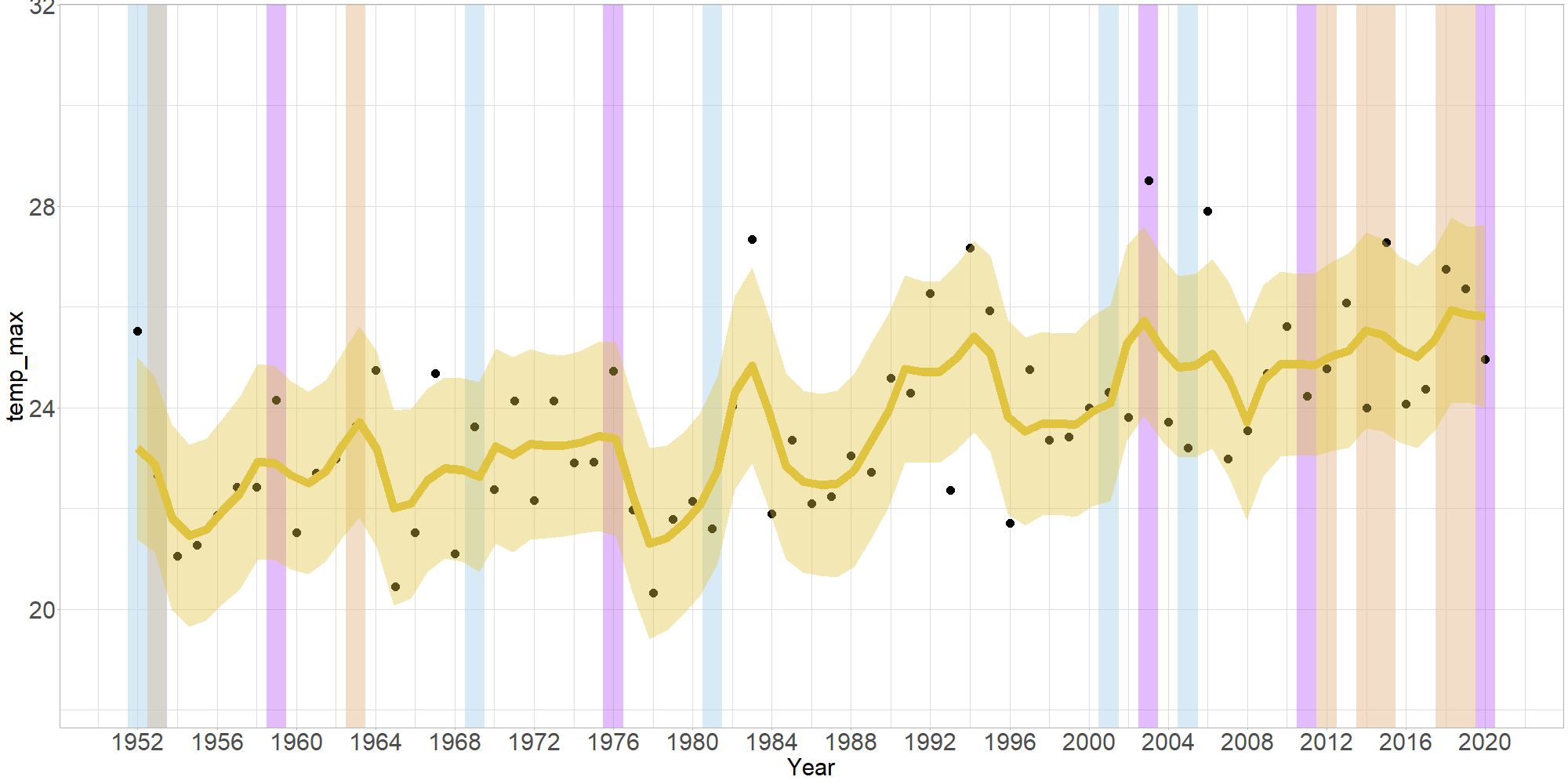}};
			
			\node[inner sep=0pt] (Col1) at (-16, 0)  {\includegraphics[width=0.67\textwidth]{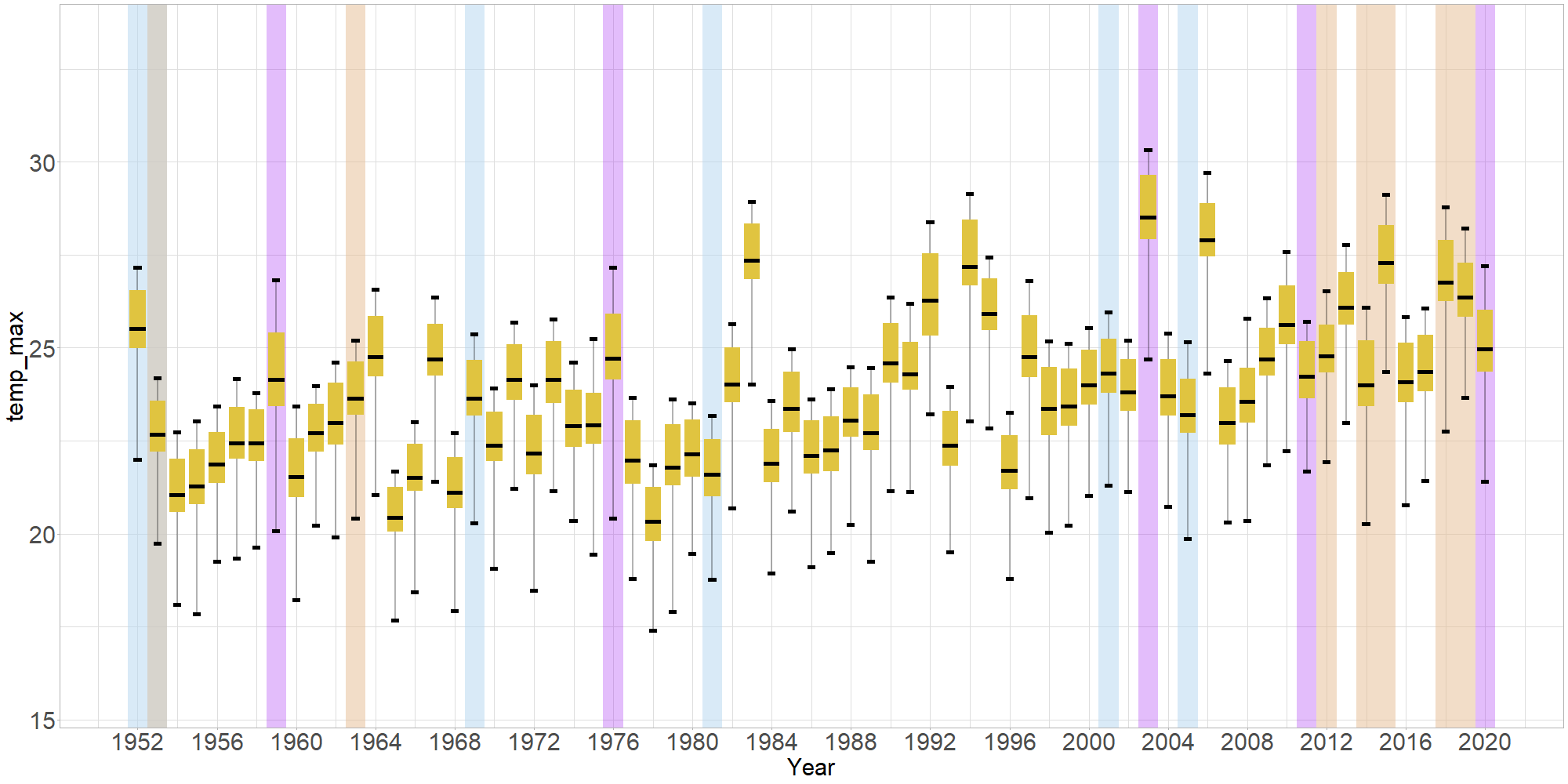}};
			
		\end{tikzpicture}		
	\end{figure}
\end{landscape}

\begin{landscape}	
	\begin{figure}
		\centering
		\begin{tikzpicture}
			
			\node[inner sep=0pt] (Col1) at (-26, 7)  {\includegraphics[width=0.67\textwidth]{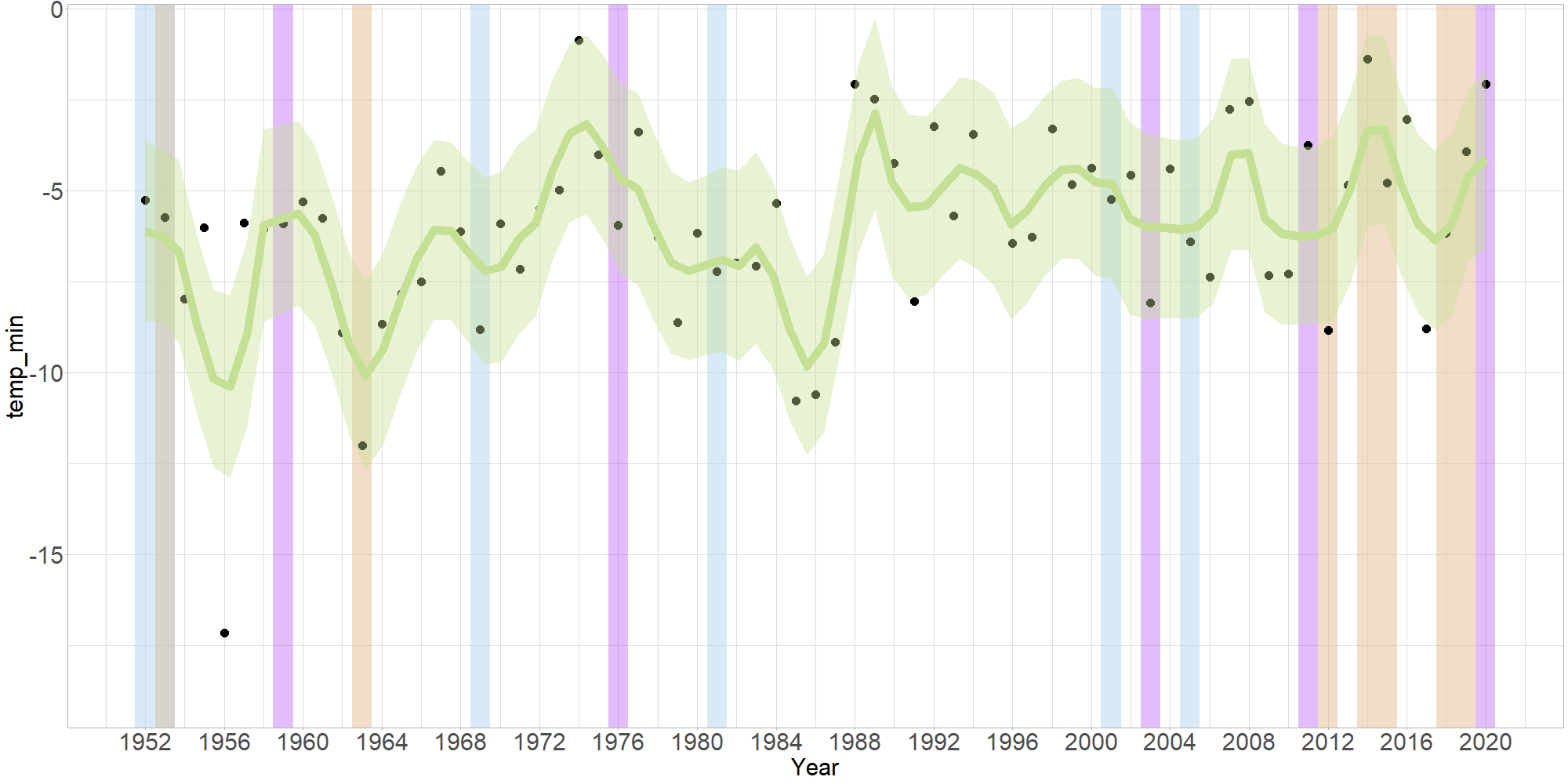}};
			
			\node[inner sep=0pt] (Col1) at (-16, 7)  {\includegraphics[width=0.67\textwidth]{temp_mean_box_plot.png}};

			
			\node[inner sep=0pt] (Col1) at (-26, 0)  {\includegraphics[width=0.67\textwidth]{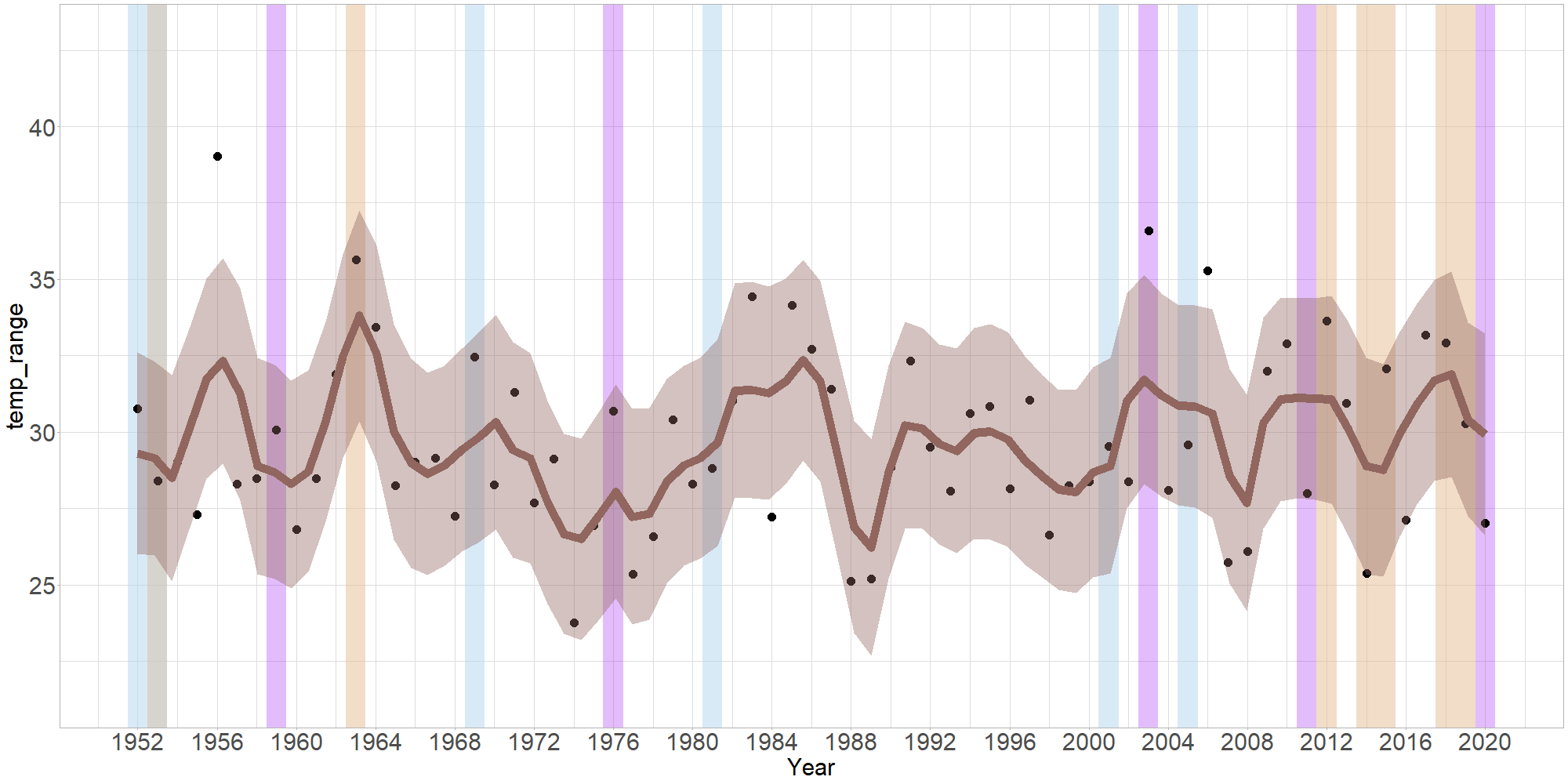}};
			
			\node[inner sep=0pt] (Col1) at (-16, 0)  {\includegraphics[width=0.67\textwidth]{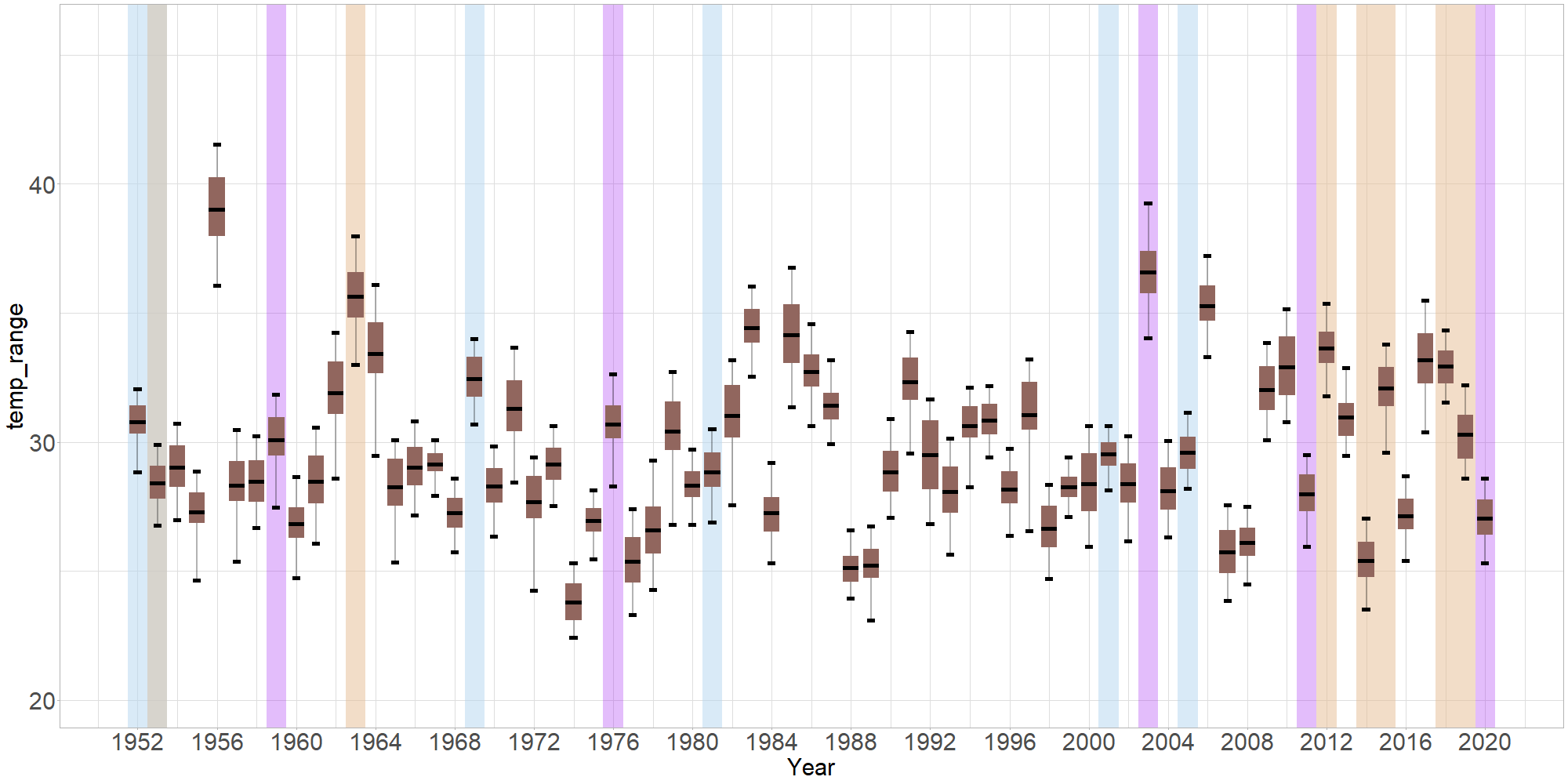}};
			
		\end{tikzpicture}		
	\end{figure}
\end{landscape}	
\begin{landscape}	
	\begin{figure}
		\centering
		\begin{tikzpicture}
			
			\node[inner sep=0pt] (Col1) at (-26, 7)  {\includegraphics[width=0.67\textwidth]{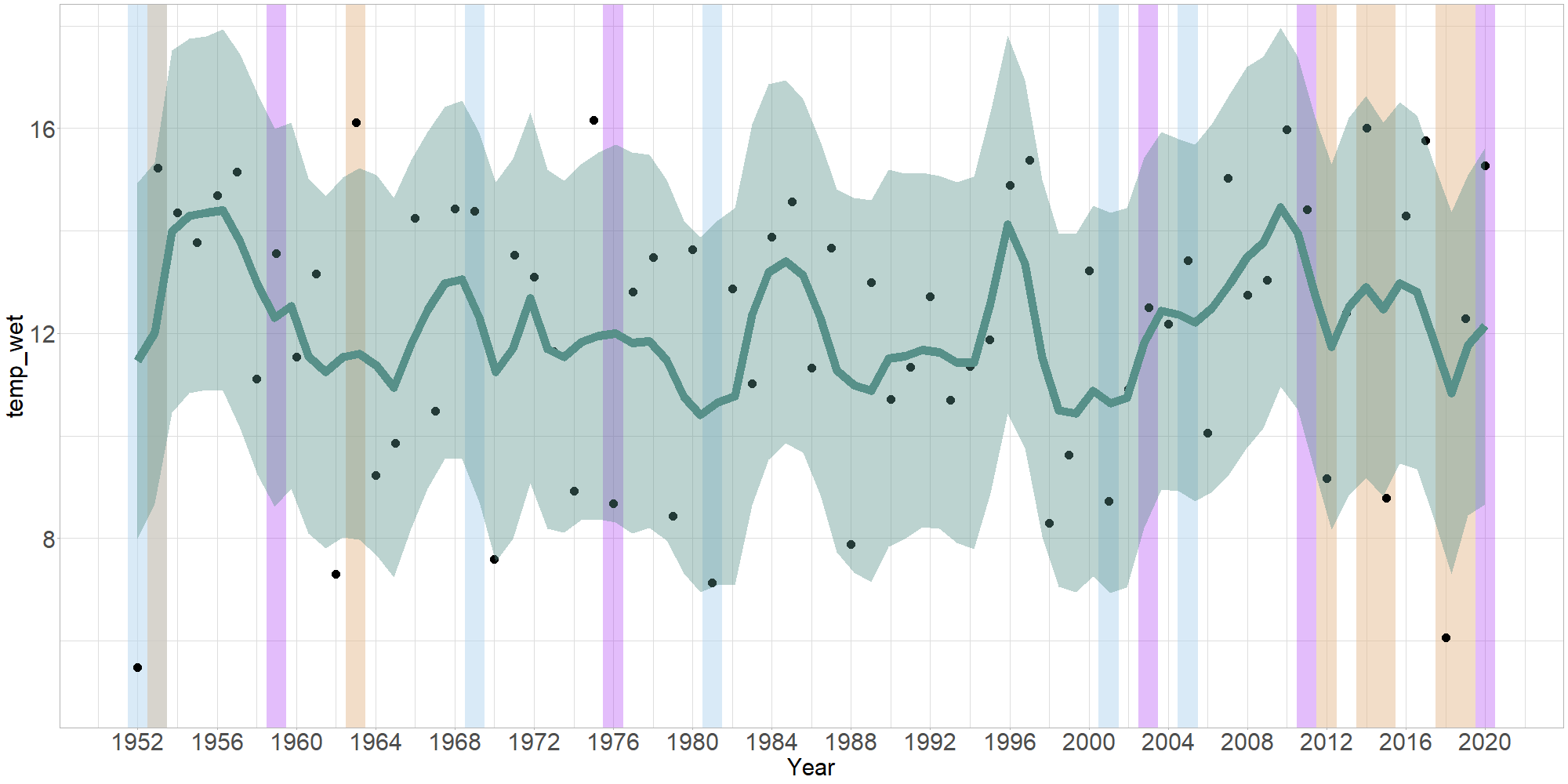}};
			
			\node[inner sep=0pt] (Col1) at (-16, 7)  {\includegraphics[width=0.67\textwidth]{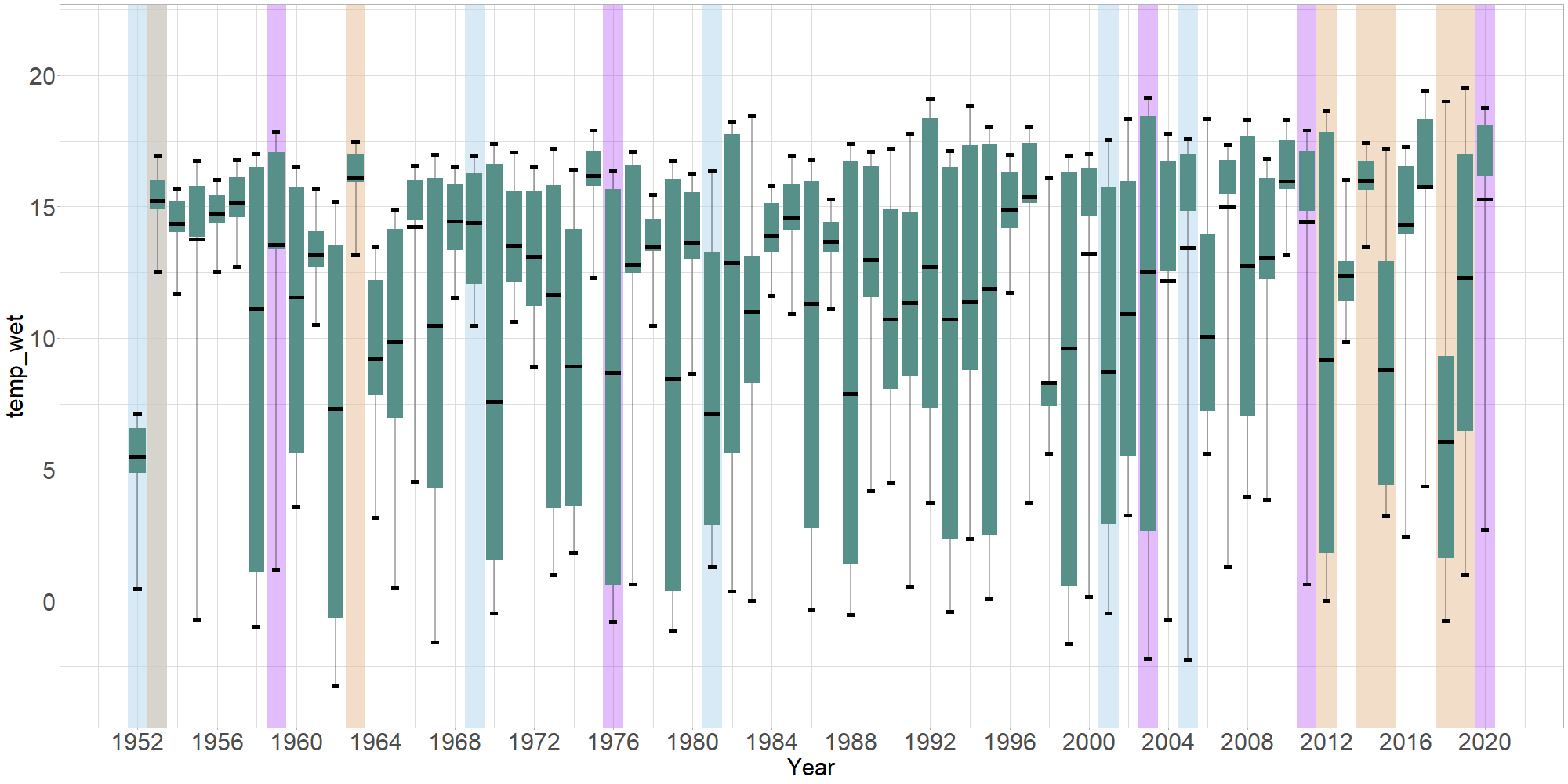}};

			
			\node[inner sep=0pt] (Col1) at (-26, 0)  {\includegraphics[width=0.67\textwidth]{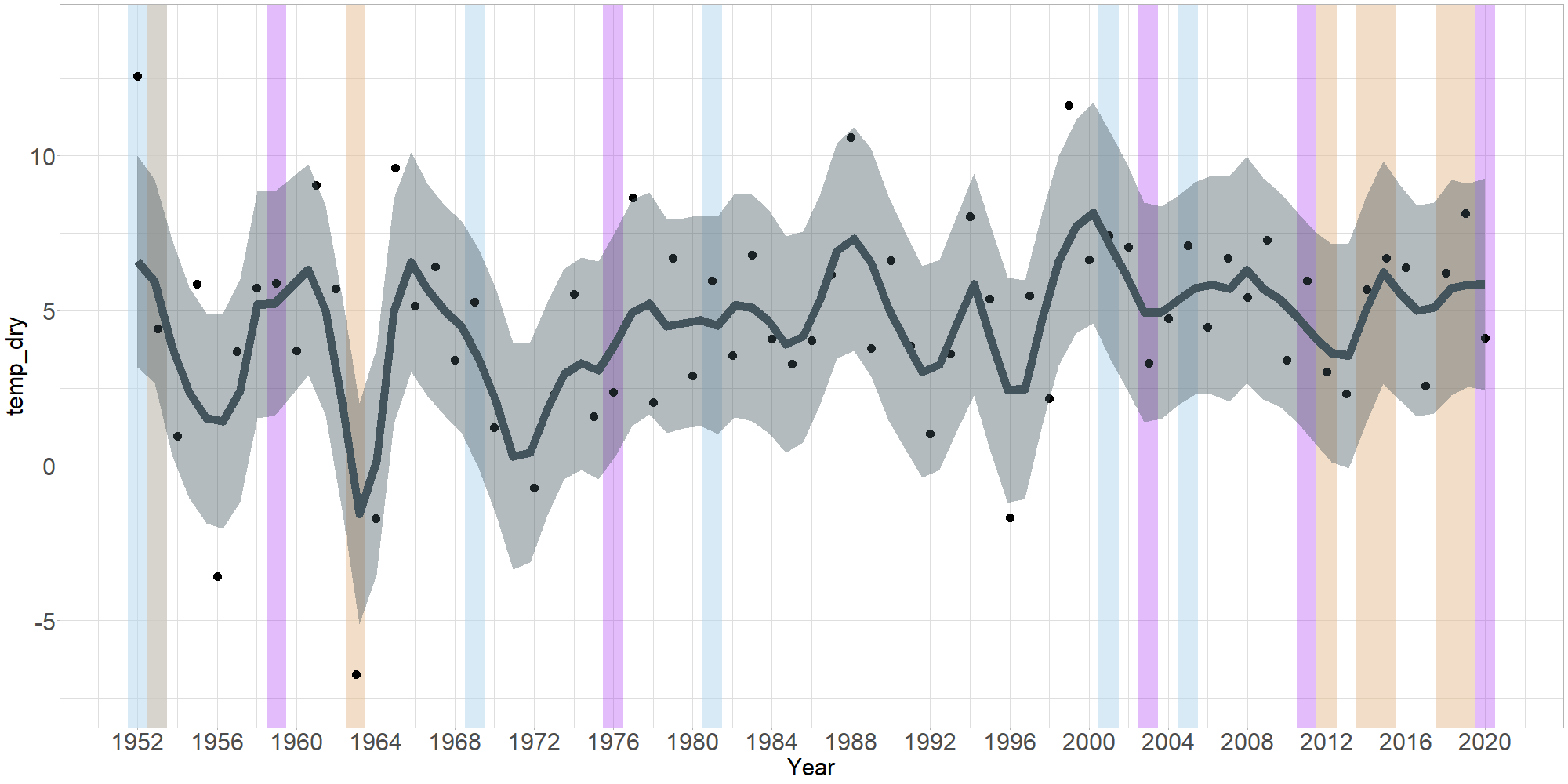}};
			
			\node[inner sep=0pt] (Col1) at (-16, 0)  {\includegraphics[width=0.67\textwidth]{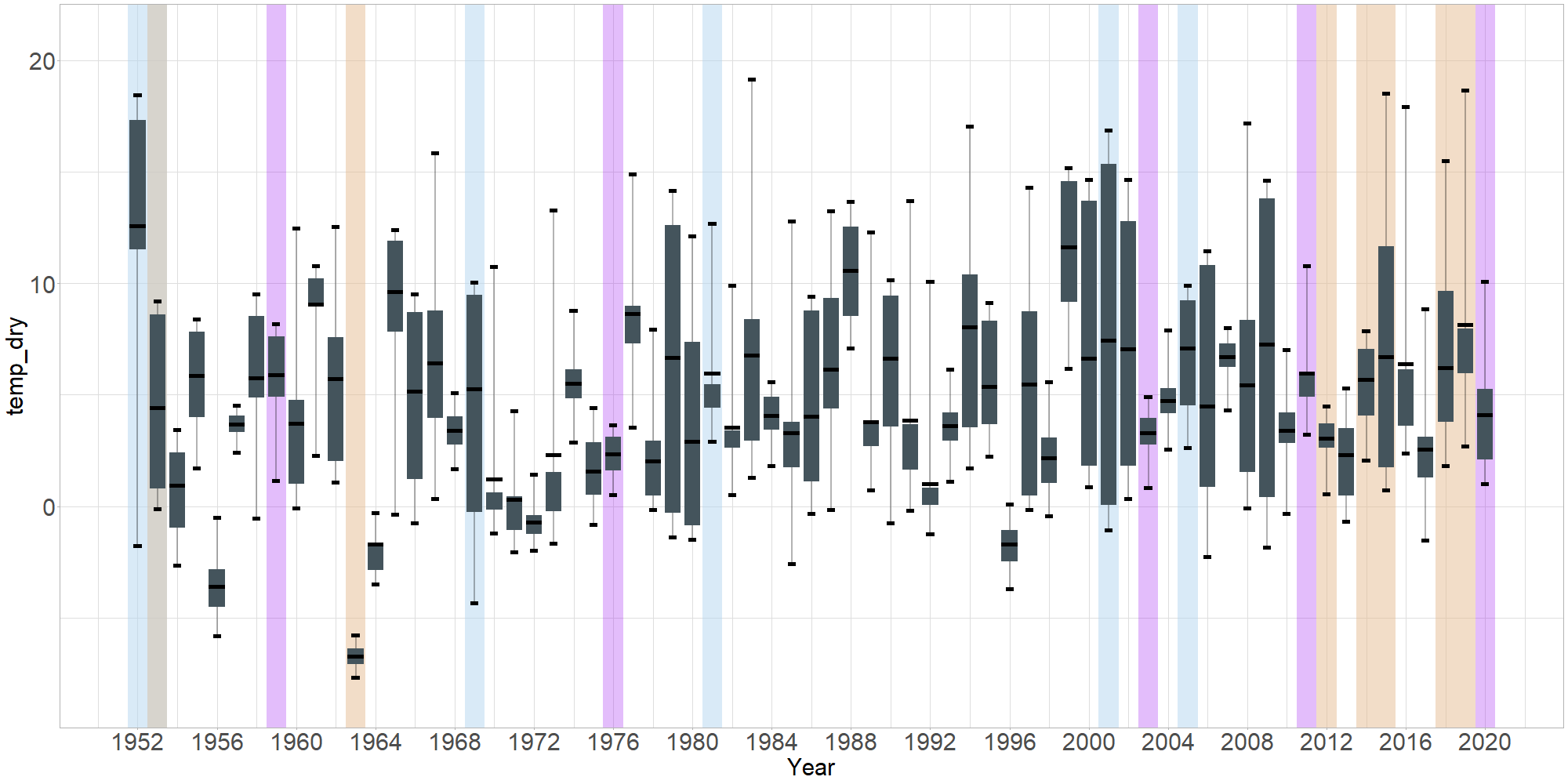}};
			
		\end{tikzpicture}		
	\end{figure}
\end{landscape}	
\begin{landscape}	
	\begin{figure}
		\centering
		\begin{tikzpicture}
			
			\node[inner sep=0pt] (Col1) at (-26, 7)  {\includegraphics[width=0.67\textwidth]{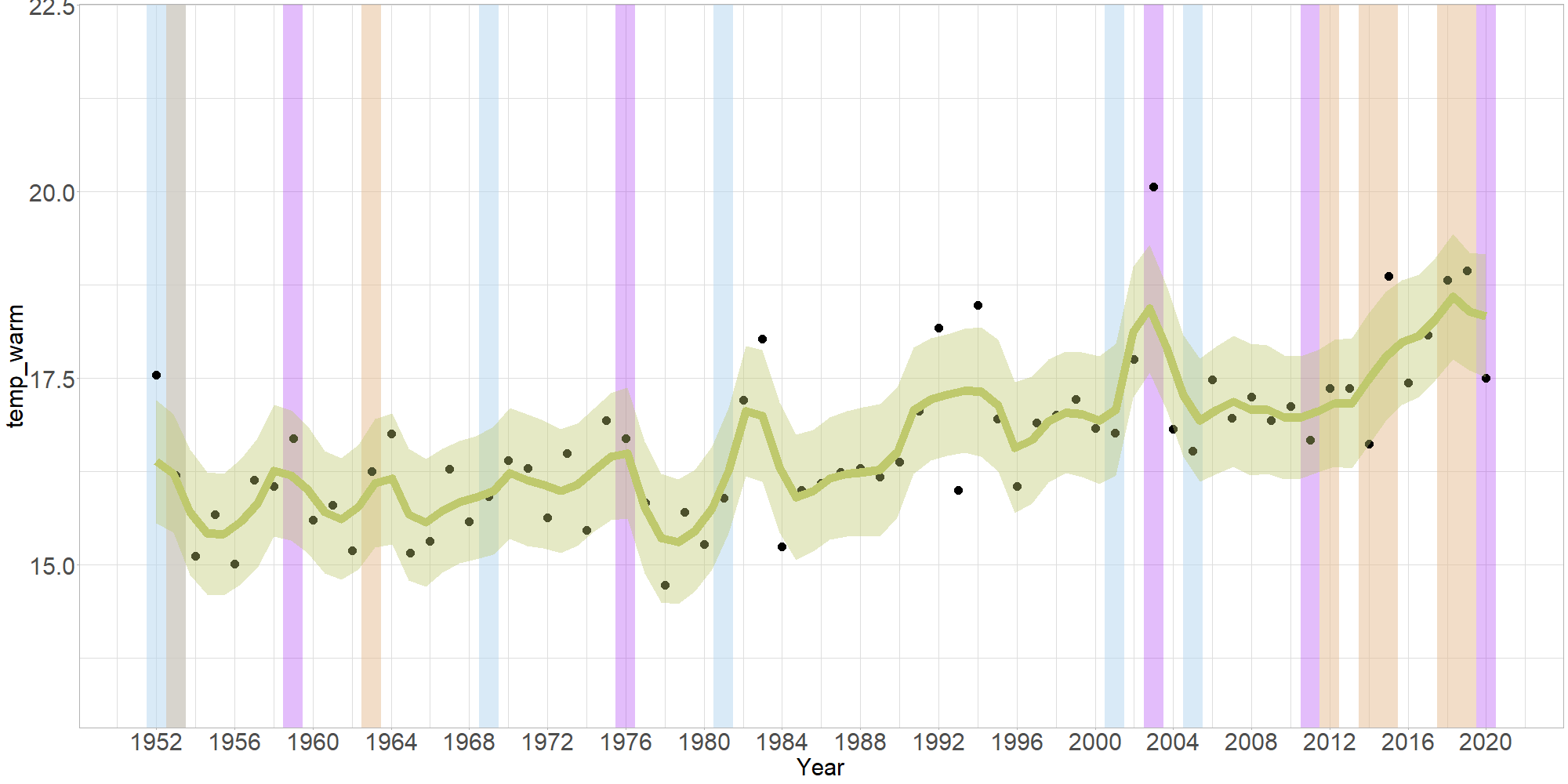}};
			
			\node[inner sep=0pt] (Col1) at (-16, 7)  {\includegraphics[width=0.67\textwidth]{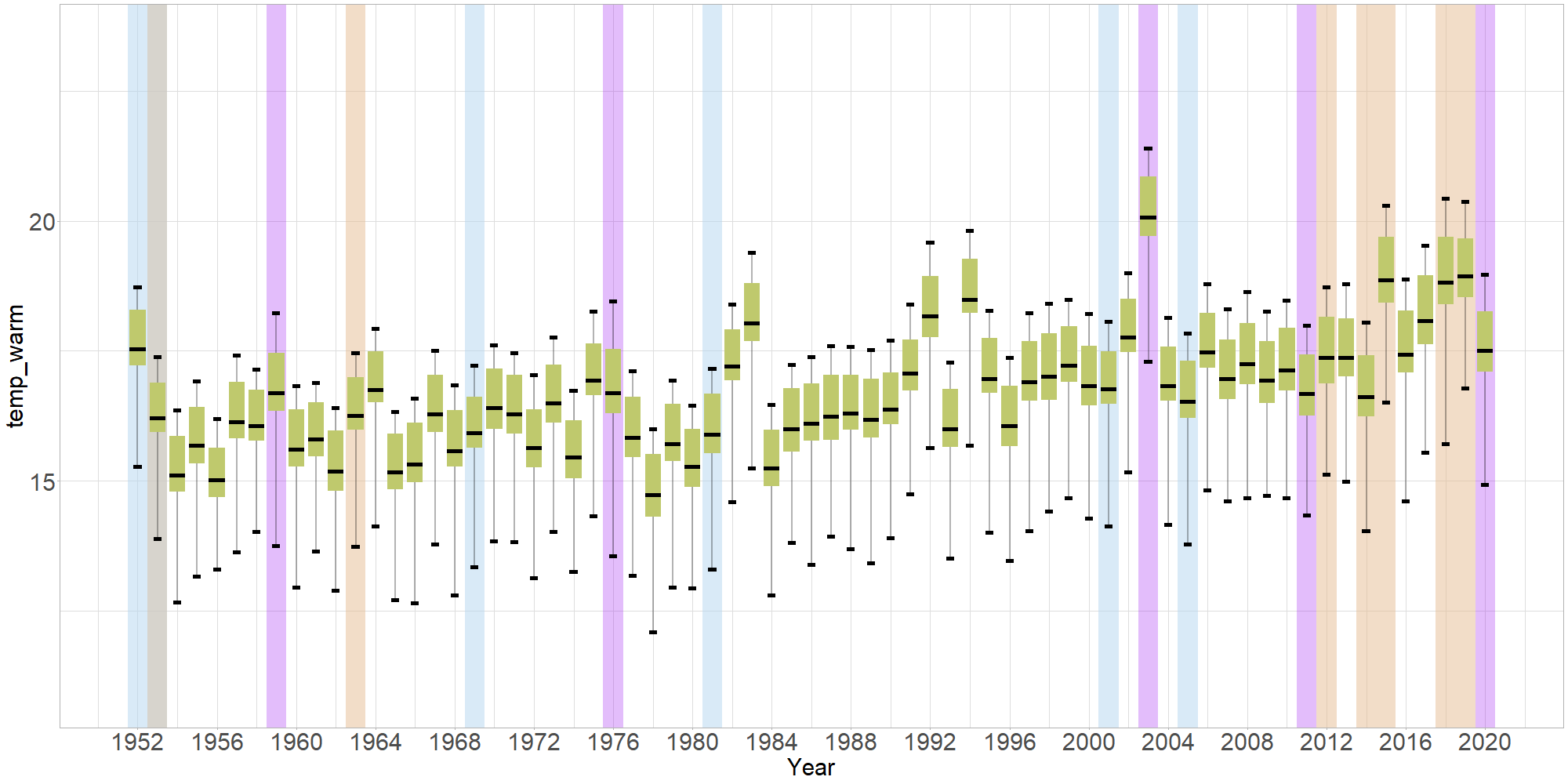}};

			
			\node[inner sep=0pt] (Col1) at (-26, 0)  {\includegraphics[width=0.67\textwidth]{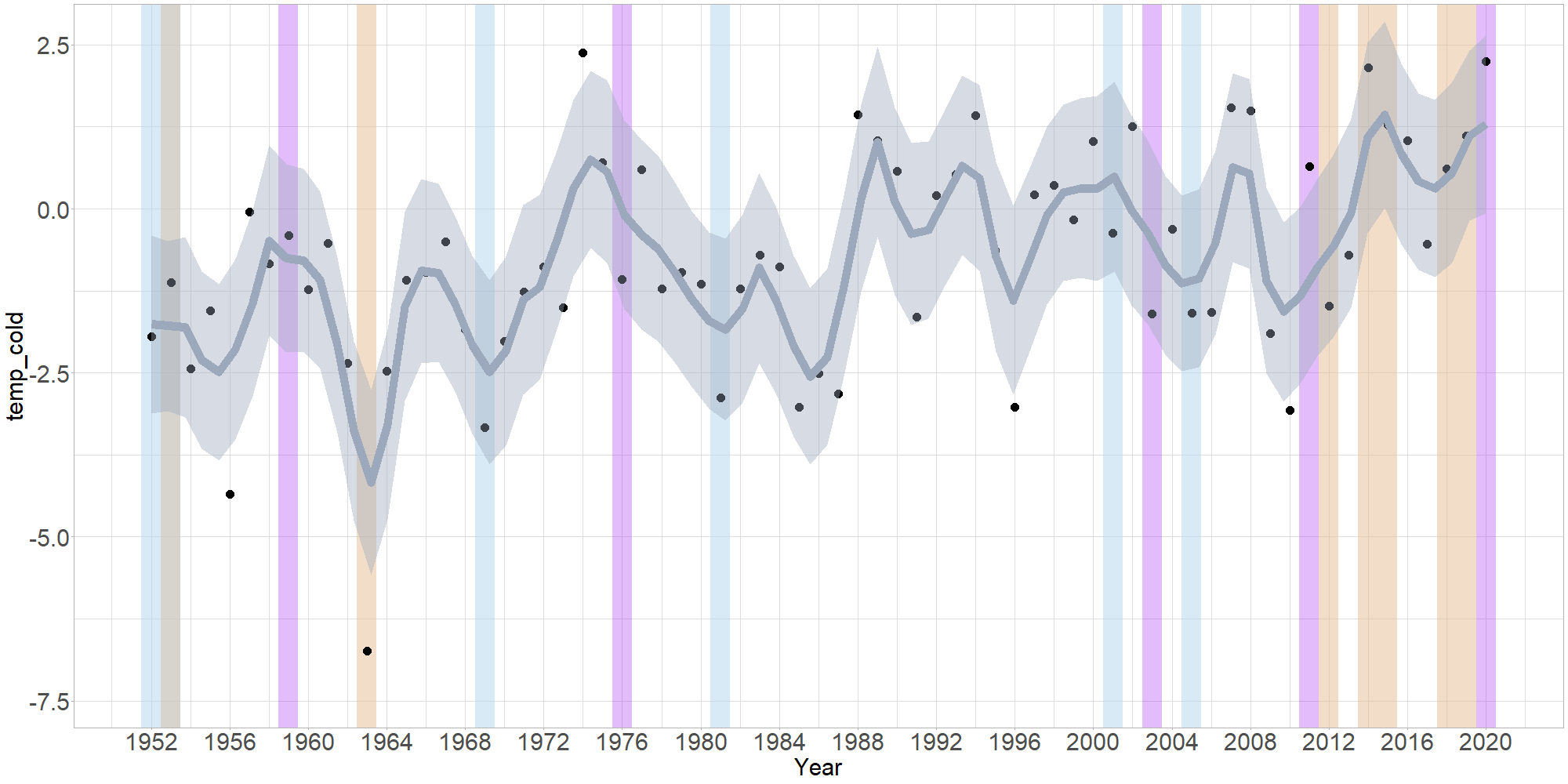}};
			
			\node[inner sep=0pt] (Col1) at (-16, 0)  {\includegraphics[width=0.67\textwidth]{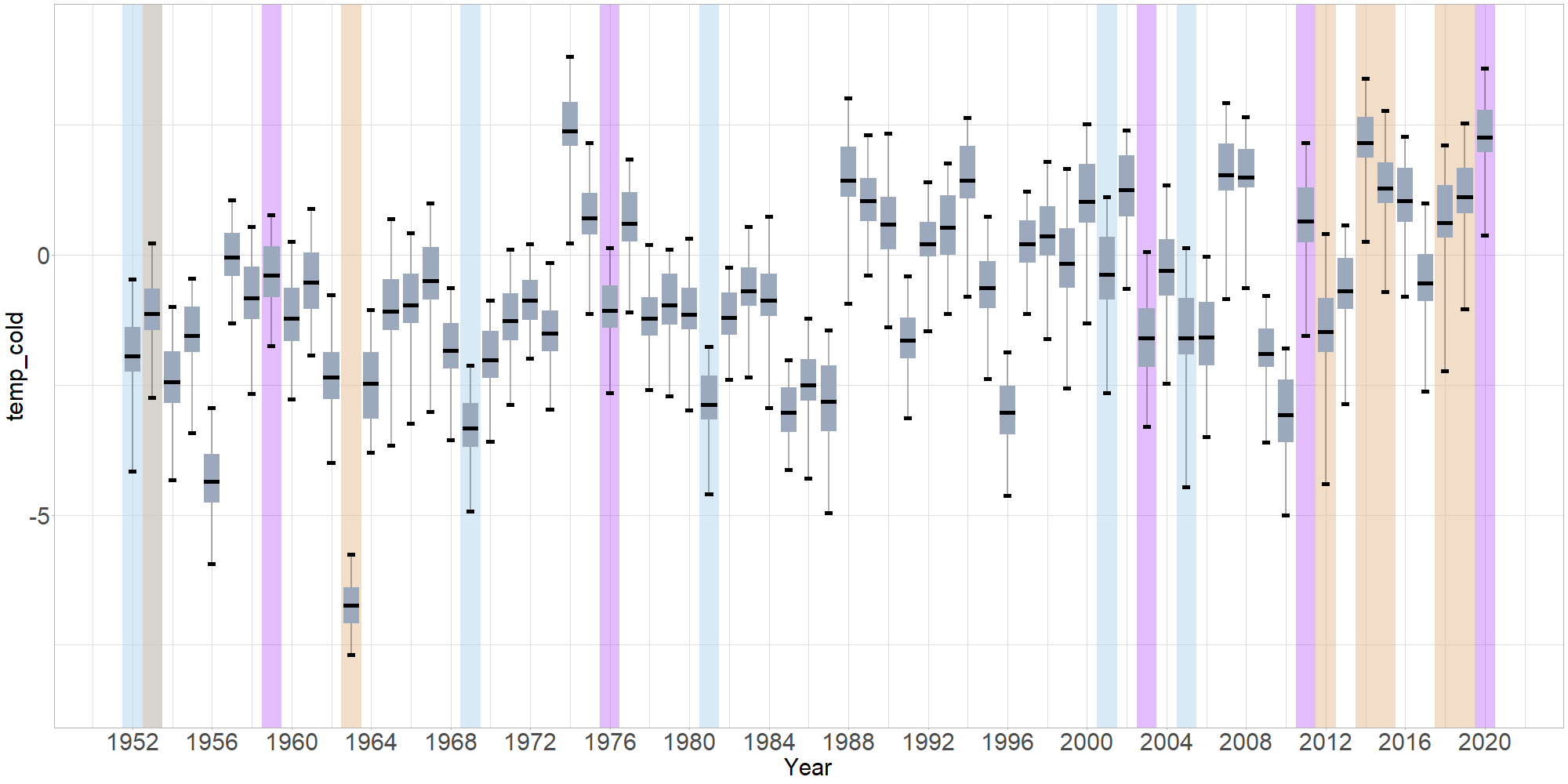}};
			
		\end{tikzpicture}		
	\end{figure}
\end{landscape}	
\begin{landscape}	
	\begin{figure}
		\centering
		\begin{tikzpicture}
			
			\node[inner sep=0pt] (Col1) at (-26, 7)  {\includegraphics[width=0.67\textwidth]{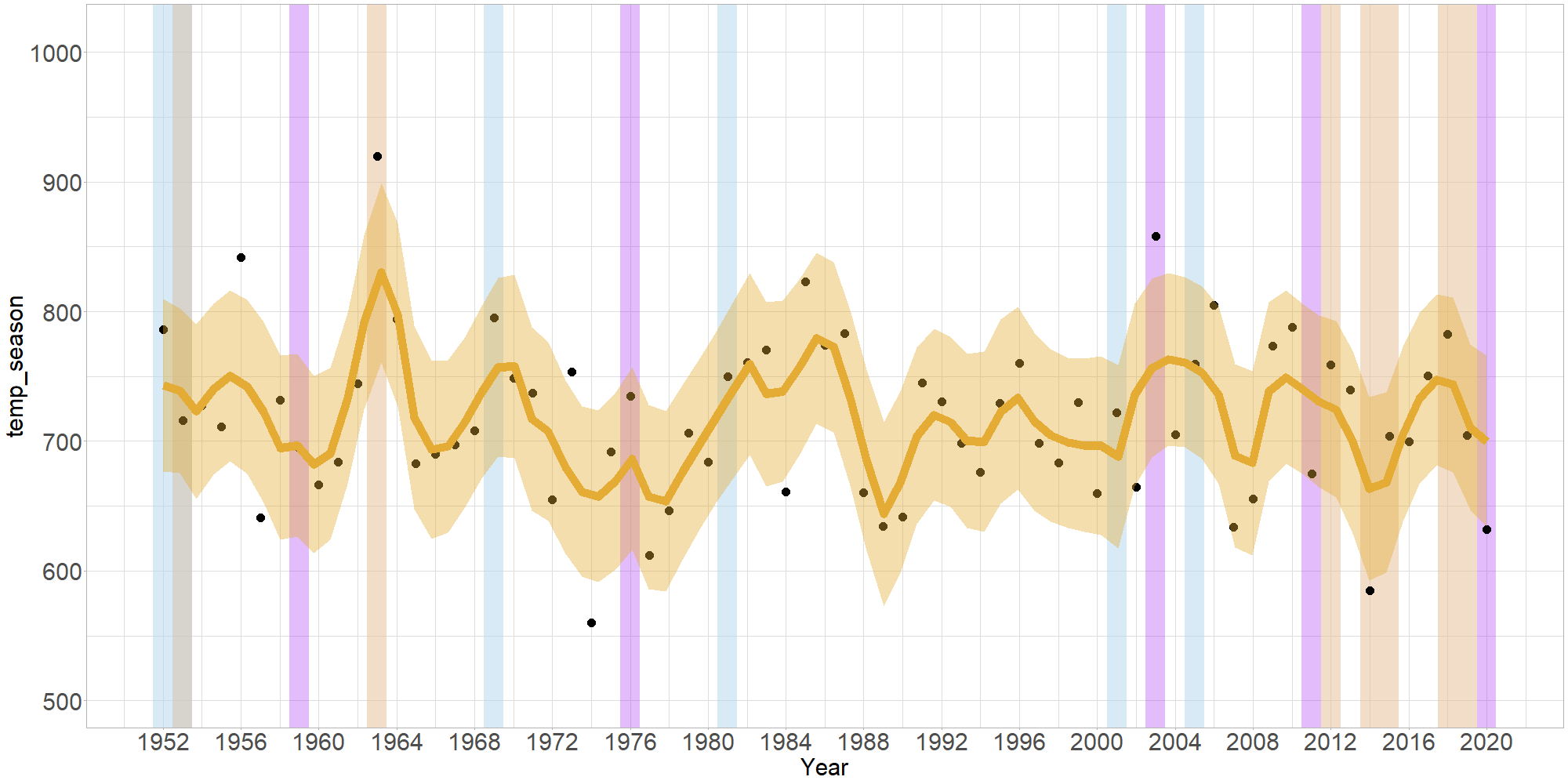}};
			
			\node[inner sep=0pt] (Col1) at (-16, 7)  {\includegraphics[width=0.67\textwidth]{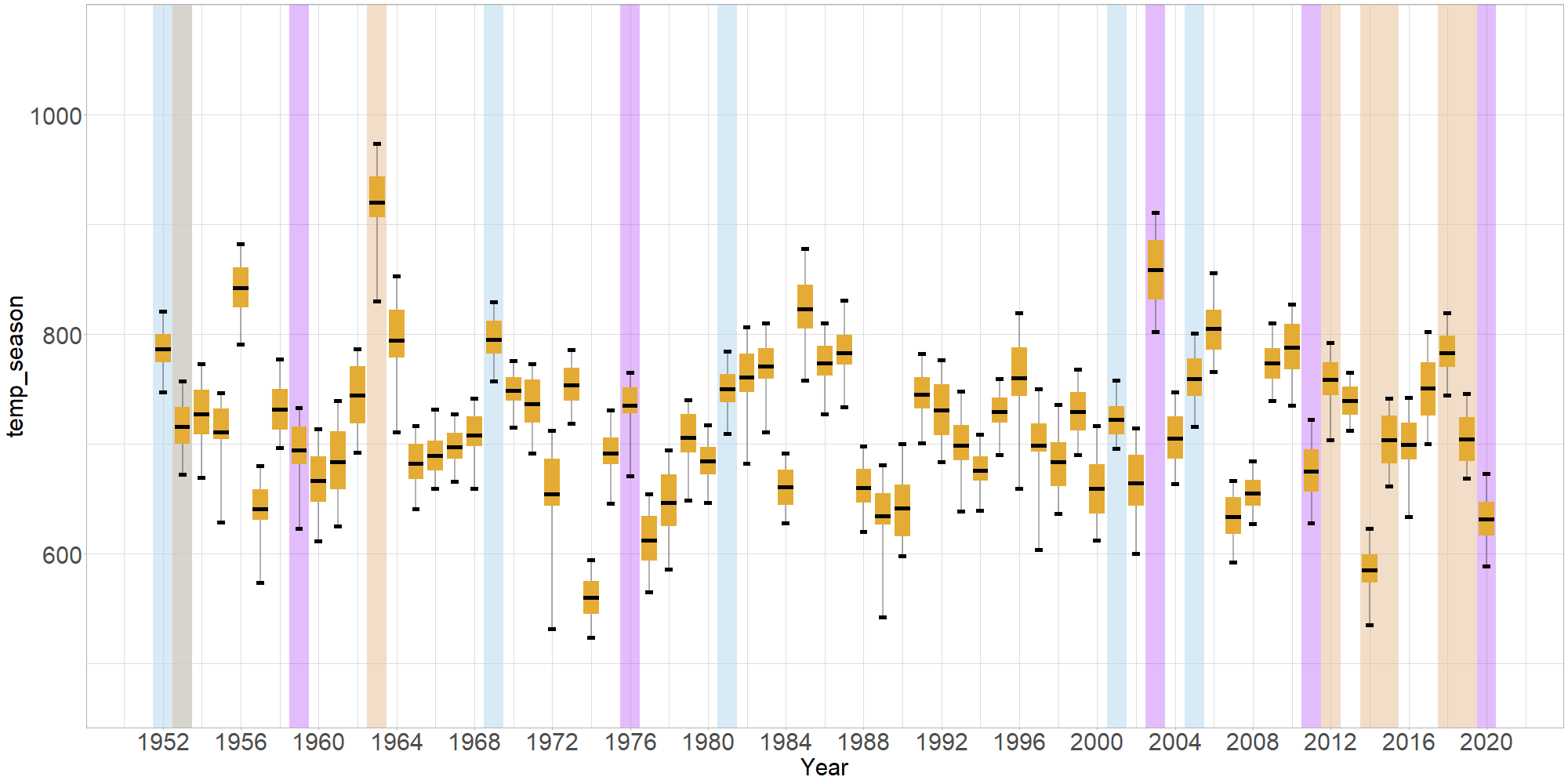}};

			
			\node[inner sep=0pt] (Col1) at (-26, 0)  {\includegraphics[width=0.67\textwidth]{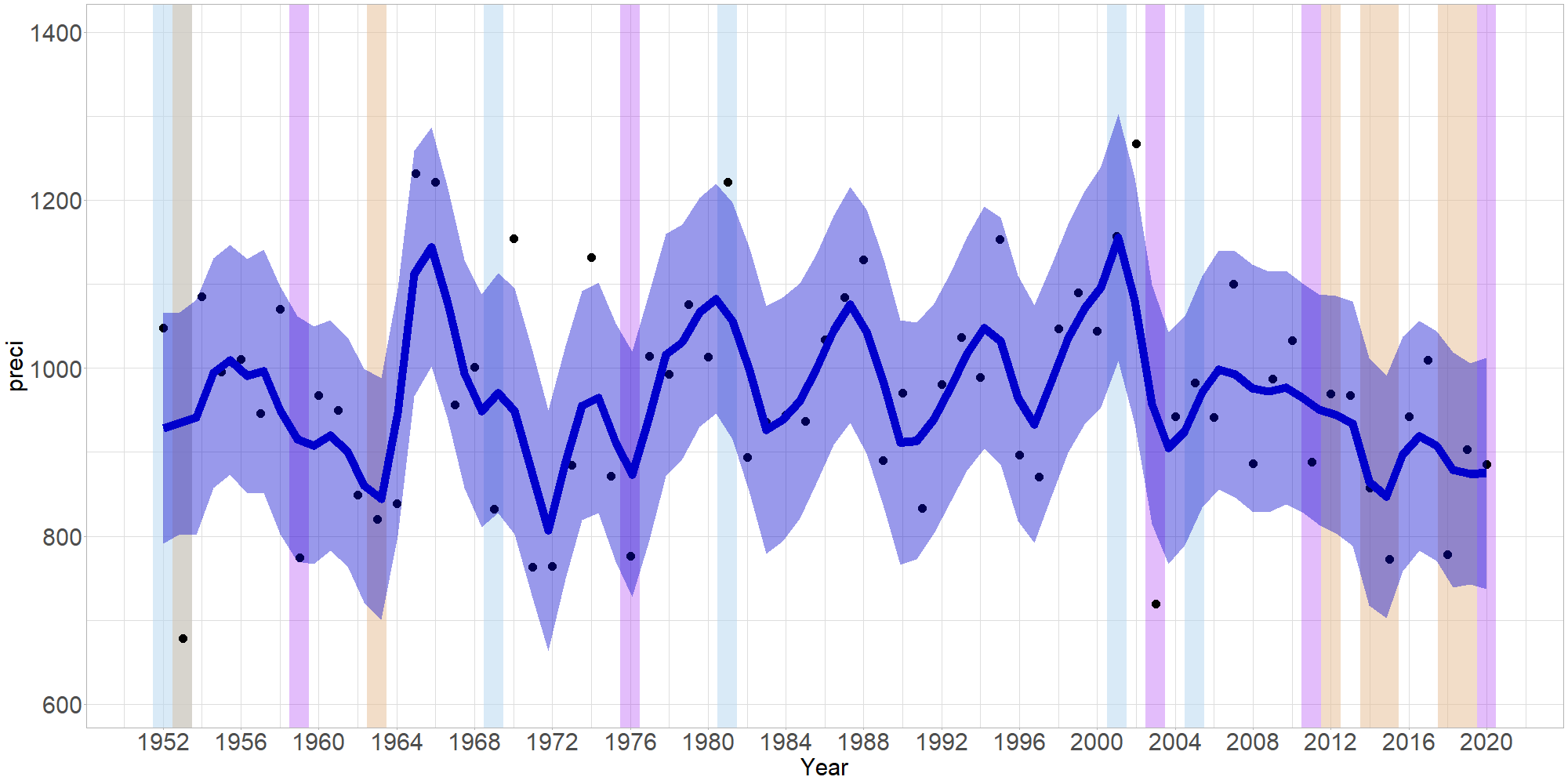}};
			
			\node[inner sep=0pt] (Col1) at (-16, 0)  {\includegraphics[width=0.67\textwidth]{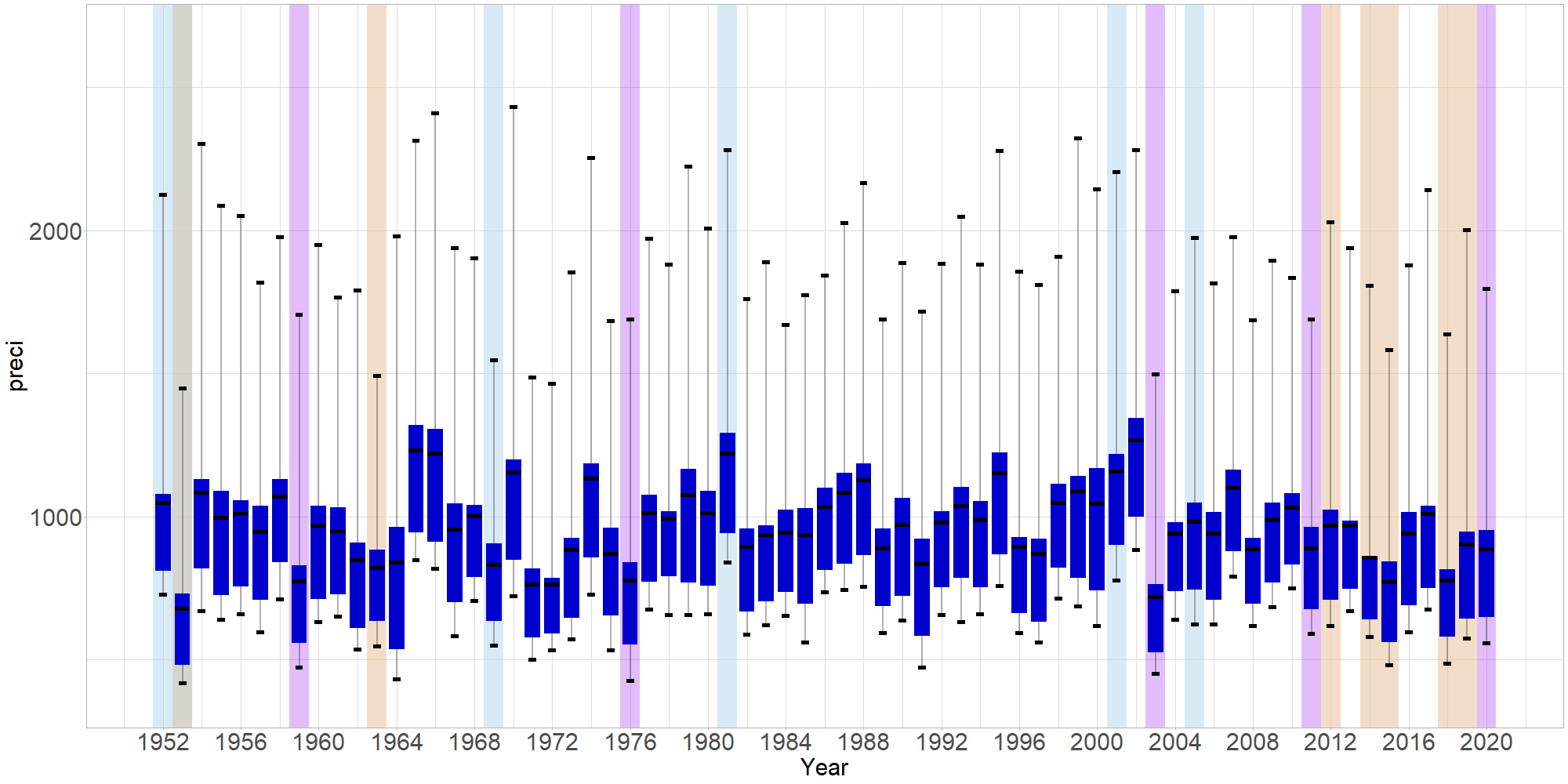}};
			
		\end{tikzpicture}		
	\end{figure}
\end{landscape}	
\begin{landscape}	
	\begin{figure}
		\centering
		\begin{tikzpicture}
			
			\node[inner sep=0pt] (Col1) at (-26, 7)  {\includegraphics[width=0.67\textwidth]{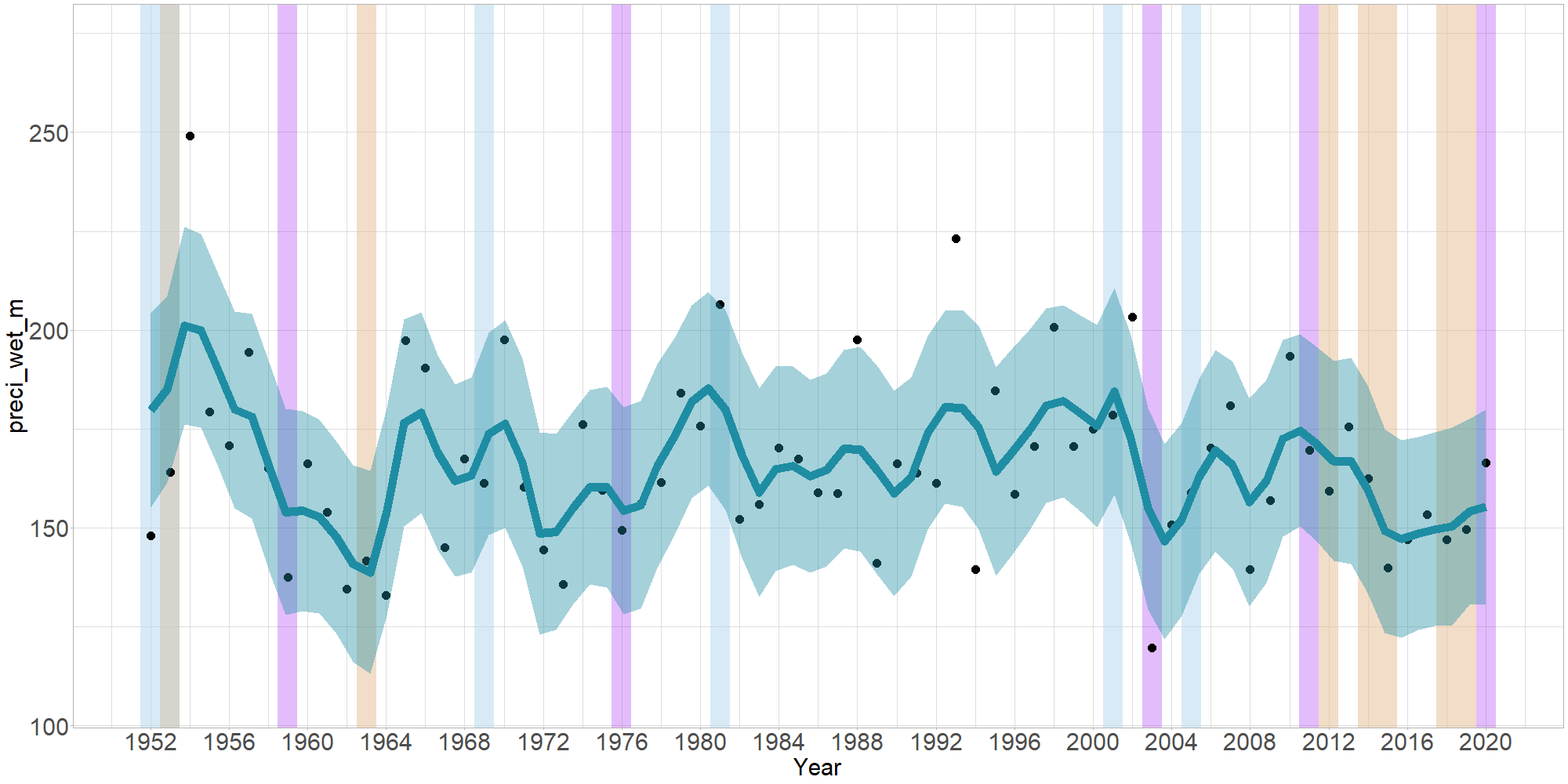}};
			
			\node[inner sep=0pt] (Col1) at (-16, 7)  {\includegraphics[width=0.67\textwidth]{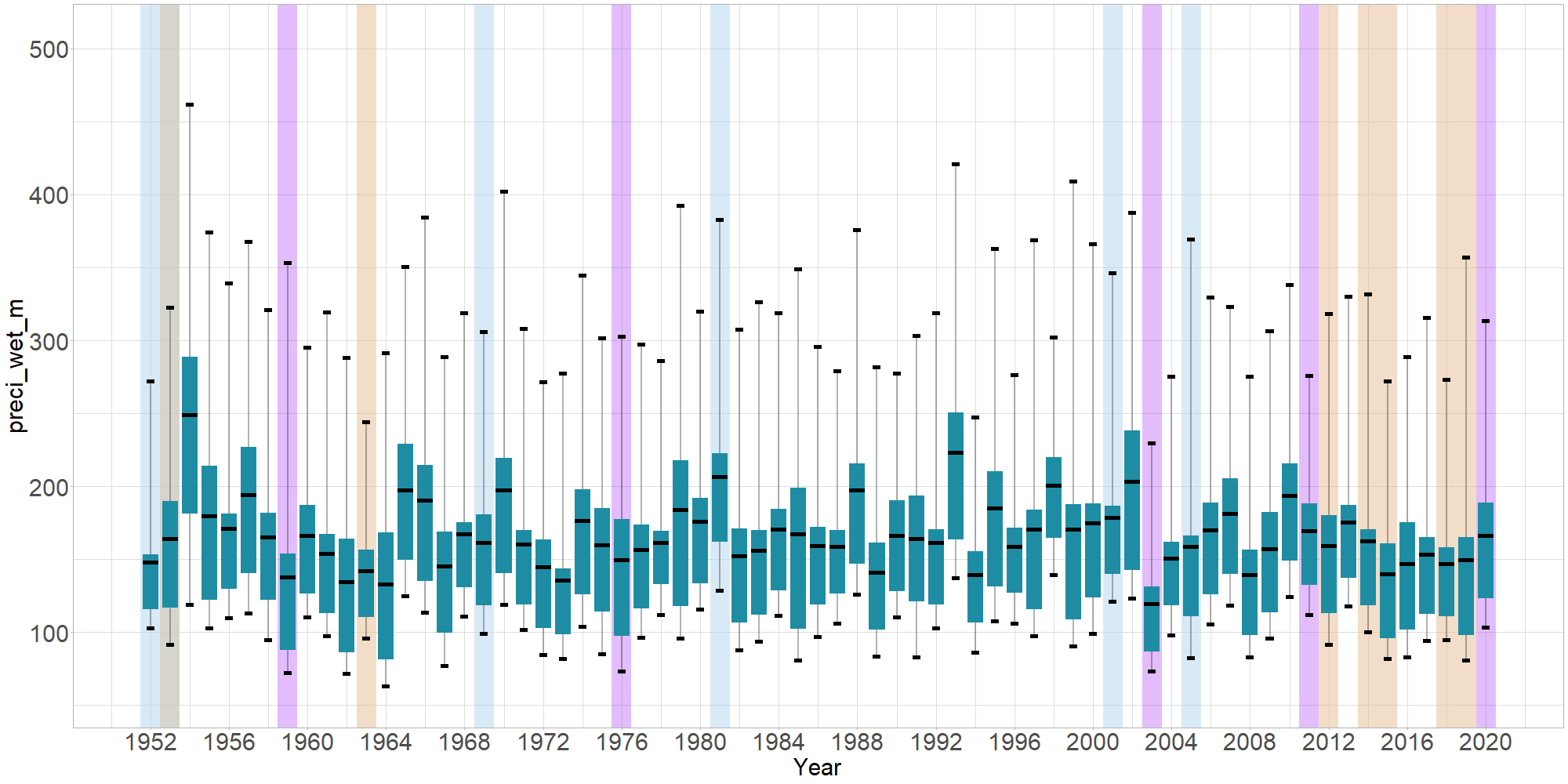}};

			
			\node[inner sep=0pt] (Col1) at (-26, 0)  {\includegraphics[width=0.67\textwidth]{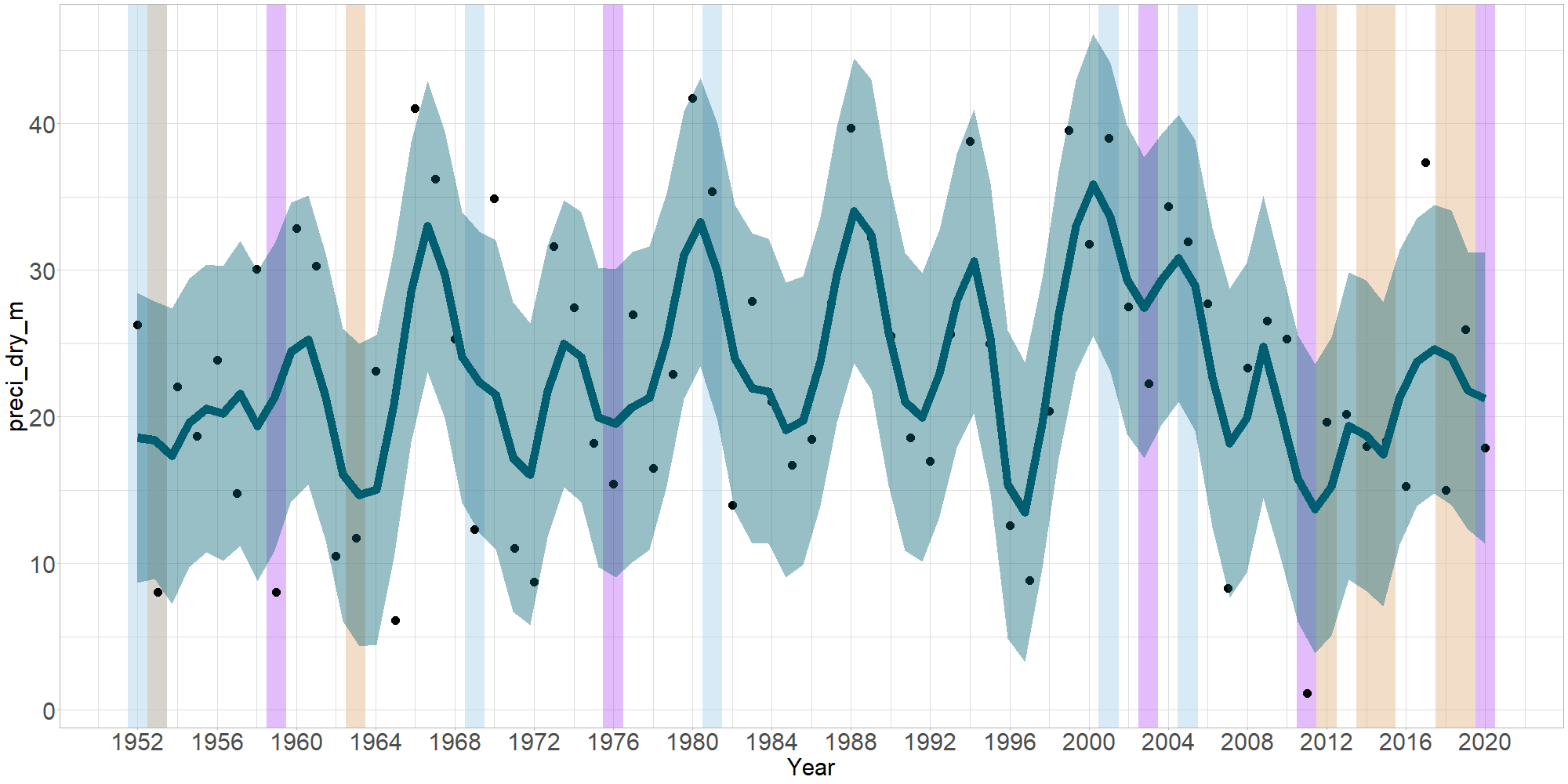}};
			
			\node[inner sep=0pt] (Col1) at (-16, 0)  {\includegraphics[width=0.67\textwidth]{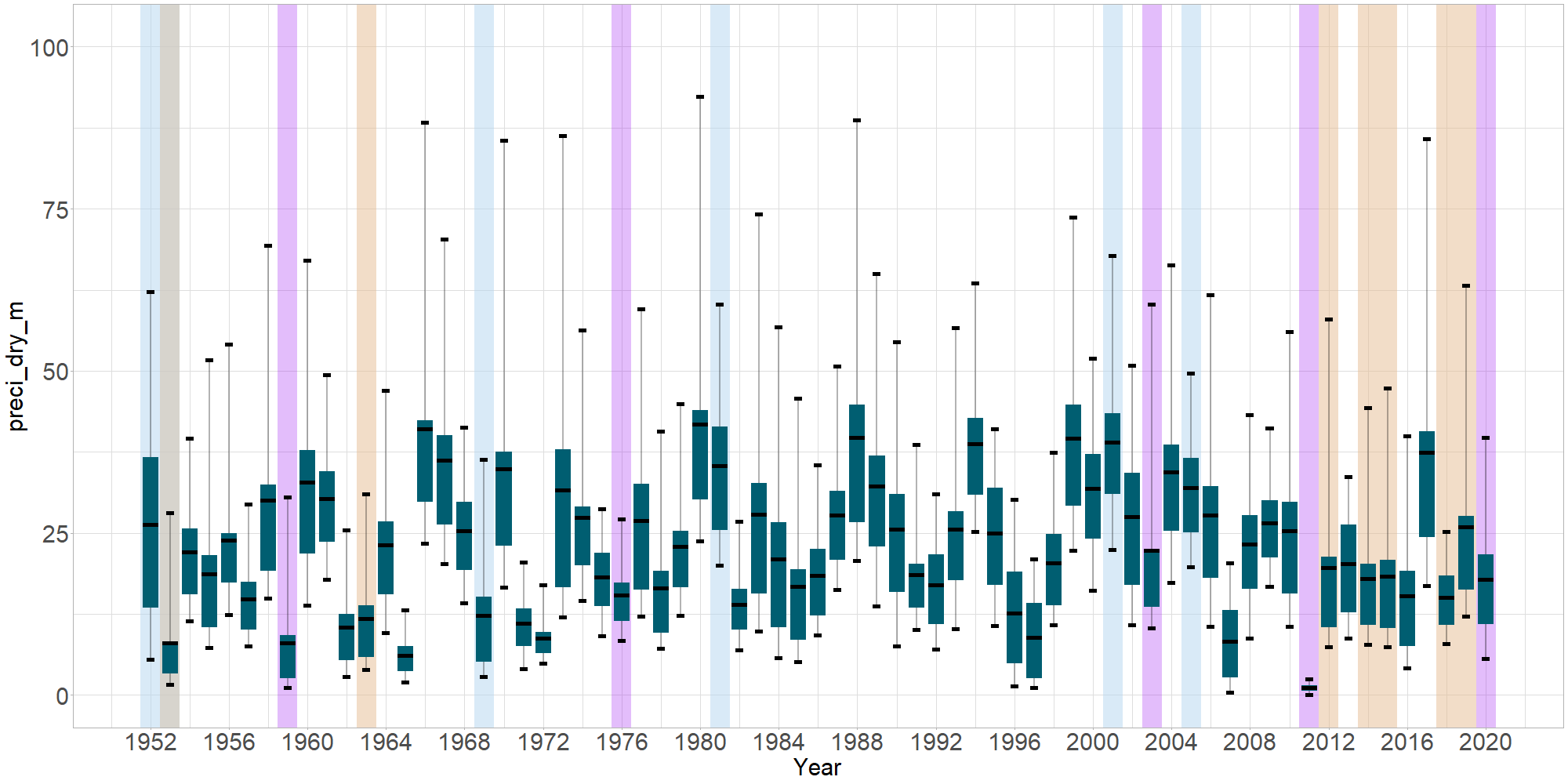}};
			
		\end{tikzpicture}		
	\end{figure}
\end{landscape}	
\begin{landscape}	
	\begin{figure}
		\centering
		\begin{tikzpicture}
			
			\node[inner sep=0pt] (Col1) at (-26, 7)  {\includegraphics[width=0.67\textwidth]{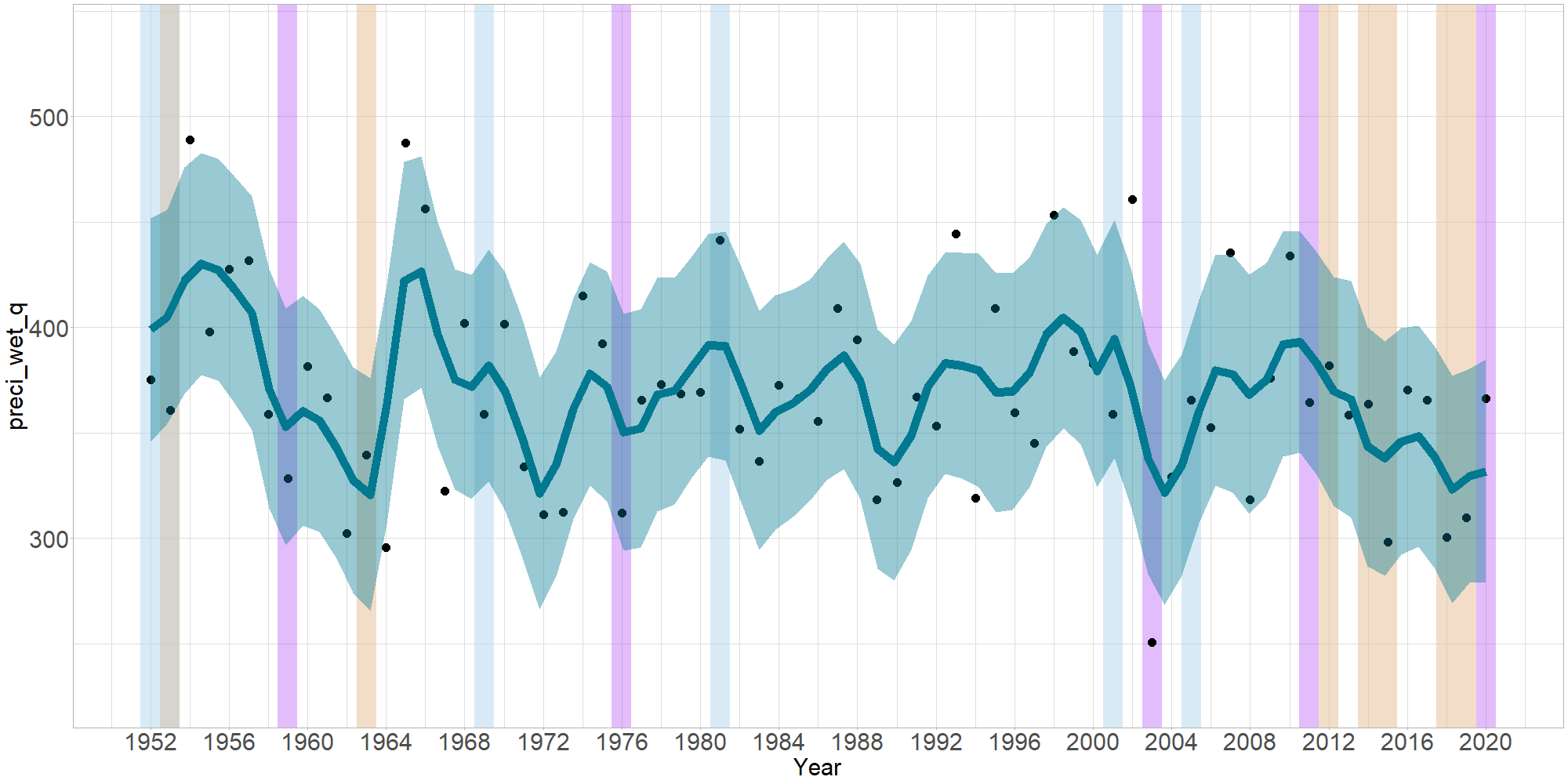}};
			
			\node[inner sep=0pt] (Col1) at (-16, 7)  {\includegraphics[width=0.67\textwidth]{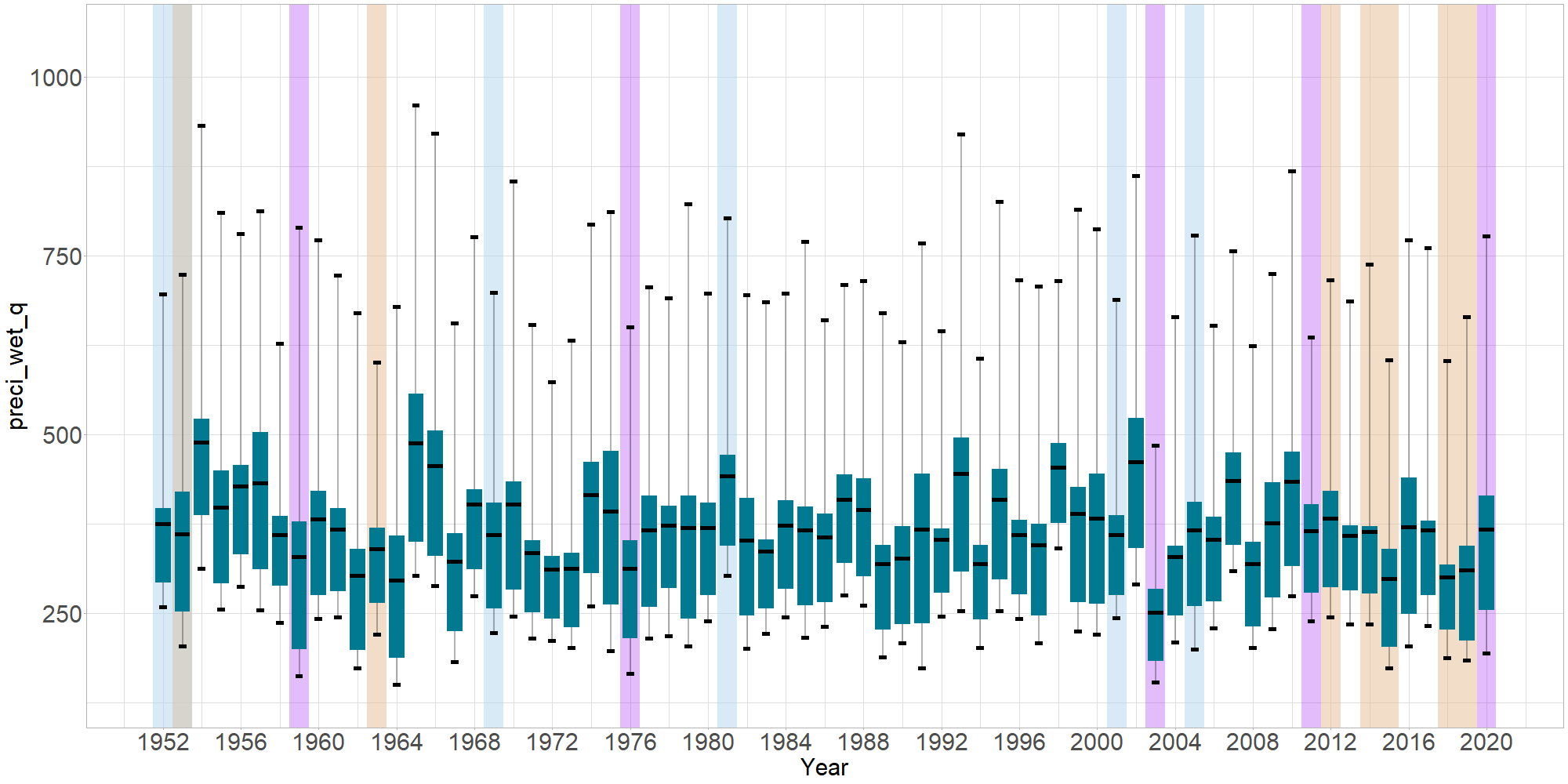}};

			
			\node[inner sep=0pt] (Col1) at (-26, 0)  {\includegraphics[width=0.67\textwidth]{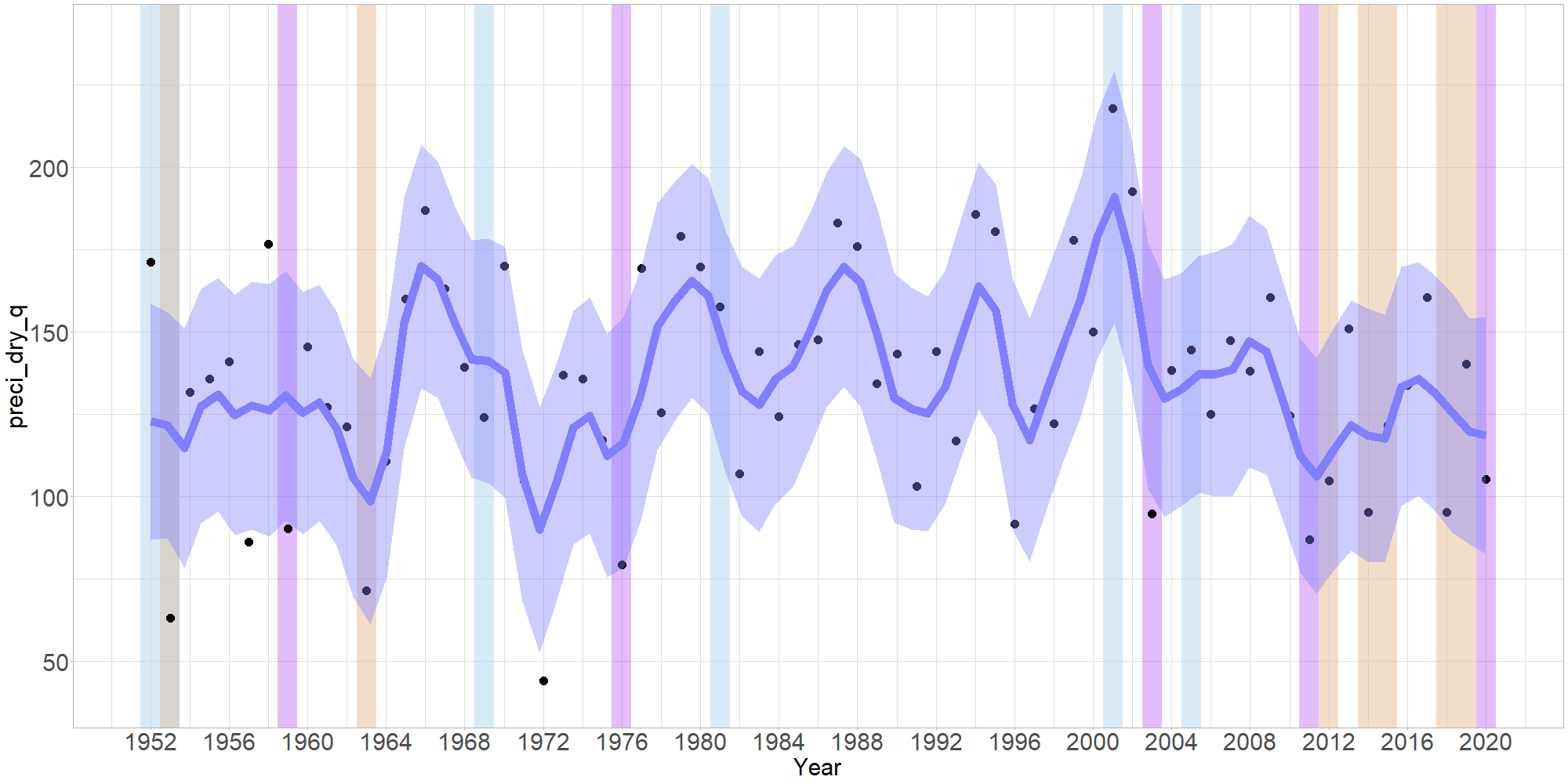}};
			
			\node[inner sep=0pt] (Col1) at (-16, 0)  {\includegraphics[width=0.67\textwidth]{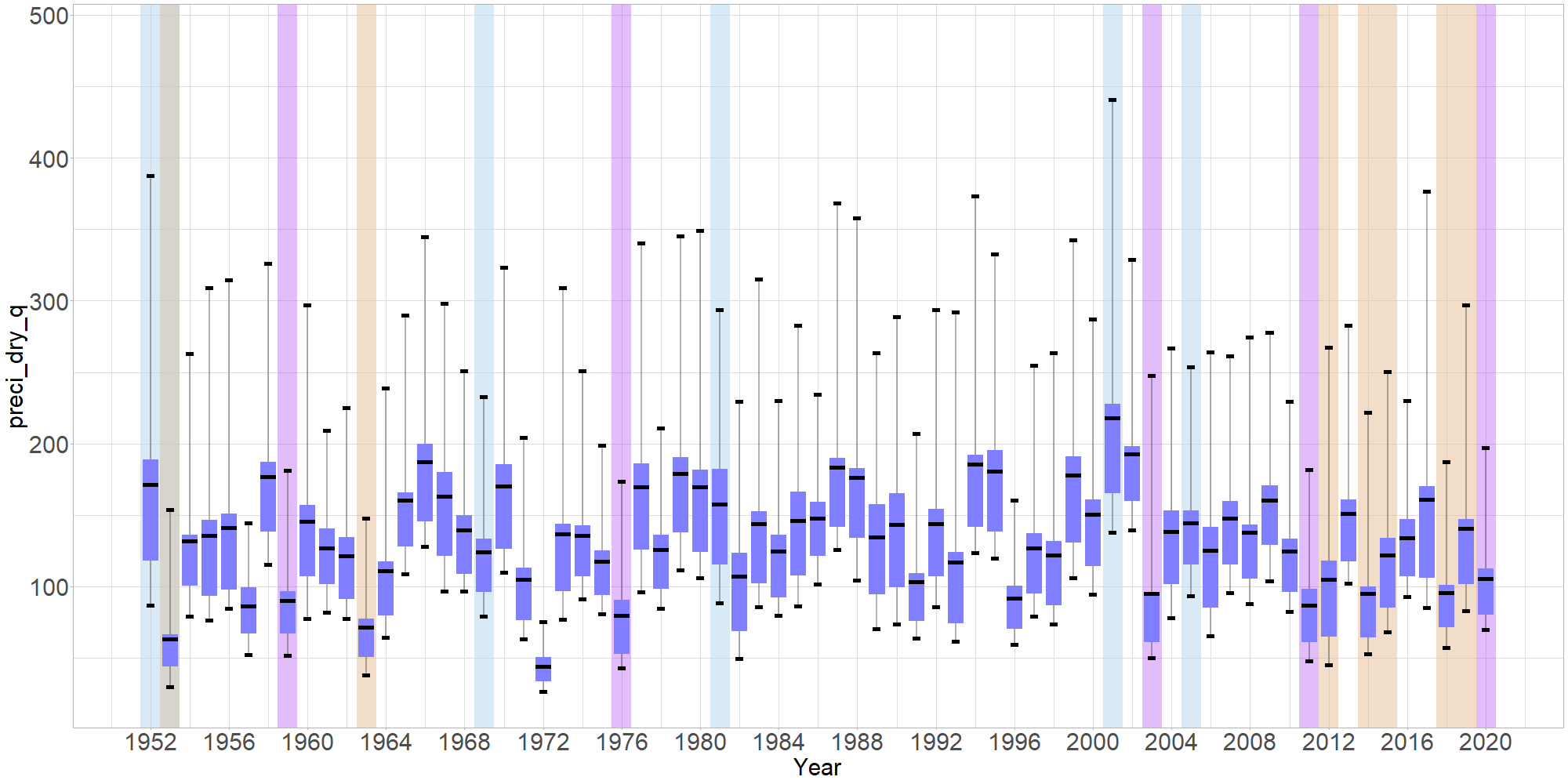}};
			
		\end{tikzpicture}		
	\end{figure}
\end{landscape}	
\begin{landscape}	
	\begin{figure}
		\centering
		\begin{tikzpicture}
			
			\node[inner sep=0pt] (Col1) at (-26, 7)  {\includegraphics[width=0.67\textwidth]{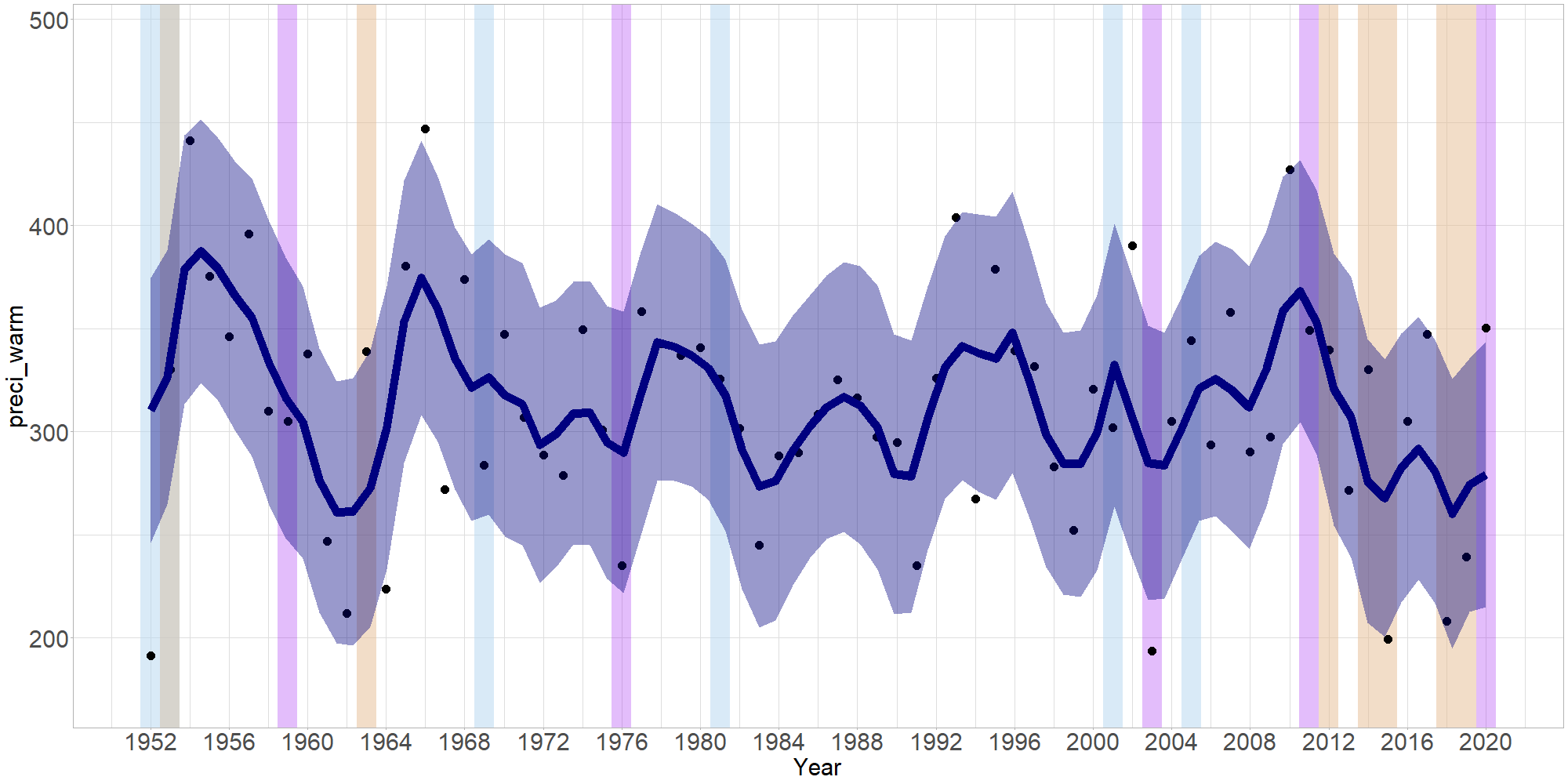}};
			
			\node[inner sep=0pt] (Col1) at (-16, 7)  {\includegraphics[width=0.67\textwidth]{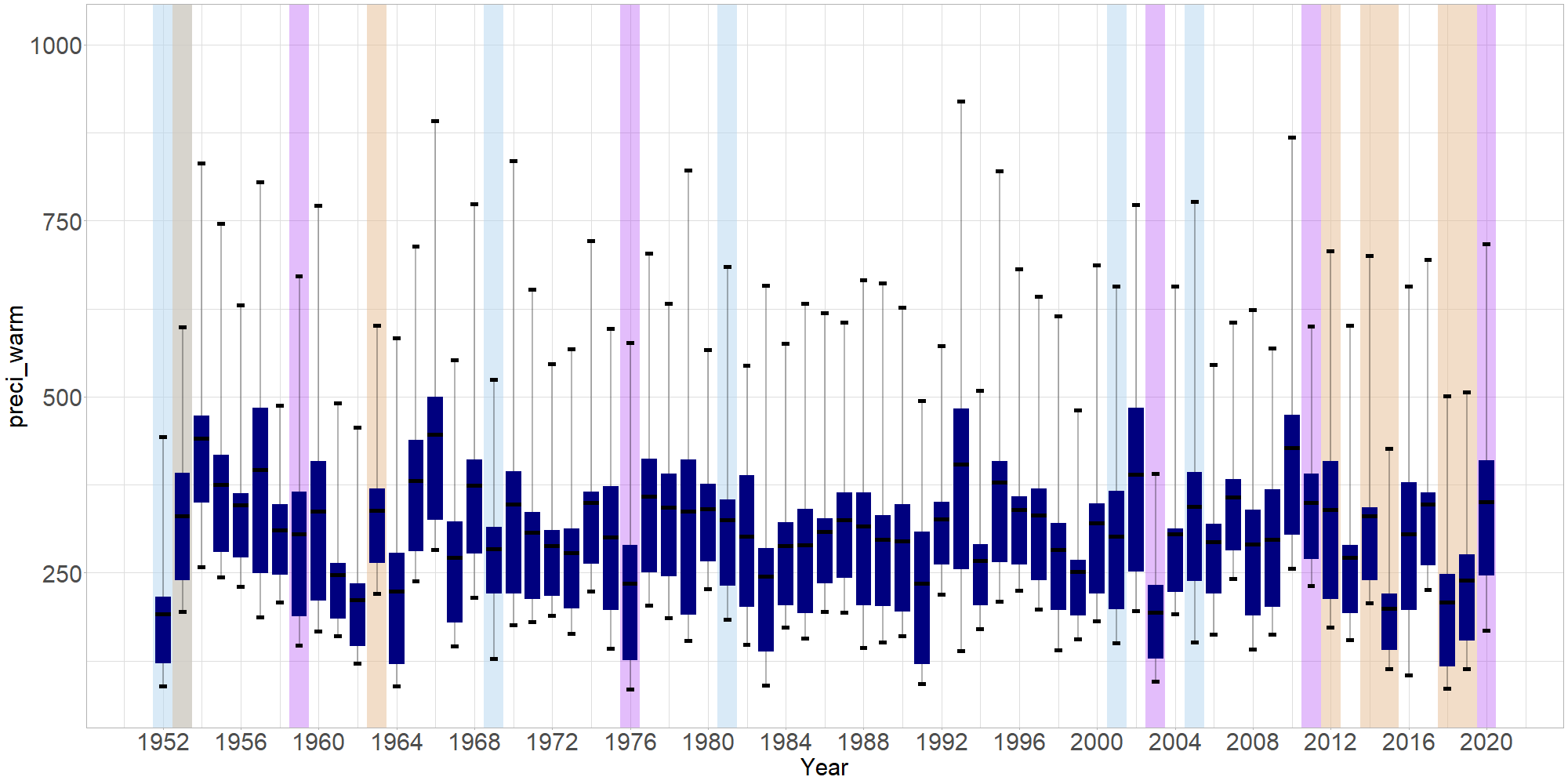}};

			
			\node[inner sep=0pt] (Col1) at (-26, 0)  {\includegraphics[width=0.67\textwidth]{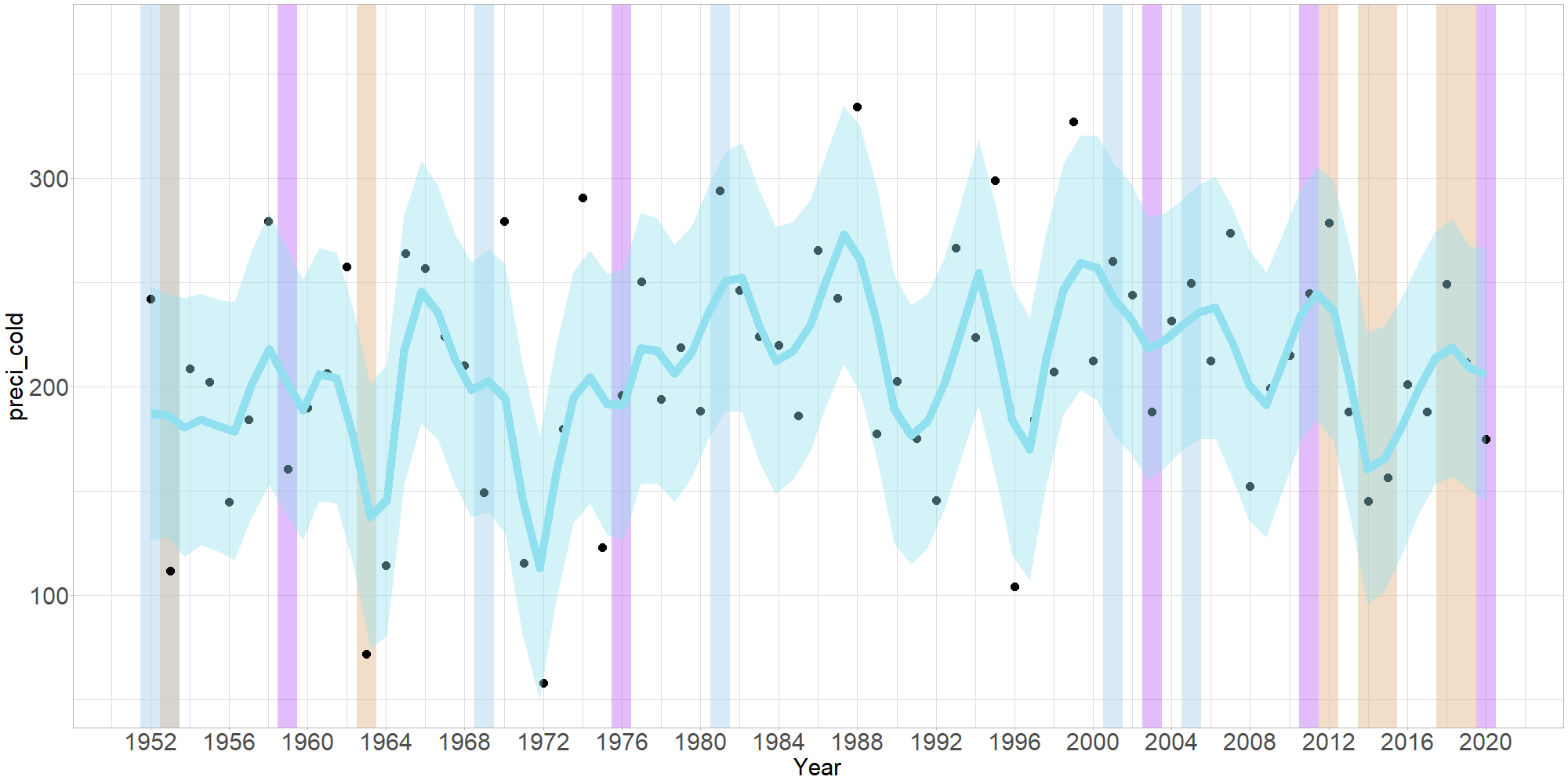}};
			
			\node[inner sep=0pt] (Col1) at (-16, 0)  {\includegraphics[width=0.67\textwidth]{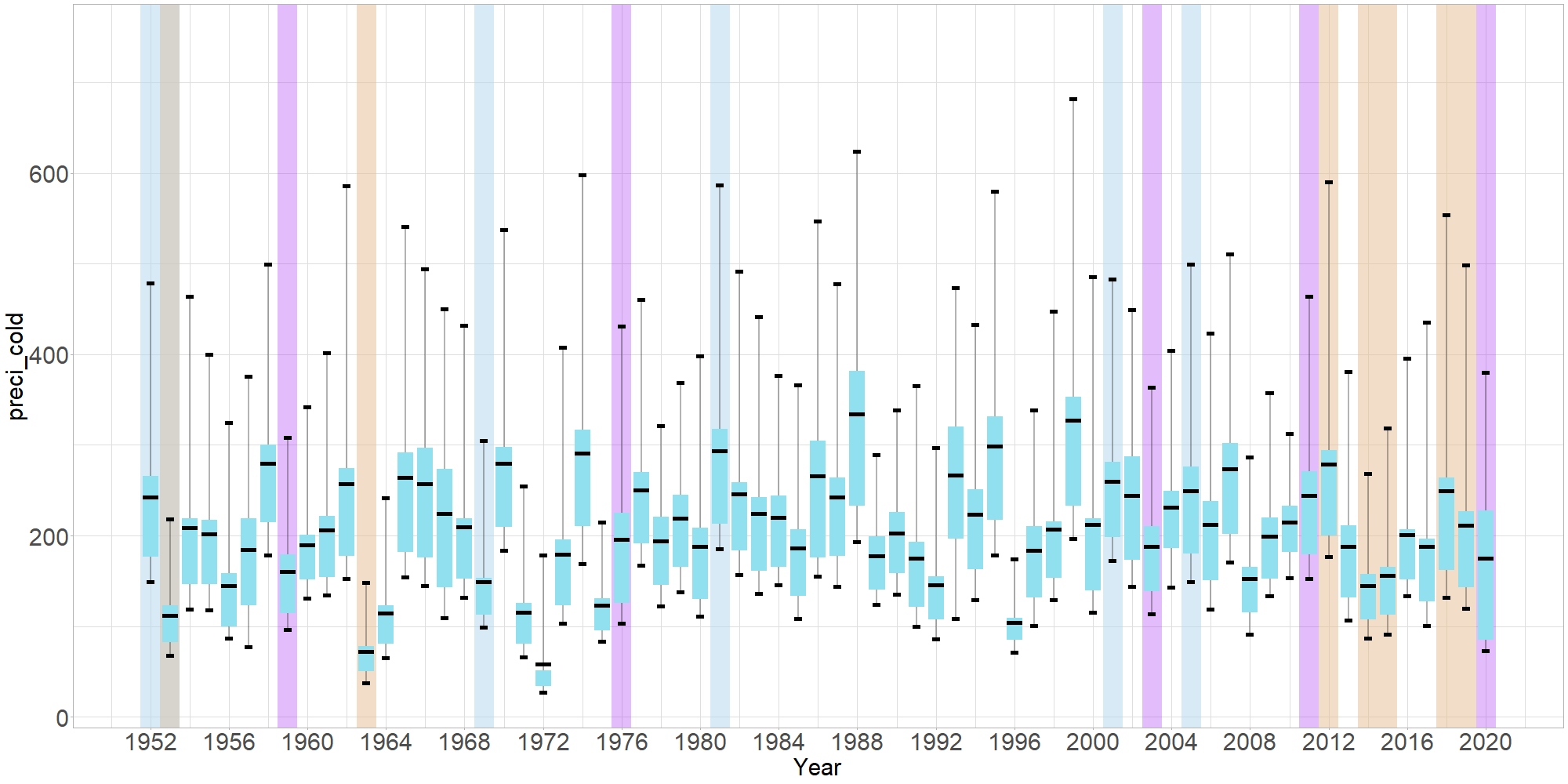}};
			
		\end{tikzpicture}		
	\end{figure}
\end{landscape}	



\clearpage 

\section{Additiuonal model outputs}

\begin{figure}
	\centering
	\begin{tikzpicture}[scale=0.9] 
		
		\node[rotate=90] (a) at (-32, 10)  {Frost};
		
		\node[inner sep=0pt] (Col1) at (-25, 10)  {\includegraphics[width=0.8\textwidth]{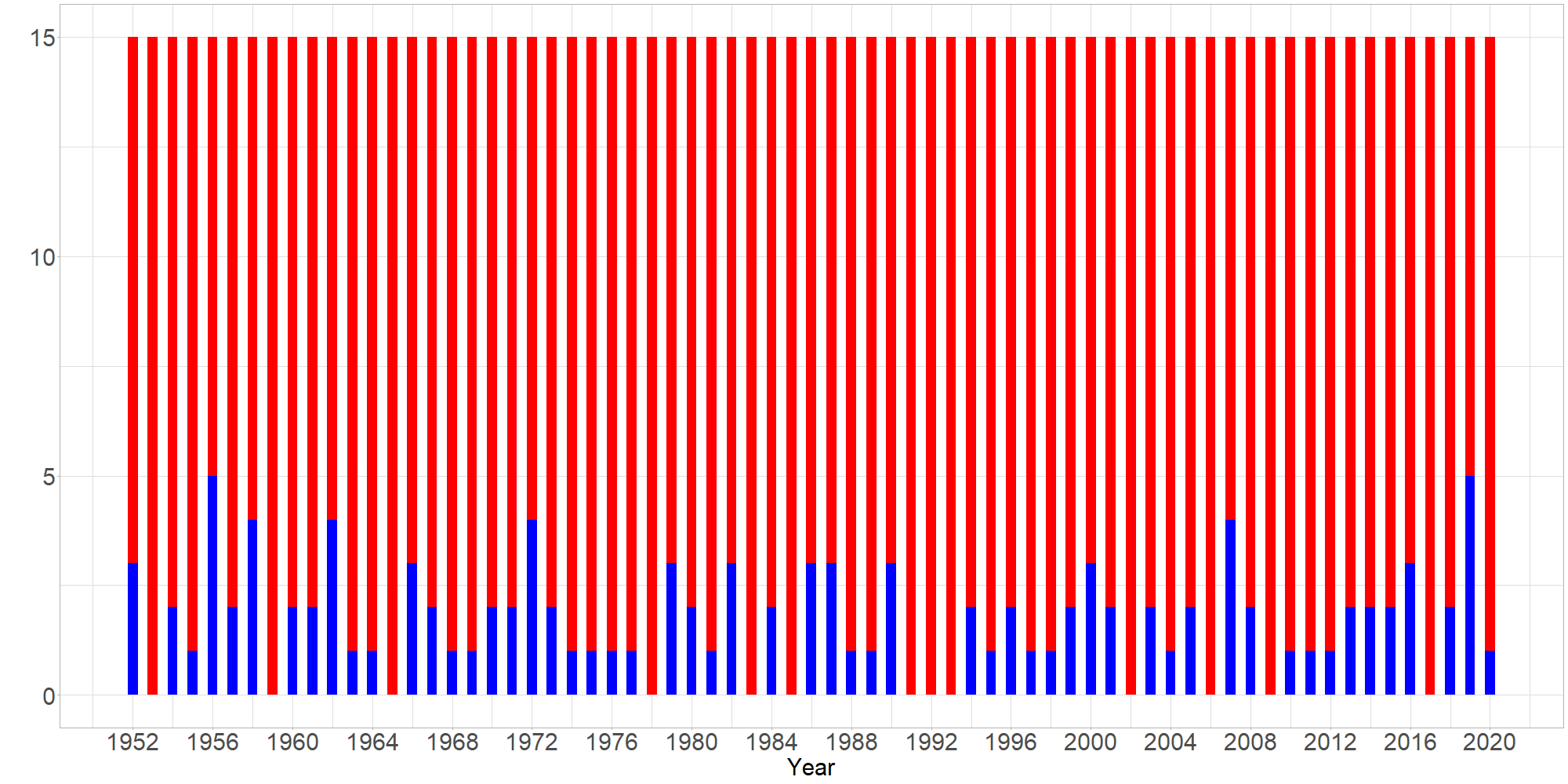}};
		\node[rotate=90] (a) at (-32, 3)  {Drought};
		
		\node[inner sep=0pt] (Col1) at (-25, 3)  {\includegraphics[width=0.8\textwidth]{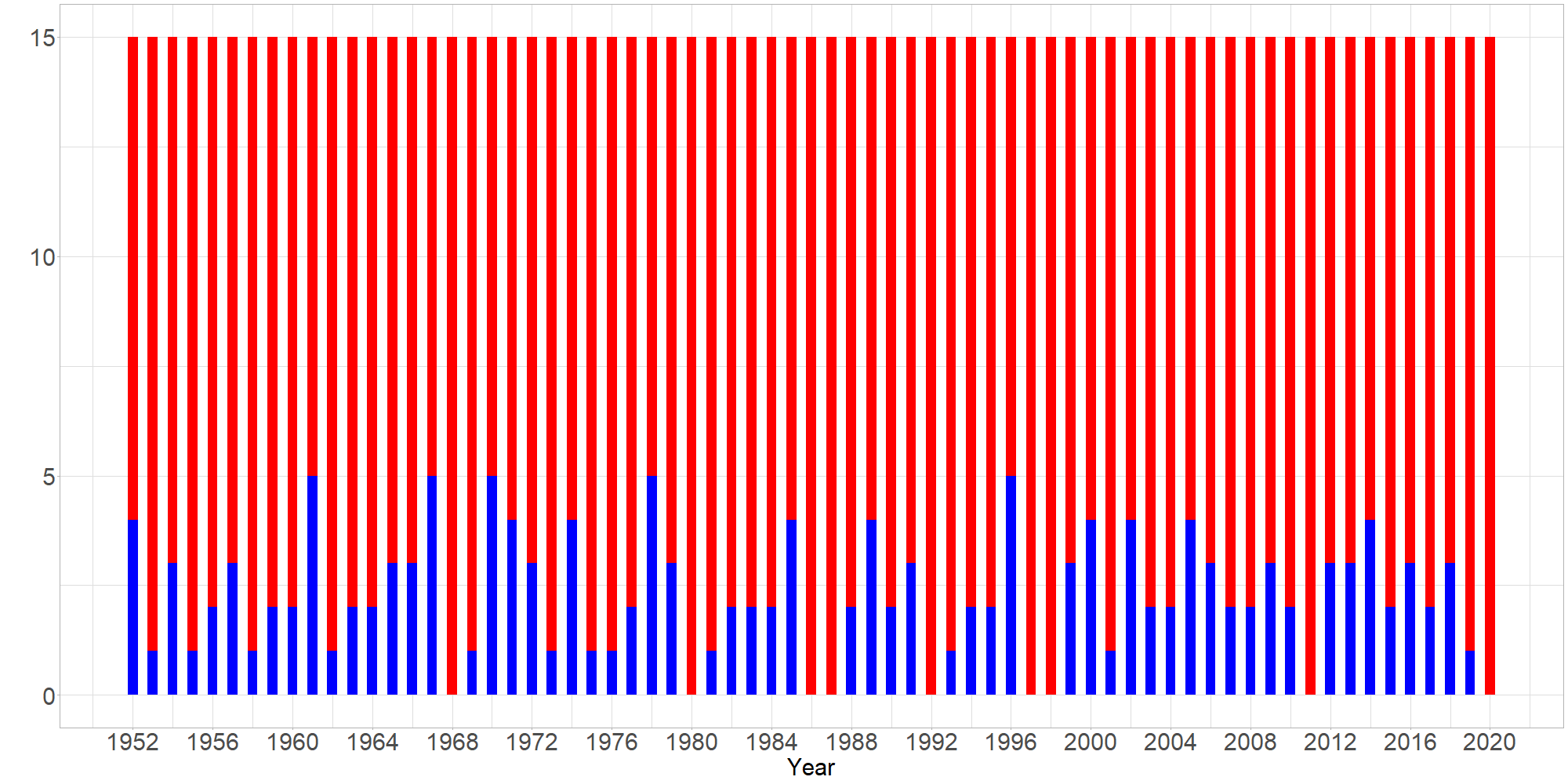}};
		
		\node[rotate=90] (a) at (-32, -4)  {Frost+Drought};
		
		\node[inner sep=0pt] (Col1) at (-25, -4)  {\includegraphics[width=0.8\textwidth]{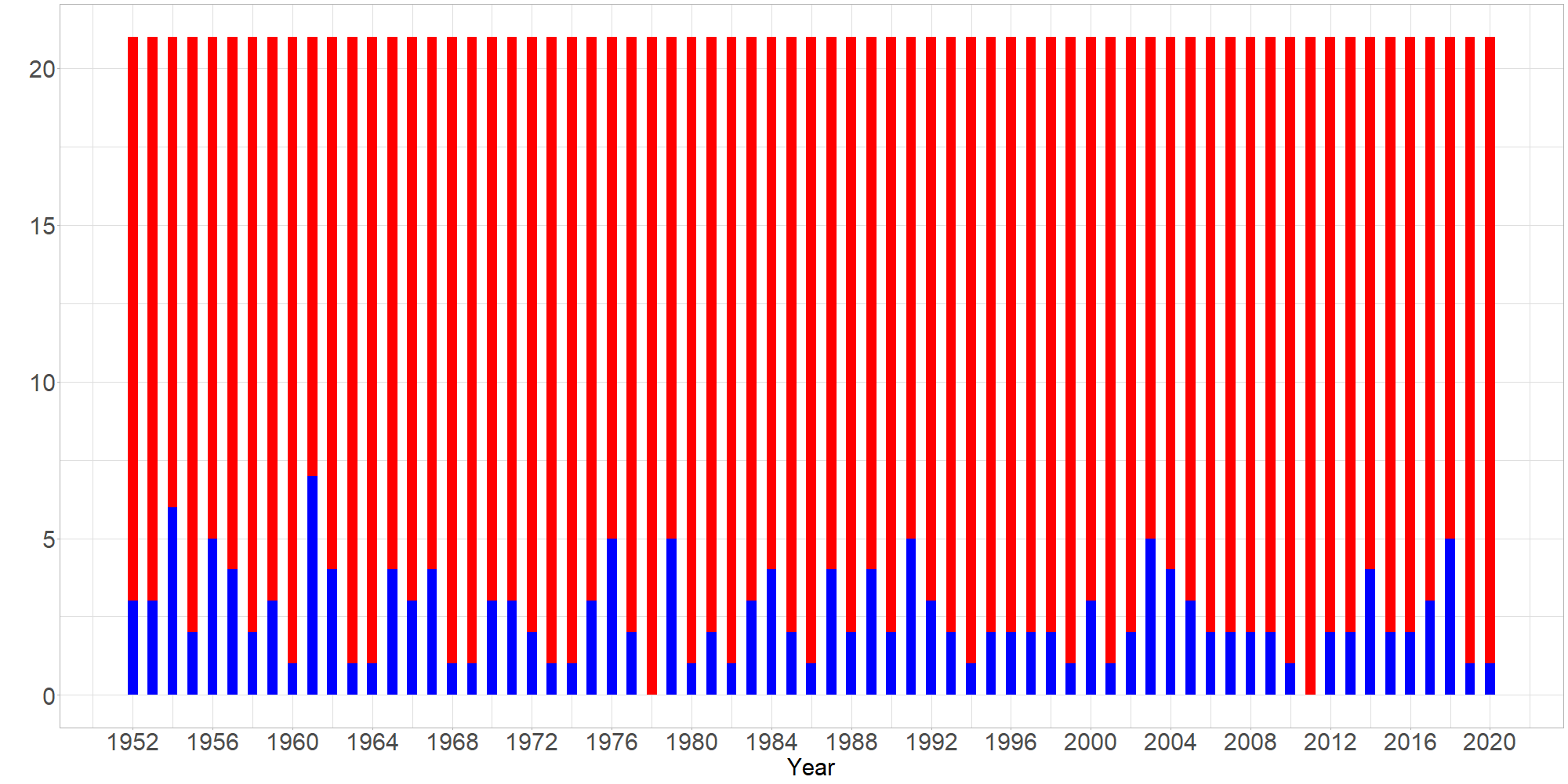}};
		
		\node[inner sep=0pt] (Col1) at (-25, -8)  {\includegraphics[width=0.37\textwidth]{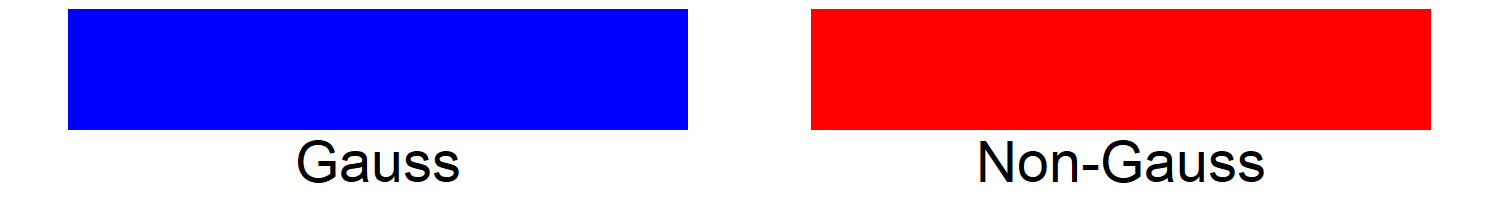}};
	\end{tikzpicture}
\caption{Count of the fitted Gaussian pair copulas, shown in blue and non-Gaussian pair copulas (including rotations), shown in red.}    
\label{climvine:gausscount}
\end{figure}

 In Figure \ref{climvine:gausscount} we show how many times, in each model for each year, Gaussian copula (rotations included) is been fitted and how many time a non-Gaussian copula is fitted (the choices are: Clayton, Gumbel, Frank, Joe, and their rotations as well). 
The non-Gaussian fitted pair copulas are shown with red color, while the Gaussian pair copulas are shown with blue color. We observe that the red color is much more pronounced in all 3 models.

In addition, the \textbf{optimal order} for each model is defined as the order in which the first element of the order corresponds to the predictor that
appeared the most in the first position over the 69 models, then the second element
is defined as the element that appeared the most in the second position among the
elements not chosen as first and so on. The optimal orders for each model are given in Table \ref{otimalordersclimvine}, together with  how many times the chosen predictor is selected to be in a given position for all 5 possible positions in the order. If for some positions in the order 2 predictors appear the same number of times in a particular position, for example in the optimal order for $\hat{\mathcal{D}}_{{frost}_{t}}$  both  \texttt{temp\_range} and  \texttt{temp\_season} appear in the fourth position in the order 7 times each out of the 69 possible orders.
\begin{table}
\caption{\label{otimalordersclimvine} The optimal orders for each model over all years, together with the count of appearances of the predictor in a certain position in the order. }
    \begin{tabular}{|c|r|r|r|r|r|}
 		\hline
		Model & 1 & 2 & 3 & 4 & 5\\
		\hline
		\multirow{2}{*}{$\hat{\mathcal{D}}_{{frost}_{t}}$}  &  temp\_warm(11) &  lon(17) &  lat(10) &  temp\_range,  & preci\_season(7)\\
		& & & &temp\_season (7) &  \\
		\hline
		$\hat{\mathcal{D}}_{{drought}_{t}}$ & lat(14) & lon (11) & preci\_warm(10)  & preci\_wet\_q(7) & elevation(13) \\
		\hline 
		\multirow{2}{*}{$\hat{\mathcal{Y}}_{t}$}	&  lat (17) & lon (14) & preci\_warm,& isotherm(6) & temp\_season (8)\\
		& & &  preci\_wet\_q(6) & &  \\
		\hline	
    \end{tabular}
\end{table}
